\documentclass[a4paper,12pt,twoside]{article}
\usepackage[pdftex,colorlinks,bookmarksnumbered,bookmarksopen,citecolor=red,urlcolor=red]{hyperref}
\usepackage[utf8]{inputenc}
\usepackage{graphicx}
\usepackage{a4wide}
\usepackage{textcomp}
\usepackage{float}
\usepackage{sidecap}
\usepackage[font=small,labelfont=bf]{caption}
\usepackage[export]{adjustbox}
\usepackage{multirow}
\usepackage{tabulary}
\usepackage{siunitx}
\usepackage{ulem}

\usepackage{natbib}

\usepackage[caption = false]{subfig}
\usepackage{amsmath,mathtools,amssymb,amsfonts}
\usepackage[only,llbracket,rrbracket,interleave]{stmaryrd}
\usepackage{gensymb}
\allowdisplaybreaks

\usepackage{booktabs,array,dcolumn}
\newcommand{\PreserveBackslash}[1]{\let\temp=\\#1\let\\=\temp}
\newcolumntype{C}[1]{>{\PreserveBackslash\centering}p{#1}}
\newcolumntype{R}[1]{>{\PreserveBackslash\raggedleft}p{#1}}
\newcolumntype{L}[1]{>{\PreserveBackslash\raggedright}p{#1}}

\usepackage{xcolor}

\newcommand{\hl}[1]{{\color{black} #1\color{black}}}

\newcommand{\bm}[1]{\text{\boldmath $#1$\unboldmath}}
\newcommand{\abs}[1]{\lvert#1\rvert}
\newcommand{\norm}[1]{\lVert#1\rVert}

\newcommand{\bcdot}{\operatorname{\bm{\cdot}}}
\newcommand{\vect}[1]{\bm{#1}}
\newcommand{\mat}[1]{\mathbf{#1}}
\newcommand{\bvect}[1]{\mathbf{#1}}
\newcommand{\bmat}[1]{\mathbf{#1}}

\newcommand{\Div}{{\bm{\nabla}\bcdot\,}}
\newcommand{\Grad}{\bm{\nabla}}
\newcommand{\GradS}{\bm{\nabla^{\texttt{S}}}}

\newcommand{\pd}[2]{\frac{\partial{#1}}{\partial{#2}}}

\newcommand{\RR}{\mathbb{R}}

\newcommand{\eltwo}{\ensuremath{\mathcal{L}_2}}
\newcommand{\elinf}{\ensuremath{\mathcal{L}_\infty}}

\newcommand{\nsd}{\ensuremath{\texttt{n}_{\texttt{sd}}}}

\newcommand{\numel}{\ensuremath{\texttt{n}_{\texttt{el}}}}
\newcommand{\nface}{\ensuremath{\texttt{n}_{\texttt{fa}}}}
\newcommand{\nfaceE}{\ensuremath{\nface^e}}

\newcommand{\tras}{^{{T}}}
\newcommand{\tr}{\operatorname{tr}}
\newcommand{\bu}{\bm{U}}
\newcommand{\bv}{\bm{v}}

\newcommand{\bn}{\bm{n}}

\newcommand{\bbAll}{\bm{B}}
\newcommand{\bb}{\widehat{\bbAll}}

\newcommand{\bhu}{\widehat{\bu}}
\newcommand{\bhv}{\widehat{\bv}}
\newcommand{\bhT}{\widehat{T}}

\newcommand{\bF}{\bm{F}}

\newcommand{\bG}{\bm{G}}

\newcommand{\btau}{\boldsymbol{\tau}}
\newcommand{\bq}{\bm{q}}

\newcommand{\bsigma}{\boldsymbol{\sigma}^d}
\newcommand{\beps}{\boldsymbol{\varepsilon}^{\! d}}
\newcommand{\bphi}{\boldsymbol{\phi}}

\newcommand{\buE}{\bu_{\! e}}

\newcommand{\bepsE}{\beps_{\! e}}
\newcommand{\bphiE}{\bphi_{\! e}}

\newcommand{\bUinf}{\bm{U}_{\! \infty}}
\newcommand{\Tend} {\textrm{t}_{\texttt{end}}}

\newcommand{\hlamax} {\widehat{\lambda}_{\texttt{max}}}
\newcommand{\Id}[1]{\bmat{I}_{#1}}

\newcommand{\dev} {\ensuremath{\mathcal{D}}}

\newcommand{\hu}{\widehat{\bvect{U}}}

\newcommand{\uE}{\bvect{U}_{\! e}}

\newcommand{\Rey}{\small{\textsl{Re}}}
\newcommand{\Pra}{\small{\textsl{Pr}}}
\newcommand{\Ma}{\small{\textsl{M}}}
\newcommand{\vinf}{v_{\! \infty}}
\newcommand{\rhoinf}{\rho_{\! \infty}}
\newcommand{\muinf}{\mu_{\! \infty}}
\newcommand{\Minf}{\Ma_{\! \infty}}

\newcommand{\cp}{c_p}
\newcommand{\cv}{c_v}
\newcommand{\Tref}{\ensuremath{T_{\texttt{ref}}}}

\newcommand{\setAe}{\ensuremath{\mathcal{A}_e}}
\newcommand{\setEe}{\ensuremath{\mathcal{E}_e}}
\newcommand{\setIe}{\ensuremath{\mathcal{I}_e}}

\newcommand{\re}{\bvect{R}_{\! e}}
\newcommand{\hre}{\widehat{\bvect{R}}_{\! e}}
\newcommand{\Ce}{\mat{C}_{\! e}}

\newcommand{\sAngE}{s_e^{\text{a}}}
\newcommand{\sVolE}{s_e^{\text{v}}}
\newcommand{\numbl}{\ensuremath{\texttt{n}_{\texttt{bl}}}}

\usepackage{scalerel}
\usepackage{stackengine}
\stackMath
\def\hatgap{0pt}
\def\subdown{-2pt}
\newcommand\reallywidehat[2][]{
	\renewcommand\stackalignment{l}
	\stackon[\hatgap]{#2}{
		\stretchto{\scalerel*[\widthof{$#2$}]{\kern-.6pt\bigwedge\kern-.6pt}{\rule[-\textheight/2]{1ex}{\textheight}}}{0.5ex}_{\smash{\belowbaseline[\subdown]{\scriptstyle#1}}}}}

\graphicspath{{figsPDF/} {figsPNG/}}
\begin{document}

\title{\vspace{-50pt}Benchmarking the face-centred finite volume method for compressible laminar flows}	

\author{Jordi Vila-P\'erez, Matteo Giacomini and Antonio Huerta}

\author{
	\renewcommand{\thefootnote}{\arabic{footnote}}
	Jordi Vila-P\'erez\footnotemark[1], \
	Matteo Giacomini\footnotemark[2]\textsuperscript{\, ,}\footnotemark[3]\textsuperscript{\, ,}*{\!},\
	Antonio Huerta\footnotemark[2]\textsuperscript{\, ,}\footnotemark[3]
}

\date{}
\maketitle

\renewcommand{\thefootnote}{\arabic{footnote}}
\footnotetext[1]{Department of Aeronautics and Astronautics, Massachusetts Institute of Technology, Cambridge, 02139, Massachusetts, USA.}
\footnotetext[2]{Laboratori de C\`alcul Num\`eric (LaC\`aN), ETS de Ingenier\'ia de Caminos, Canales y Puertos, Universitat Polit\`ecnica de Catalunya, Barcelona, Spain.}
\footnotetext[3]{Centre Internacional de M\`etodes Num\`erics en Enginyeria (CIMNE), Barcelona, Spain.\\[0.5ex]
* Corresponding author: Matteo Giacomini. \textit{E-mail:} \texttt{matteo.giacomini@upc.edu}
}

\vspace{-30pt}

%------------------------------------------------------------------------------------------------------------
%------------------------------------------------------------------------------------------------------------
%------------------------------------------------------------------------------------------------------------
\begin{abstract}
%\vspace{-30pt}
\mbox{}\par\vspace{-\baselineskip}
\paragraph*{Purpose -} 
This study aims to assess the robustness and accuracy of the face-centred finite volume (FCFV) method for the simulation of compressible laminar flows in different regimes, using numerical benchmarks.
\mbox{}\par\vspace{-\baselineskip}
\paragraph*{Design/methodology/approach -}
The work presents a detailed comparison with reference solutions published in the literature --when available-- and numerical results computed using a commercial cell-centred finite volume software.
\mbox{}\par\vspace{-\baselineskip}
\paragraph*{Findings -}
The FCFV scheme provides first-order accurate approximations of the viscous stress tensor and the heat flux, insensitively to cell distortion or stretching. The strategy demonstrates its efficiency in inviscid and viscous flows, for a wide range of Mach numbers, also in the incompressible limit. In purely inviscid flows, non-oscillatory approximations are obtained in the presence of shock waves. In the incompressible limit, accurate solutions are computed without pressure correction algorithms. The method shows its superior performance for viscous high Mach number flows, achieving physically admissible solutions without carbuncle effect and predictions of quantities of interest with errors below $5 \%$. 
\mbox{}\par\vspace{-\baselineskip}
\paragraph*{Originality/value -}
The FCFV method accurately evaluates, for a wide range of compressible laminar flows, quantities of engineering interest, such as drag, lift and heat transfer coefficients, on unstructured meshes featuring distorted and highly stretched cells, with an aspect ratio up to ten thousand. The method is suitable to simulate industrial flows on complex geometries,  relaxing the requirements on mesh quality introduced by existing finite volume solvers and alleviating the need for time-consuming manual procedures for mesh generation to be performed by specialised technicians. 
\mbox{}\par\vspace{-\baselineskip}
\paragraph*{Keywords -}
CFD, finite volume, face-centred, Ansys Fluent, numerical benchmarks, compressible flows
\end{abstract}

%------------------------------------------------------------------------------------------------------------
%------------------------------------------------------------------------------------------------------------
%------------------------------------------------------------------------------------------------------------
\section{Introduction}
\label{sc:Introduction}

Despite the growing interest towards high-order methods for computational fluid dynamics (CFD)~\citep{NASA-HO-14,HighOrder-Review-13,Vincent-HWV-14,Abgrall-AR-17}, finite volume (FV) approaches still represent the \emph{de facto} standard in the simulation of industrial flows, aerodynamics, heat and mass transfer problems~\citep{Chalot-17}. Such a success, stemming from the inherent conservation properties of these techniques, their suitability for a wide range of flow regimes and their efficiency in simulating large-scale engineering systems, is testified by their widespread implementation in commercial, industrial and open-source software, including Ansys Fluent~\citep{FluentManual}, OpenFOAM~\citep{jasak2009openfoam}, SU2~\citep{SU2-manual}, NASA CFL3D~\citep{CFL3D:06}, NASA FUN3D~\citep{FUN3D:19},  FLITE~\citep{Morgan1991,sorensen2003a,sorensen2003b} and TAU~\citep{gerhold2005overview}, just to name a few.

The FV rationale relies on an integral formulation of the partial differential equations under analysis, obtained via the appropriate positioning of the unknowns in the computational mesh and the definition of suitable inter-cell fluxes to transfer the information among neighbouring cells.  The position of the unknowns yields a first classification of FV approaches into the vertex-centred finite volume (VCFV) method, which defines the degrees of freedom of the solution at the mesh nodes, and the cell-centred finite volume (CCFV) rationale, locating the unknowns at the centroid of each cell~\citep{Eymard2000,Morton2007,Leveque2013,Barth-BHO-17,Cardiff-CD-21}. More recently,  a new FV paradigm, the so-called face-centred finite volume (FCFV) method, was introduced by positioning the unknowns of the system at the barycentre of each face and eliminating all the degrees of freedom within the cells by means of a hybridisation procedure~\citep{RS-SGH:2018_FCFV1,RS-SGH:2019_FCFV2,RS-VGSH:20,MG-RS-20,VilaPerez2022}.  To transmit the necessary information across the interface between two neighbouring cells, suitable numerical fluxes are then defined based on approximate Riemann solvers~\citep{LeVeque-92,Toro2009,Hesthaven-17}.

One of the most appealing properties of the FCFV paradigm is its capability to handle general unstructured meshes featuring different cell types --triangles and quadrilaterals in 2D; tetrahedra, hexahedra, prisms and pyramids in 3D--, with possibly distorted and highly stretched cells. Such flexibility is especially relevant in the simulation of industrial problems with complex geometries for which traditional FV schemes require dedicated mesh generation procedures, particularly expensive in terms of man-hours of specialised technicians,  to achieve grids with sufficient quality to be suitable for CFD simulations. Nonetheless, in order to devise a competitive FV solver, robustness and accuracy in different flow conditions are also critical. Hence, in this work, an extensive numerical study is presented to benchmark the approximation properties of the FCFV method in the simulation of compressible laminar flows.  Of course,  a thorough examination of the performance of the method in the presence of turbulent effects is crucial for industrial applications but it lies outside the scope of this contribution. Indeed, this work presents a detailed comparison with existing reference solutions published in the literature and the results provided by Ansys Fluent CCFV solvers, with particular attention to the robustness and the accuracy of the FCFV scheme in five challenging problems, involving:
\renewcommand{\theenumi}{P\arabic{enumi}}
\begin{enumerate}\itemsep 0ex
\item convergence properties in the presence of distorted cells;
\item viscous laminar flows, with highly stretched meshes in the boundary layer region;
\item purely inviscid flows, possibly featuring discontinuous solutions; 
\item nearly incompressible viscous laminar flows, with velocity-pressure coupling effects;
\item viscous laminar flows at high Mach number, with strong bow shocks.
\end{enumerate}
The above cases are devised to address the most common difficulties faced by FV solvers in the context of flow problems and to submit the FCFV method to a stress test in order to evaluate its performance in a wide variety of scenarios.  Both qualitative analyses of the physical variables and quantitative computations of aerodynamic coefficients are presented.
In particular, test P1 stems from the observation that traditional VCFV and CCFV methods achieve second-order accurate approximations by means of a flux reconstruction procedure. Nonetheless, this step is strongly dependent on the quality of the mesh and the accuracy of both approaches is known to deteriorate in the presence of distorted cells, leading to suboptimal approximations of the gradient of the solution~\citep{Diskin2010,Diskin2011}. On the contrary,  the FCFV method provides first-order accuracy of the gradient of the solution without any reconstruction step and it is thus insensitive to cell distortion. Test P2 further extends this mesh sensitivity analysis by considering a set of viscous laminar flows where highly stretched cells, with an aspect ratio between $10^3$ and $10^4$, naturally arise in the discretisation of the boundary layer region.
In addition, purely inviscid flows ranging from subsonic to supersonic conditions (test P3) are explored to assess the capability of the FCFV method to represent both smooth and discontinuous solutions.  The main difficulty in these problems is to maintain physically admissible solutions with positivity-preserving techniques and to avoid nonphysical oscillations in the vicinity of steep gradients and discontinuities. This is particularly challenging in the absence of the regularisation effect of the viscous terms of the Navier-Stokes equations and the FCFV method achieves such a result without resorting to artificial viscosity or slope limiters techniques~\citep{Cockburn-CS-89,Persson-PP:2006,Casoni-HCP-12}.
Furthermore, test P4 analyses the low Mach number case and the challenge to construct stable approximations of the velocity-pressure coupling arising in nearly incompressible flows. Contrary to traditional VCFV and CCFV schemes, the FCFV paradigm is locking-free in the incompressible limit and it does not require the introduction of specific pressure correction algorithms such as the semi-implicit method for pressure linked equations (SIMPLE)~\citep{Patankar1972} or the pressure-implicit splitting operator (PISO)~\citep{Issa1986}.
Finally, the robustness of the FCFV method in the presence of strong bow shocks is studied for high Mach number flows (test P5), where traditional FV schemes tend to suffer from a loss of accuracy due to the carbuncle phenomenon and the resulting solution is extremely sensitive to the quality of the computational mesh~\citep{Elling2009,Kitamura2012}.

The article is organised as follows. Section~\ref{sc:equations} introduces the compressible Navier-Stokes equations and section~\ref{sc:FCFV} briefly reviews the formulation of the FCFV approximation for compressible flows. In section~\ref{sc:Results},  an extensive set of benchmark tests is presented to address the five challenging problems identified above. Special emphasis is devoted to the comparison of the FCFV outcomes with reference solutions available in the literature as well as the CCFV results yielded by the commercial CFD software Ansys Fluent.  Finally,  the conclusions of this study are summarised in section~\ref{sc:Conclusions}.

%------------------------------------------------------------------------------------------------------------
%------------------------------------------------------------------------------------------------------------
%------------------------------------------------------------------------------------------------------------
\section{Compressible Navier-Stokes equations}
\label{sc:equations}

Let $\Omega \subset \RR^{\nsd}$ be an open, bounded and connected domain in $\nsd$ dimensions, with boundary $\partial \Omega$. Let $\Tend>0$ be a final time of interest. The unsteady nondimensional compressible Navier-Stokes equations, expressed in conservation form, are given by
\begin{align} 	\label{eq:NScompact}
	\pd{\bu}{t} + \Div \left( \bF (\bu)- \bG (\bu, \Grad \bu) \right) = \vect{0} \qquad  \text{in } \Omega \times (0,\Tend],
\end{align}
with appropriate boundary and initial conditions. Here, $\bu \in \RR^{\nsd + 2}$ is the vector of conserved quantities and $\bF$ and $\bG$ are the advection and diffusion flux tensors, respectively, given by
\begin{equation}\label{eq:NSterms}
	\bu = \begin{Bmatrix} \rho\\ \rho\bv\\ \rho E \end{Bmatrix}\!,
	\quad
	\bF (\bu) = \begin{bmatrix} \rho \bv\tras\\
		\rho\bv \otimes\bv + p \Id{\nsd}\\
		(\rho E + p) \bv\tras \end{bmatrix} \!,
	\quad
	\bG (\bu, \Grad \bu)= \begin{bmatrix} \vect{0}\\
		\bsigma \\
		(\bsigma \bv + \bq )\tras \end{bmatrix}.
\end{equation}
In these expressions, $\rho$ denotes the density, $\bv$ the velocity field, $E$ the total specific energy, $p$ the pressure, $\bsigma$ the viscous stress tensor and $\bq$ the heat flux vector, whereas $\Id{\nsd}$ stands for the $\nsd$-dimensional identity matrix.

Under the assumption of ideal gas,  the relation $\gamma p  = (\gamma - 1) \rho T$ stands,  $T$ being the temperature field and $\gamma = \cp / \cv$ being the ratio of specific heats at constant pressure, $\cp$, and constant volume, $\cv$. \hl{In particular, $\gamma = 1.4$ is selected for all test cases in this work, to account for a balanced diatomic gas mixture such as air.} For a calorically perfect gas, it also holds that $p = (\gamma - 1) \rho \left( E -  \norm{\bv}^2/2 \right)$. In addition, for a Newtonian fluid under Stokes' hypothesis and employing Fourier's law of heat conduction, the viscous stress tensor $\bsigma$ and the heat flux vector $\bq$ are given by
\begin{equation}\label{eq:StressTensorHeatFlux}
	\bsigma = \frac{\mu}{\Rey} \left(2 \GradS \bv - \frac{2}{3} (\Div \bv) \Id{\nsd} \right), \quad \quad \bq = \frac{\mu}{\Pra \Rey} \Grad T,
\end{equation}
where $\mu$ is the dynamic viscosity and $\GradS := (\Grad + \Grad^T)/2$ denotes the symmetric part of the gradient operator. Moreover, the nondimensional Mach, Reynolds and Prandtl numbers are defined as
\begin{equation} \label{eq:non-dimensionalQuantities}
	\Minf = \frac{\vinf}{c_\infty}, \qquad \Rey = \frac{\rhoinf \vinf L}{\muinf}, \qquad \Pra = \frac{\cp \muinf}{\kappa},
\end{equation}
where $c = \sqrt{\gamma p/\rho}$ stands for the speed of sound, $L$ is a characteristic length, $\kappa$ refers to the thermal conductivity and \hl{the subscript $\infty$ denotes the free-stream reference values.  In particular,  $\Pra=0.71$ is selected for air at normal conditions of temperature and pressure~\citep{Schlichting-BL:2016}. This setup is employed to reproduce relevant benchmarks of compressible laminar flows published in the literature and featuring the aforementioned value of the Prandtl number (cf. section~\ref{sc:Results}).}
Finally,  the Sutherland's law is employed to model the nondimensional dynamic viscosity, namely
\begin{equation}\label{eq:SutherlandLaw}
	\mu = \left(\frac{T}{T_\infty}\right)^{3/2} \frac{T_\infty + S}{T + S},
\end{equation}
$S = S_0 T_\infty/\Tref$ being the Sutherland's constant, with $S_0 = \SI{110}{\kelvin}$ and $\Tref = \SI{273}{\kelvin}$, whereas $T_\infty = 1/\left[(\gamma - 1) \Minf^2\right]$ denotes the nondimensional free-stream temperature.

%------------------------------------------------------------------------------------------------------------
%------------------------------------------------------------------------------------------------------------
%------------------------------------------------------------------------------------------------------------
\section{FCFV approximation of compressible flows}
\label{sc:FCFV}

The FCFV method constructs an approximation of the compressible Navier-Stokes equations using a mixed formulation.  Introducing two additional \emph{mixed} variables, namely, the deviatoric strain rate tensor $\beps$ and the gradient of temperature $\bphi$, equation~\eqref{eq:NScompact} is rewritten as the system of first-order equations
\begin{subequations} \label{eq:mixedVariables} 
	\begin{align}
		\beps - \left(2 \GradS \bv - \frac{2}{3} (\Div \bv) \Id{\nsd}\right)  &= \bm{0} & \text{in } &\Omega \times (0,\Tend],  \label{eq:mixedVariablesA} \\
		\bphi - \Grad T &= \bm{0} & \text{in } &\Omega \times (0,\Tend],  \\
		\pd{\bu}{t} +\Div \left( \bF (\bu) - \bG (\bu, \beps, \bphi) \right) &= \bm{0}  & \text{in } &\Omega \times (0,\Tend].
	\end{align}
\end{subequations}

It is worth noticing that the viscous stress tensor and the heat flux vector can be easily retrieved from the above mixed variables through simple linear expressions, that is 
	\begin{equation} \label{eq:ViscousTermsNewMixed} 
		\bsigma = \frac{\mu}{\Rey} \beps, \qquad
		\bq = \frac{\mu}{\Rey \Pra} \bphi.
	\end{equation}

For the sake of readability,  equation~\eqref{eq:mixedVariablesA} will be henceforth represented as $\beps = \dev \GradS \bv$, with the linear operator $\dev$ defined as
	\begin{equation} \label{eq:deviatoric}
		\dev \bm{W}  = \left( \bm{W} + \bm{W}\tras \right) - \frac{2}{3} \tr (\bm{W}) \Id{}.
	\end{equation}
From equation~\eqref{eq:deviatoric}, it follows that the deviatoric strain rate tensor can thus be expressed as a function of the symmetric part of the gradient of the velocity. Moreover, Voigt notation is employed to store only the non-redundant components of the second-order tensor as detailed in~\citep{VilaPerez2022,Sevilla-SGKH:2018,Giacomini-GKSH:2018,Tutorial-GSH:2020,MG-SGH:20,AlS-SKGWH:20,JVP_HDG-VGSH:20}.

%------------------------------------------------------------------------------------------------------------
\subsection{Integral formulation}

A partition of the domain $\Omega$ in $\numel$ disjoint cells $\Omega_e$, namely $\Omega = \bigcup_{e=1}^{\numel} \Omega_e$, is considered for the discretisation. The boundary $\partial \Omega_e$ of the cell $\Omega_e$ is obtained as the union of its $\nfaceE$ faces $\Gamma_{e\!, j}$, that is, $\partial \Omega_e :=\bigcup_{j=1}^{\nfaceE} \Gamma_{e\!, j}$. In addition, the internal interface $\Gamma$ is defined as $\Gamma:=\left[ \bigcup_{e=1}^{\numel}\partial\Omega_e \right] \setminus \partial \Omega$.

The FCFV solver is devised in two stages~\citep{RS-SGH:2018_FCFV1} introducing an additional \emph{hybrid} unknown $\bhu$, which denotes the vector of conservative variables on the mesh faces $\Gamma \cup \partial \Omega$. In the first step,  the system of equations~\eqref{eq:mixedVariables} is discretised using a constant approximation of the conservative and mixed variables $(\bu,\beps,\bphi)$ in each cell $\Omega_e$, $e = 1,\dotsc,\numel$, to express them as functions of $\bhu$. To this end, given the initial condition $\bu = \bu^0$ at time $t=0$ and the boundary condition $\bu = \bhu$ on $\partial\Omega_e$,  the divergence theorem is applied to the system of equations~\eqref{eq:mixedVariables} yielding the integral form of the FCFV local problem
\begin{subequations} \label{eq:IntegralLocal}
  \begin{align}
    \int_{\Omega_e} \beps \, d\Omega- \int_{\partial \Omega_e} \dev \bhv \otimes \bn \, d\Gamma &= \vect{0}, \label{eq:IntegralLocal_eps}\\
    \int_{\Omega_e} \bphi \, d\Omega - \int_{\partial \Omega_e} \bhT \bn \, d\Gamma &= \vect{0}, \label{eq:IntegralLocal_phi}\\
    \int_{\Omega_e} \pd{\bu}{t}  \, d\Omega  + \int_{\partial \Omega_e}  \left( \reallywidehat{ \bF(\bu) \bn} - \reallywidehat{\bG(\bu, \beps, \bphi) \bn} \right) \, d\Gamma &= \vect{0}, \label{eq:IntegralLocal_U}
  \end{align}
\end{subequations}
where $\bn$ is the outward normal vector to the cell face. In the above equations,  the velocity $\bhv$ and the temperature $\bhT$ on the cell faces forming $\partial \Omega_e$ are computed from the vector $\bhu$ of the conservative variables on the faces. Moreover,  $\reallywidehat{ \bF \bn}$ and $\reallywidehat{\bG \bn}$ denote the convection and the diffusion numerical fluxes, respectively.

The solution of equations~\eqref{eq:IntegralLocal} allows to eliminate the cell unknowns associated with $(\bu,\beps,\bphi)$ by expressing them in terms of the face unknowns $\bhu$. The second step of the FCFV method requires the computation of $\bhu$, which is discretised employing a constant value at the barycentre of the faces. The FCFV global problem thus prescribes the continuity of the normal fluxes on the internal faces $\Gamma$ and the boundary conditions on $\partial \Omega$,  namely
\begin{equation} \label{eq:IntegralGlobal}
	\sum_{e = 1}^{\numel} \left\lbrace \int_{\partial \Omega_e \setminus \partial \Omega}  \left( \reallywidehat{ \bF(\bu) \bn} - \reallywidehat{\bG(\bu, \beps, \bphi) \bn}  \right) \, d\Gamma + \int_{\partial \Omega_e \cap \partial \Omega} \bb(\bu,\bhu,\beps,\bphi) \, d\Gamma \right\rbrace = \vect{0},
\end{equation}
$\bb$ being the trace operator imposing the boundary conditions on $\partial \Omega$.

%------------------------------------------------------------------------------------------------------------
\subsubsection{Inter-cell numerical fluxes}

The traces of the FCFV convective and diffusive inter-cell numerical fluxes~\citep{RS-SGH:2018_FCFV1,RS-SGH:2019_FCFV2,RS-VGSH:20,MG-RS-20,VilaPerez2022} are expressed as
\begin{subequations} \label{eq:NumericalFluxes}
	\begin{align}
		\reallywidehat{ \bF(\bu) \bn} &=  \bF(\bhu) \bn + \btau^a(\bhu) \, (\bu - \bhu), \label{eq:fluxF}\\
		\reallywidehat{\bG(\bu, \beps, \bphi) \bn} &= \bG(\bhu, \beps, \bphi) \bn - \btau^d \, (\bu - \bhu) ,\label{eq:fluxG}
	\end{align}
\end{subequations}
where $ \btau^a$ and $ \btau^d$ denote the stabilisation tensors associated with the convective and the viscous effects, respectively.

Different approximate Riemann solvers can be devised for the FCFV method depending on the definition of $ \btau^a$~\citep{JVP_HDG-VGSH:20}. In particular, the HLL~\citep{Harten-HLL:1983} and HLLEM~\citep{Einfeldt1988,Einfeldt1991} Riemann solvers employed in this study yield the stabilisation tensors
\begin{equation}\label{eq:RS-HLLtype}
	\btau^a_{\text{HLL}} = s^+ \bmat{I}_{\nsd + 2}, \qquad \qquad \btau^a_{\text{HLLEM}} = s^+ \bm{\theta} (\bhu),
\end{equation}
where $s^+ := \max (0,\bhv\cdot \bn + \widehat{c})$ is an estimate of the largest wave speed of the Riemann problem and the matrix $\bm{\theta}(\bhu) = \mat{R} \bm{\Theta} \mat{R}^{-1}$ is constructed starting from the spectral decomposition of the Jacobian $\bm{A}_n(\bhu):=[\partial \bF(\bhu) / \partial \bhu] \bn$ of the convective fluxes in the normal direction to the cell face. In this context, $ \mat{R}$ denotes the matrix of the right eigenvectors of $\bm{A}_n(\bhu) = \mat{R} \mat{\Lambda} \mat{R}^{-1}$,  $\mat{\Lambda}$ stands for the diagonal matrix of the corresponding eigenvalues, whereas $\bm{\Theta}$ is the diagonal matrix given by $\bm{\Theta} = \text{diag}\left( 1, \widehat{\theta} \bmat{1}_{\nsd} , 1 \right)$, with $\widehat{\theta} =  \abs{\bhv \cdot \bn}/\hlamax$~\citep{Rohde2001}.
Finally, the stabilisation tensor $\btau^d$ of the viscous fluxes is defined as the diagonal matrix
\begin{equation} \label{eq:diffStabilisation}
	\btau^d = \frac{1}{\Rey}  \, \text{diag} \left(  0,  \vect{1}_{\nsd}, \frac{1}{(\gamma - 1) \Minf^2 \Pra} \right) ,
\end{equation}
where $\bmat{1}_{\nsd}$ is a vector of ones of dimension $\nsd$.

%------------------------------------------------------------------------------------------------------------
\subsubsection{Imposition of the boundary conditions}

The trace boundary operator $\bb(\bu,\bhu,\beps,\bphi)$ is responsible for introducing appropriate boundary conditions in the integral equation~\eqref{eq:IntegralGlobal}. Equation~\eqref{eq:boundaryConditions} details the expression of this operator for the most common types of boundary conditions in aerodynamic applications involving compressible flows, namely
\begin{subequations} \label{eq:boundaryConditions}
	\begin{align}
		\bb &= \bm{A}_n^+(\bhu) (\bu - \bhu) + \bm{A}_n^-(\bhu) (\bUinf - \bhu),  & \text{(far-field boundary)},  \label{eq:boundaryConditionsA}\\
		\bb &= \lbrace \rho, \, \rho \bv\tras,\,  p_{\text{out}}/(\gamma - 1) + \rho \norm{\bv}^2/2 \rbrace\tras - \bhu\tras, & \text{(subsonic/pressure outlet)},  \\
		\bb &= \lbrace \rho - \widehat{\rho},\,  \widehat{\rho \bv}\tras,\,  \widehat{\kappa} \bphi \bn - \tau^d_{\rho E} (\rho E - \widehat{\rho E}) \rbrace\tras, & \text{(adiabatic wall)},  \label{eq:boundaryConditionsC}\\
		\bb &= \lbrace \rho,\,  \bm{0}_{\nsd}\tras,\,  \rho T_{\text{w}}/\gamma \rbrace\tras - \bhu\tras, & \text{(isothermal wall)},  \\
		\bb &= \lbrace \rho,\,  \left[ (\Id{\nsd} - \bn\otimes \bn) \rho \bv \right]\tras,\,  \rho E \rbrace\tras - \bhu\tras, & \text{(inviscid/symmetry wall)}.
	\end{align}
\end{subequations}

In these expressions, far-field boundary conditions (also used for subsonic inlets, supersonic inlets and supersonic outlets) exploit the spectral decomposition of the matrix $\bm{A}_n(\bhu)$ via its positive and negative characteristics given by $\bm{A}_n^\pm := (\bm{A}_n \pm \abs{\bm{A}_n})/2$, see equation~\eqref{eq:boundaryConditionsA}. In addition,  $\bUinf$,  $p_{\text{out}}$ and $T_{\text{w}}$ represent the free-stream data for the conservative variables, the outlet pressure and the wall temperature, respectively, whereas $\bm{0}_{\nsd}$ stands for an $\nsd$-dimensional vector of zeros. Finally, for the adiabatic wall in equation~\eqref{eq:boundaryConditionsC},  the thermal conductivity $\widehat{\kappa} = \widehat{\mu}/(\Rey \Pra)$ is computed as a function of the hybrid state $\bhu$ and the stabilisation coefficient $\tau^d_{\rho E}$ for the energy equation is given by $\tau^d_{\rho E} = 1/\left[ \Rey (\gamma - 1) \Minf^2 \Pra \right]$.

%------------------------------------------------------------------------------------------------------------
\subsection{The FCFV solver}
\label{ssc:FCFVdiscretisation}

To construct the FCFV solver for the compressible Navier-Stokes equations,  the integral equations~\eqref{eq:IntegralLocal} and~\eqref{eq:IntegralGlobal} are approximated using a quadrature rule employing one integration point per cell/face.

Let $\setAe$ be the set of all the faces of cell $\Omega_e$,  $\setIe$ the set of its internal faces and $\setEe$ the set of its external faces, that is, 
\begin{equation}
	\setAe := \lbrace 1,\dotsc,\nfaceE \rbrace, \qquad \setIe := \lbrace j \in \setAe \mid \Gamma_{e\!, j} \cap \Gamma \neq \emptyset \rbrace , \qquad \setEe := \setAe \setminus \setIe.
\end{equation}
Furthermore,  denote by $\chi_{\setIe}$ and $\chi_{\setEe}$ the indicator functions associated with the sets $\setIe$ and $\setEe$, respectively.

From~\eqref{eq:IntegralLocal}, the local problems of the FCFV solver yield: given the initial condition $\buE = \buE^0$ at $t=0$ and the trace variable $\bhu_{\! j}$ on the faces $\Gamma_{e\!, j}, \ j=1,\ldots,\nfaceE$, compute $(\buE,\bepsE,\bphiE)$ in each cell $\Omega_e$, $e = 1,\dotsc,\numel$ such that
\begin{subequations} \label{eq:FCFVLocal}
	\begin{align}
		&\abs{\Omega_e} \bepsE =  \sum_{j \in \setAe} \abs{\Gamma_{e\!, j}} \dev \widehat{\bv}_{\! j} \otimes \bn_j , \label{eq:FCFVLocal_eps}\\
		&\abs{\Omega_e} \bphiE = \sum_{j \in \setAe} \abs{\Gamma_{e\!, j}} \bhT_{\! j} \bn_j , \label{eq:FCFVLocal_phi}\\
		& \begin{aligned} \int_{\Omega_e} \pd{\buE}{t}  \, d\Omega +  \sum_{j \in \setAe} \abs{\Gamma_{e\!, j}} & \left\lbrace \bF(\bhu_{\! j}) \bn_j - \bG(\bhu_{\! j}, \bepsE, \bphiE) \bn_{\! j} \right. \\
			& \left. + \left( \btau^a(\bhu_{\! j}) + \btau^d\right) \, (\buE - \bhu_{\! j}) \right\rbrace = \vect{0} .
		\end{aligned} \label{eq:FCFVLocal_U}
%
%		& \begin{aligned} \int_{\Omega_e} \pd{\buE}{t} &  \, d\Omega +  \sum_{j \in \setAe} \abs{\Gamma_{e\!, j}} \left( \btau^a(\bhu_{\! j}) + \btau^d\right) \,\buE \\
%			&= - \sum_{j \in \setAe} \abs{\Gamma_{e\!, j}} \left\lbrace \bF(\bhu_{\! j}) \bn_j - \bG(\bhu_{\! j}, \bepsE, \bphiE) \bn_{\! j} - \left( \btau^a(\bhu_{\! j}) + \btau^d\right) \,  \bhu_{\! j} \right\rbrace .
%		\end{aligned} \label{eq:FCFVLocal_U}
	\end{align}
\end{subequations}

It is worth noticing that this step is independent cell-by-cell, computationally inexpensive and it can be easily performed in parallel.  More precisely, equations~\eqref{eq:FCFVLocal_eps} and~\eqref{eq:FCFVLocal_phi} only involve a scaled identity matrix, whereas equation~\eqref{eq:FCFVLocal_U} requires to solve for $\uE$ one linear system of dimension $\nsd+2$, namely, 
\begin{equation} \label{eq:LocalSemiDiscreteCompact_U} 
		\abs{\Omega_e} \frac{d \uE}{d t} + \Ce(\hu) \uE = \re(\hu) ,
\end{equation}
where $\uE$ and $\hu$ are the vectors containing the conservative variables at the centroid of the cell and at the barycentres of the faces, respectively,  whereas $\Ce$ and $\re$ are the matrix and  vector obtained by the FCFV discretisation of
\begin{subequations}
\begin{align}
\Ce(\hu) & := \sum_{j \in \setAe} \abs{\Gamma_{e\!, j}} \left( \btau^a(\hu_{\! j}) + \btau^d\right) , \\
\re(\hu) & := - \sum_{j \in \setAe} \abs{\Gamma_{e\!, j}} \left\lbrace \bF(\hu_{\! j}) \bn_j - \bG(\hu_{\! j},\bepsE(\hu),\bphiE(\hu)) \bn_{\! j} - \left( \btau^a(\hu_{\! j}) + \btau^d\right) \hu_{\! j} \right\rbrace .
\end{align}
\end{subequations}
Finally, an appropriate time integration scheme is required to derive the fully-discrete form of the local problems~\citep{Hairer-book,Jaust-JS:2014,Sanjay2020,Sevilla-21}. Since the present study focuses on steady-state flows,  this term is henceforth neglected. Alternatively, a time marching scheme based on an artificial pseudo-time could be considered as a relaxation approach to speed-up the convergence of the nonlinear global problem detailed below.

From the $\numel$ local problems~\eqref{eq:FCFVLocal}, the expressions of the primal and mixed variables in each cell $\Omega_e$ as functions of the hybrid unknowns on the faces $\Gamma_{e\!, j}, \ j=1,\ldots,\nfaceE$ are exploited to rewrite the FCFV global problem. More precisely, from~\eqref{eq:IntegralGlobal} it follows that the second step of the FCFV solver is: for all $i \in \setAe$, compute $\bhu_{\! i}$ that satisfies
\begin{multline} \label{eq:FCFVGlobal}
	\sum_{e = 1}^{\numel} \abs{\Gamma_{e\!, i}} \left\lbrace \left[ \bF(\bhu_{\! i}) \bn_i - \bG(\bhu_{\! i}, \bepsE, \bphiE) \bn_{\! i} + \left( \btau^a(\bhu_{\! i}) + \btau^d\right) \, (\buE - \bhu_{\! i}) \right] \chi_{\setIe}(i) \right. \\
	\left. + \, \bb(\bu,\bhu,\beps,\bphi) \chi_{\setEe}(i) \right\rbrace = \vect{0}.
\end{multline}
Equation~\eqref{eq:FCFVGlobal} can thus be reformulated using only the face unknowns yielding the nonlinear problem
\begin{equation} \label{eq:GlobalSemiDiscreteCompact} 
	\sum_{e = 1}^{\numel} \hre(\hu) = \bmat{0} ,
\end{equation}
$\hre$ being the residual vector obtained from the FCFV discretisation of the contribution of cell $\Omega_e$ to the global problem. The resulting equation~\eqref{eq:GlobalSemiDiscreteCompact} is linearised using the Newton-Raphson algorithm. In this context, it is important to observe that the expressions of $\buE$, $\bepsE$ and $\bphiE$ in terms of $\bhu$, yielded by the FCFV local problems in each cell, are highly nonlinear, whence the computation of the Jacobian matrix for equation~\eqref{eq:GlobalSemiDiscreteCompact} entails the differentiation of both the local and global FCFV operators appearing in equations~\eqref{eq:FCFVLocal} and~\eqref{eq:FCFVGlobal}.

%------------------------------------------------------------------------------------------------------------
%------------------------------------------------------------------------------------------------------------
\section{Numerical benchmarks for the FCFV method}
\label{sc:Results}

This section presents a comprehensive set of numerical benchmarks of viscous and inviscid compressible flows to showcase the capability of the FCFV method to provide accurate and robust results in a wide variety of flow conditions. The performance of the FCFV scheme is compared to reference solutions published in the literature and to the first and second-order CCFV results provided by the commercial CFD software Ansys Fluent~\citep{FluentManual}. 

%------------------------------------------------------------------------------------------------------------
\subsection{Convergence study for the compressible Taylor-Couette flow}

The first test case considers a compressible Taylor-Couette flow~\citep{Welsh2014,Manela2007},  describing the motion of a viscous fluid confined between two rotating cylinders with isothermal walls. This example has been subject of thorough study in the literature~\citep{Chandrasekhar1981,Chossat1994,Sevilla2020,HDGlab-GSH-20} since, under certain conditions, the flow develops a series of instabilities that break the symmetry of the problem and generate an unsteady behaviour~\citep{Hatay1993,Kao1992}. The setup considered in this study avoids the instabilities and consists of a flow at low Reynolds number, with angular symmetry and no radial velocity, that is, $v_r = 0$. The exact solution can be obtained analytically upon integration of the Navier-Stokes equations in cylindrical coordinates, once the corresponding terms and angular derivatives are dismissed.  This leads to the system of differential equations
\begin{subequations}\label{eq:Couette_analytical}
\begin{align}
\frac{\partial p}{\partial r}  - \frac{\gamma}{\gamma-1}\frac{p v_\theta^2}{r T}&= 0, \\
\frac{\partial}{\partial r} \left(\frac{\partial v_\theta}{\partial r} + \frac{v_{\theta}}{r}\right) &= 0, \\
\frac{1}{r} \frac{\partial}{\partial r} \left(r \frac{\partial T}{\partial r}\right) + \Pra \left[r \frac{\partial}{\partial r}  \left(\frac{v_\theta}{r}\right)\right]^2 &= 0,
\end{align}
\end{subequations}
with boundary conditions
\begin{subequations}\label{eq:Couette_BC}
\begin{align}
v_\theta &= \Omega_0 R_0,  \ T = T_0, \quad \text{at } r = R_0 , \\
v_\theta &= \Omega_1 R_1, \ T = T_1,  \quad \text{at } r = R_1,
\end{align}
\end{subequations}
\hl{where $\Omega_0$ and $\Omega_1$ stand for the angular velocities of the inner and outer cylinders of radius $R_0$ and $R_1$, respectively}.
It is worth noticing that the continuity equation for this system is automatically satisfied.  The solution $(v_\theta, T, p)$ of the boundary value problem~\eqref{eq:Couette_analytical}--\eqref{eq:Couette_BC} reads as
	\begin{align}\label{eq:Couette_analyticalSol}
		v_\theta &= c_1 r + \frac{c_2}{r}, \quad &T &= \alpha + \beta \log (r) - \frac{c_2^2 \Pra}{r^2}, \quad &p &= \frac{1}{\gamma \Ma_{\! \infty}} \exp \left( \frac{-\gamma}{\gamma-1} \int_{r}^{R1} \frac{v_\theta^2}{\zeta T} d\zeta\right),
	\end{align}
where the constants $c_1$, $c_2$, $\alpha$ and $\beta$ are defined as
\begin{subequations}\label{eq:Couette_constants}
\begin{align}
	c_1 &= \frac{\Omega_1 R_1^2 - \Omega_0 R_0^2}{R_1^2 - R_0^2}, \quad \quad c_2 = (\Omega_0-\Omega_1) \frac{R_1^2 R_0^2}{R_1^2 - R_0^2},\\
	\beta &= \frac{1}{\log (R_0/R_1)} \left[(T_0 - T_1) +  c_2^2 \Pra  \left(\frac{1}{R_0^2} - \frac{1}{R_1^2}\right) \right], \\
	\alpha &= T_0 - \beta \log(R_0) + \frac{c_2^2 \Pra}{R_0^2} ,
\end{align}
\end{subequations}
starting from the boundary conditions~\eqref{eq:Couette_BC}.  Note that the velocity field, expressed in cylindrical coordinates, needs to be transformed to the Cartesian reference frame, as the corresponding derivative terms. In particular, provided that $v_r=0$ and $\partial/\partial \theta = 0$, it follows that
\begin{subequations}\label{eq:Couette_transformations}
	\begin{align}
		v_x = v_\theta \frac{y}{r}, \qquad v_y = - v_\theta \frac{x}{r},\ \qquad \text{ and } \qquad\frac{\partial}{\partial \xi} = \frac{\partial }{\partial r} \frac{\partial r}{\partial \xi}, \, \text{ with } \, \frac{\partial r}{\partial \xi} = \frac{\xi}{r},
	\end{align}
\end{subequations}
with $r = \sqrt{x^2 + y^2}$.  Finally, the density field can be obtained by means of the equation of state, namely, $\rho = \gamma p /[(\gamma -1) T]$.

In the case under analysis, a domain with radii $R_0 = 1$ and $R_1 = 2$ is considered.  The flow, featuring a constant viscosity,  is defined by $\Omega_0 = 0$, $\Omega_1 = 0.5$, $T_1 = 1/\left[(\gamma-1)\Ma_{\! \infty}^2 \right]$ and $T_0 = 2T_1$. Moreover, the values of the Mach and the Reynolds numbers are set to $\Ma_{\! \infty}=0.5$ and $\Rey=100$. Note that the latter is computed using a characteristic length $L = R_1 - R_0$, with a characteristic velocity based on the conditions at the outer boundary.

\begin{figure}[!ht]
	\subfloat[Regular mesh 2 \label{fig:Couette_meshStruct1}]{\includegraphics[width=0.24\textwidth]{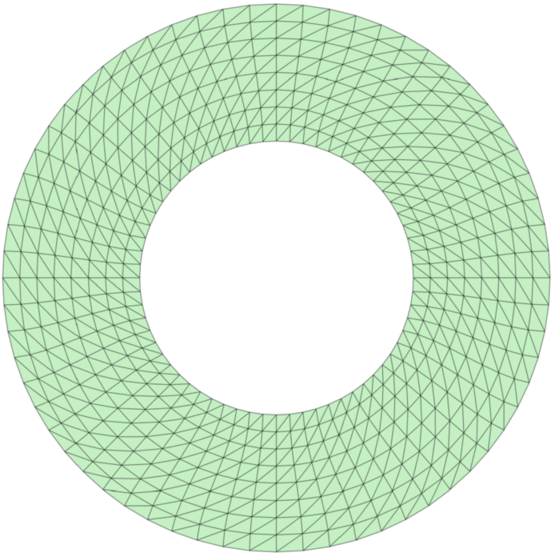}} \hfill
	\subfloat[Regular mesh 3 \label{fig:Couette_meshStruct2}]{\includegraphics[width=0.24\textwidth]{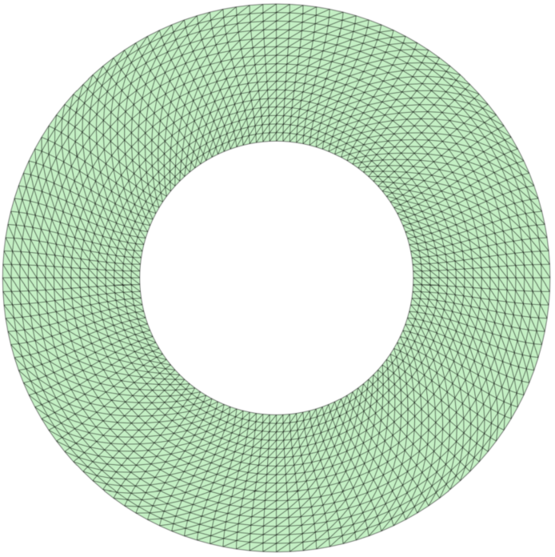}} \hfill
	\subfloat[Distorted mesh 2 \label{fig:Couette_meshDist1}]{\includegraphics[width=0.24\textwidth]{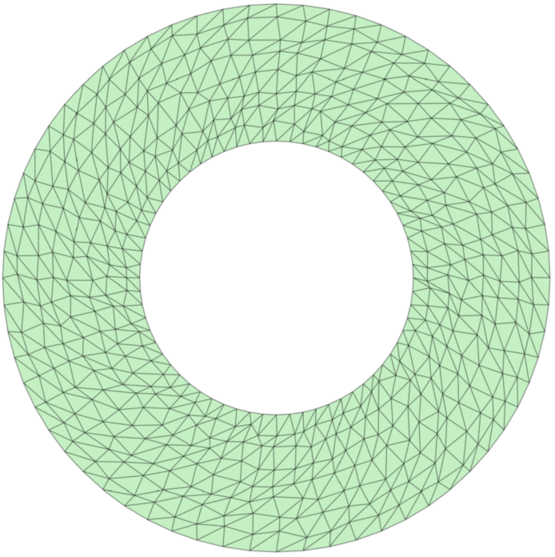}} \hfill
	\subfloat[Distorted mesh 3 \label{fig:Couette_meshDist2}]{\includegraphics[width=0.24\textwidth]{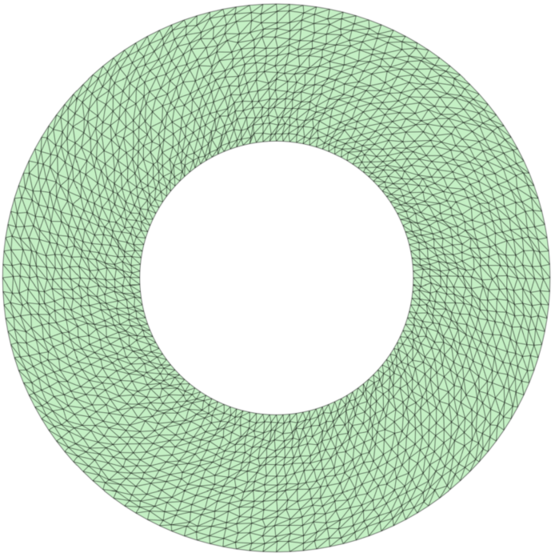}}
	\caption{Compressible Taylor-Couette flow -- Regular and distorted meshes used for the convergence study in two consecutive grid refinements.} 
	\label{fig:Couette_meshes}
\end{figure}

This example is used to evaluate the accuracy properties of the FCFV method using both regular and distorted meshes. The results are also compared to a numerical solution computed using the second-order density-based CCFV scheme available in Ansys Fluent. First, a set of structured regular grids of triangular cells is considered, see figures~\ref{fig:Couette_meshStruct1}--\ref{fig:Couette_meshStruct2}. Additionally,  a second set of meshes, displayed in figures~\ref{fig:Couette_meshDist1}--\ref{fig:Couette_meshDist2},  is employed to evaluate the sensitivity of the method to cell distortion.  The distorted meshes are constructed by randomly perturbing the position $\bm{x}_i$ of the interior nodes of the regular meshes. The resulting perturbed nodes are $\bm{\tilde{x}}_i = \bm{x}_i + \bm{r}_i$,  where $\bm{r}_i$ is a vector of dimension $\nsd$ with random components in the interval $\left[-\ell_{\min}/3, \ell_{\min}/3 \right]$,  $\ell_{\min}$ denoting the characteristic edge length of the regular mesh. 

\hl{
Note that the resulting distorted meshes may suffer from a deterioration in grid quality.  This degradation is assessed by means of two metrics measuring the skewness of the grid cells, namely the equiangle and equivolume skewness metrics. On the one hand, the equiangle skewness metric evaluates, for each triangular cell $\Omega_e$,  the maximum discrepancy between the angles of the cell and those of an equilateral triangle, namely
\begin{subequations}
	\begin{equation}
		\sAngE := \max \left(\frac{\theta_{e,\max} - \pi/3}{2\pi/3}, \frac{\pi/3 - \theta_{e,\min}}{\pi/3} \right),
	\end{equation}
where $\theta_{e,\max}$ and $\theta_{e,\min}$ denote the maximum and minimum angles of the cell $\Omega_e$.
On the other hand, the equivolume skewness measures the relative difference in area of each triangular cell $\Omega_e$ with respect to the area of the optimal triangle $\Omega_{\text{opt}}$ in the same circumcircle, that is,
	\begin{equation}
	\sVolE := \frac{|\Omega_{\text{opt}}| - |\Omega_e|}{|\Omega_{\text{opt}}|}.
\end{equation}
\end{subequations}
For both indicators, values approaching 0 are associated with high quality cells, similar to equilateral triangles. Contrarily, values approaching 1 indicate high cell skewness,  thus leading to lower quality grids. 

Figure~\ref{fig:Couette_meshesQuality} displays the cell-by-cell value of the two skewness metrics for the first two refinements of each set of meshes under analysis.  The two metrics show an important deterioration of the grid quality for distorted meshes, owing to the presence of cells with high skewness that break the uniformity of regular meshes. 
\begin{figure}[!ht]
	\subfloat[$\sAngE$, regular mesh 1 \label{fig:Couette_meshskewStruct1}]{\includegraphics[width=0.24\textwidth]{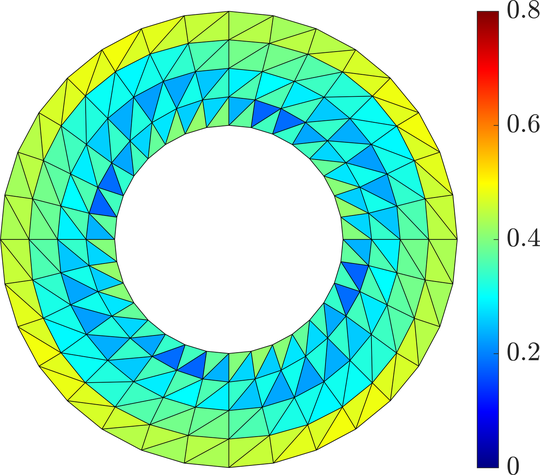}} \hfill
	\subfloat[$\sAngE$, distorted mesh 1 \label{fig:Couette_meshskewDist1}]{\includegraphics[width=0.24\textwidth]{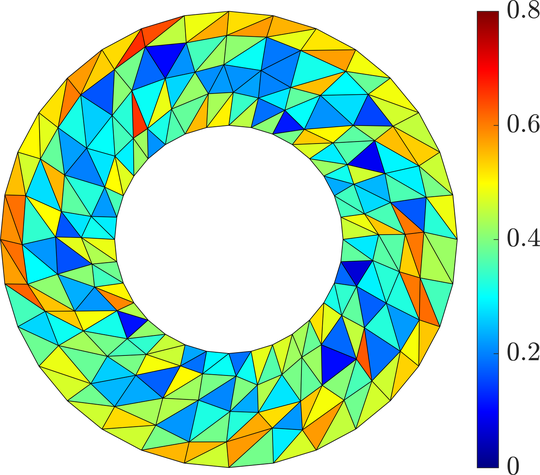}} \hfill
	\subfloat[$\sAngE$, regular mesh 2 \label{fig:Couette_meshskewStruct2}]{\includegraphics[width=0.24\textwidth]{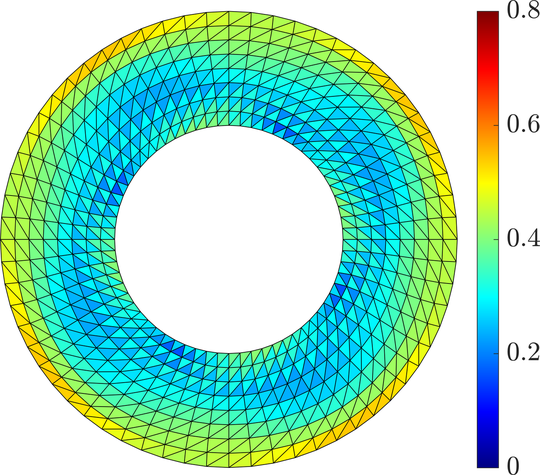}} \hfill
		\subfloat[$\sAngE$, distorted mesh 2 \label{fig:Couette_meshskewDist2}]{\includegraphics[width=0.24\textwidth]{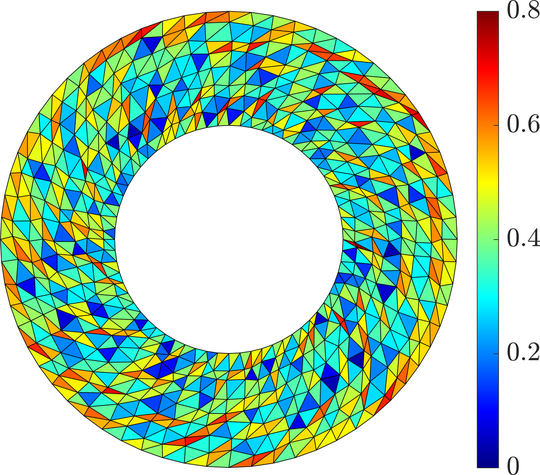}}
	
	\subfloat[$\sVolE$, regular mesh 1 \label{fig:Couette_meshequivStruct1}]{\includegraphics[width=0.24\textwidth]{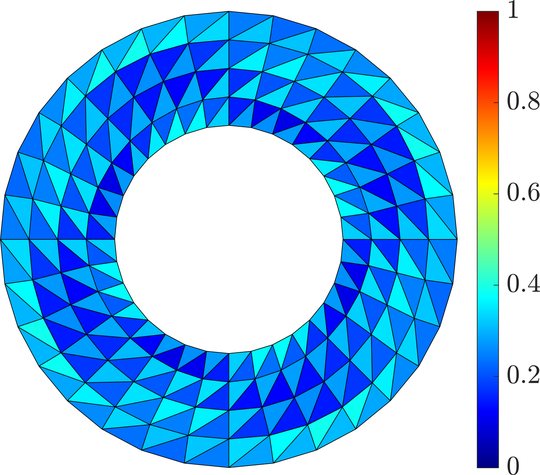}} \hfill
	\subfloat[$\sVolE$, distorted mesh 1 \label{fig:Couette_meshequivDist1}]{\includegraphics[width=0.24\textwidth]{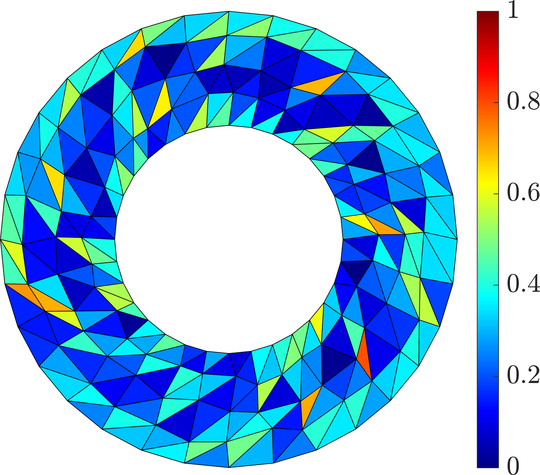}} \hfill
		\subfloat[$\sVolE$, regular mesh 2 \label{fig:Couette_meshequivStruct2}]{\includegraphics[width=0.24\textwidth]{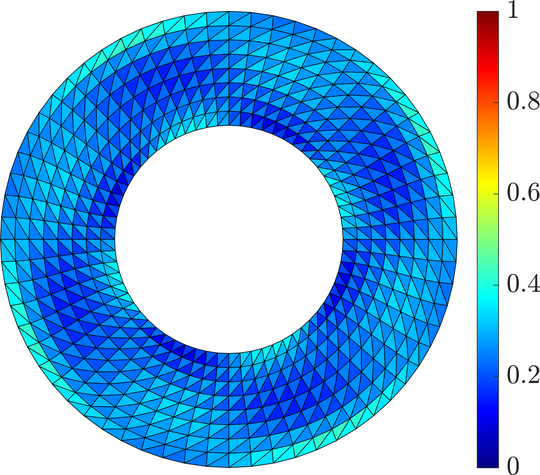}} \hfill
	\subfloat[$\sVolE$, distorted mesh 2 \label{fig:Couette_meshequivDist2}]{\includegraphics[width=0.24\textwidth]{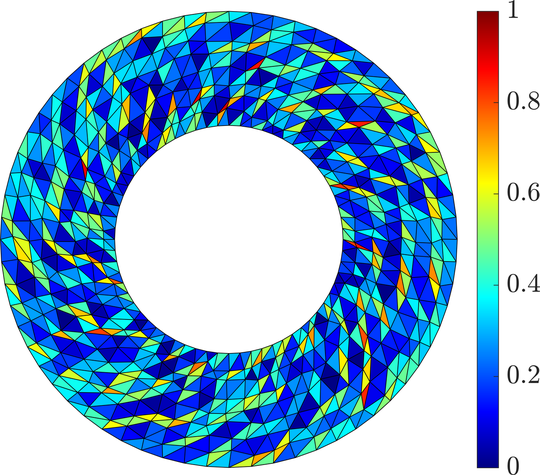}}
	
	\caption{\hl{Compressible Taylor-Couette flow --  Cell-by-cell value of the equiangle, $\sAngE$,  and equivolume, $\sVolE$,  skewness quality metrics on two refinements of the regular and distorted meshes.}} 
	\label{fig:Couette_meshesQuality}
\end{figure}
In addition,  the corresponding $\elinf$ norm of the metrics on the entire domain is reported in table~\ref{tb:meshesQualityMetrics}, for all the grids employed in the study. It is worth noticing that the maximum values of equiangle and equivolume skewness increase with grid refinement, for both regular and distorted meshes.
\begin{table}[!ht]
	\centering
	\begin{tabular}{|l||c|c|c|c|c|}
		\hline
		$\norm{\sAngE}_{\elinf}$ & Mesh 1 & Mesh 2 & Mesh 3 & Mesh 4 & Mesh 5 \\
		\hline
		Regular meshes & 0.4882 & 0.5421 & 0.5727 & 0.5884 & 0.5965 \\
		Distorted meshes & 0.6644 & 0.8002 & 0.8258 & 0.8612 & 0.9035 \\
		\hline
		$\norm{\sVolE}_{\elinf}$ & Mesh 1 & Mesh 2 & Mesh 3 & Mesh 4 & Mesh 5 \\
		\hline
		Regular meshes & 0.3892 & 0.4243 & 0.4532 & 0.4707 & 0.4800 \\
		Distorted meshes & 0.7967 & 0.9102 & 0.9685 & 0.9768 & 0.9938 \\
		\hline
	\end{tabular}
	\caption{\hl{Compressible Taylor-Couette flow -- Maximum equiangle, $\sAngE$,  and equivolume, $\sVolE$,  cell skewness on the entire domain, for each refinement of the regular and distorted meshes.}}
	\label{tb:meshesQualityMetrics}
\end{table}
}

%------------------------------------------------------------------------------------------------------------
\subsubsection{Convergence study on regular meshes}\label{sssc:Couette_convergenceStructured}

\begin{figure}[!ht]
	\subfloat[Density]{\includegraphics[width=0.24\textwidth]{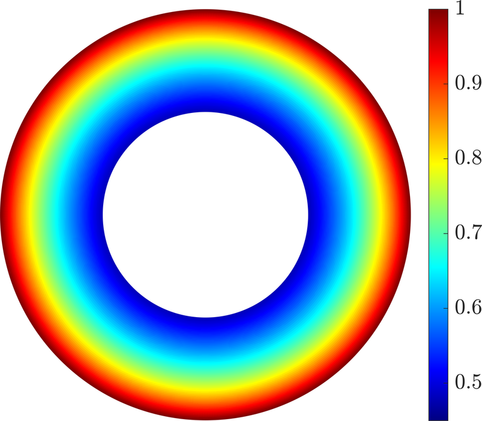}} \hfill
	\subfloat[Velocity]{\includegraphics[width=0.24\textwidth]{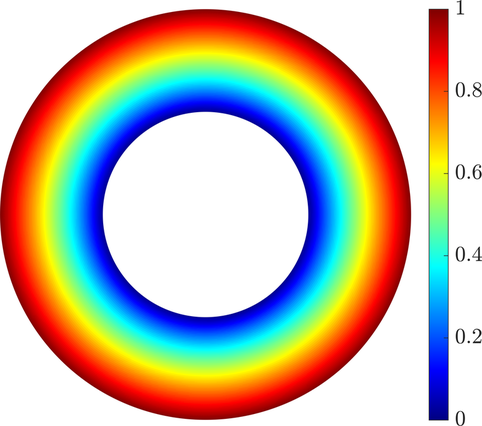}} \hfill
	\subfloat[Temperature]{\includegraphics[width=0.24\textwidth]{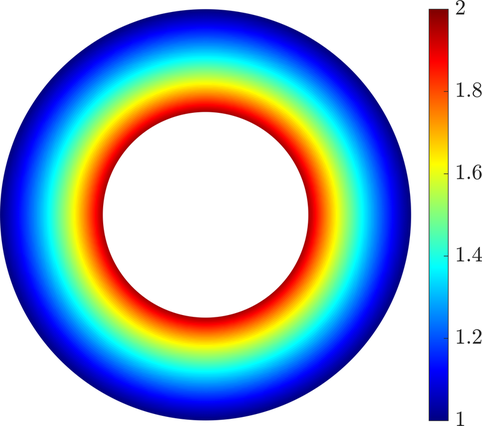}} \hfill
	\subfloat[Pressure]{\includegraphics[width=0.24\textwidth]{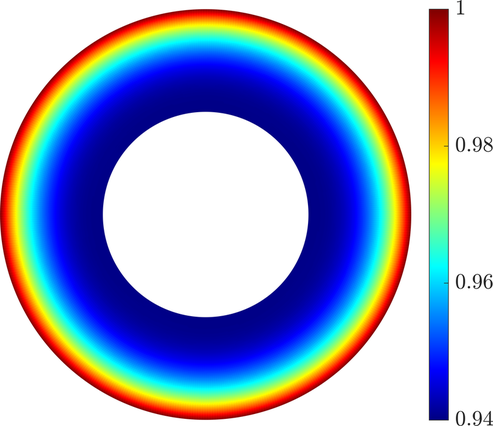}}
	\caption{Compressible Taylor-Couette flow -- Density, velocity, temperature and pressure fields obtained with the FCFV method on the fifth mesh refinement.} 
	\label{fig:Couette_solH5}
\end{figure}

The case is first solved employing regular grids composed by $16 \times 16$, $32 \times 32$, $64 \times 64$, $128 \times 128$ and $256 \times 256$ cells. The FCFV solution for the nondimensional density, velocity, temperature and pressure fields on the finest mesh is depicted in figure~\ref{fig:Couette_solH5}, showing the radial dependency of the physical quantities.

The accuracy of the FCFV approximation is evaluated by means of an $h$-convergence study of the $\eltwo(\Omega)$ error of the variables of the system and the results are also compared to the outcome of a CCFV simulation. The convergence history of the primitive variables, namely density, velocity, temperature and pressure, is reported in table~\ref{tb:convergenceStructured}.
\begin{table}[!ht]
		\centering
	\begin{tabular}{|c||cc|cc|cc|cc|}
		\hline
		\multicolumn{9}{|c|}{FCFV} \\
		\hline
		\multicolumn{1}{|c||}{} & \multicolumn{2}{c|}{$\norm{E_\rho}_{\eltwo}$} & \multicolumn{2}{c|}{$\norm{E_\bv}_{\eltwo}$} & \multicolumn{2}{c|}{$\norm{E_{T}}_{\eltwo}$} & \multicolumn{2}{c|}{$\norm{E_p}_{\eltwo}$}\\
		\textbf{$\sqrt{\numel}$} & Error & Rate & Error & Rate & Error & Rate & Error & Rate \\
		\hline
		16 & 5.13e-02 & -- & 1.24e-01 & -- & 4.96e-02 & -- & 1.20e-02 & --\\
		32 & 2.66e-02 & 1.00 & 5.74e-02 & 1.18 & 2.35e-02 & 1.14 & 7.11e-03 & 0.80\\
		64 & 1.35e-02 & 1.01 & 2.81e-02 & 1.06 & 1.14e-02 & 1.07 & 3.95e-03 & 0.87\\
		128 & 6.73e-03 & 1.02 & 1.39e-02 & 1.03 & 5.49e-03 & 1.07 & 2.22e-03 & 0.84\\
		256 & 3.36e-03 & 1.01 & 6.89e-03 & 1.02 & 2.61e-03 & 1.08 & 1.29e-03 & 0.79\\
		\hline
		\multicolumn{9}{|c|}{Fluent} \\
		\hline
		\multicolumn{1}{|c||}{} & \multicolumn{2}{c|}{$\norm{E_\rho}_{\eltwo}$} & \multicolumn{2}{c|}{$\norm{E_\bv}_{\eltwo}$} & \multicolumn{2}{c|}{$\norm{E_{T}}_{\eltwo}$} & \multicolumn{2}{c|}{$\norm{E_p}_{\eltwo}$}\\
		\textbf{$\sqrt{\numel}$} & Error & Rate & Error & Rate & Error & Rate & Error & Rate \\
		\hline
		16 & 9.00e-02 & -- & 2.87e-02 & -- & 7.32e-03 & -- & 8.66e-02 & --\\
		32 & 4.06e-02 & 1.22 & 6.58e-03 & 2.25 & 4.06e-03 & 0.90 & 3.64e-02 & 1.32\\
		64 & 1.88e-02 & 1.14 & 1.74e-03 & 1.97 & 1.53e-03 & 1.45 & 1.71e-02 & 1.13\\
		128 & 8.98e-03 & 1.08 & 4.74e-04 & 1.91 & 4.99e-04 & 1.64 & 8.37e-03 & 1.04\\
		256 & 4.38e-03 & 1.04 & 1.31e-04 & 1.86 & 1.45e-04 & 1.80 & 4.19e-03 & 1.01\\
		\hline
	\end{tabular}
	\caption{Compressible Taylor-Couette flow -- Convergence history of the primitive variables (density, velocity, temperature and pressure) on the regular meshes using the FCFV method and the second-order CCFV scheme by Ansys Fluent.} 
	\label{tb:convergenceStructured}
\end{table}
On the one hand,  the FCFV method provides first-order accuracy for density, velocity and temperature, whereas pressure converges with a slightly suboptimal rate. On the other hand, Ansys Fluent displays second-order accuracy for velocity and temperature, whereas density and pressure are approximated with first order. In addition, it is worth noticing that in this case the FCFV results slightly outperform the CCFV approximation by Ansys Fluent in terms of the accuracy of density and pressure fields.

It is important to remark that the momentum and the energy equations in the compressible Navier-Stokes system feature second-order derivatives of the velocity and the temperature. In this context, the flux reconstruction performed by CCFV methods is responsible for the accuracy gain shown by Ansys Fluent in table~\ref{tb:convergenceStructured} for velocity and temperature. Nonetheless, the mass continuity equation is a first-order partial differential equation, whence the flux reconstruction is not providing additional accuracy in the approximation of the density (nor of any derived variable such as the pressure) and its convergence is only first order. Given the critical role of the density in the computation of the conservative variables, table~\ref{tb:convergenceConservativeStructured} reports the convergence history for momentum and energy.
\begin{table}[!ht]
	\centering
	\begin{tabular}{|c||cc|cc||cc|cc|}
		\hline
		\multicolumn{1}{|c||}{} & \multicolumn{4}{c||}{$\norm{E_{\rho \bv}}_{\eltwo}$} & \multicolumn{4}{c|}{$\norm{E_{\rho E}}_{\eltwo}$}\\
		\hline
		\multicolumn{1}{|c||}{} & \multicolumn{2}{c|}{FCFV} & \multicolumn{2}{c||}{Fluent} & \multicolumn{2}{c|}{FCFV} & \multicolumn{2}{c|}{Fluent} \\
		\textbf{$\sqrt{\numel}$} & Error & Rate & Error & Rate & Error & Rate & Error & Rate  \\
		\hline
		16 & 1.20e-01 & --  & 7.59e-02 & -- & 1.25e-02 & --  & 8.64e-02 & --\\
		32 & 5.54e-02 & 1.18  & 4.07e-02 & 0.95 & 7.33e-03 & 0.82  & 3.66e-02 & 1.31\\
		64 & 2.72e-02 & 1.06  & 1.94e-02 & 1.10 & 4.05e-03 & 0.88  & 1.72e-02 & 1.13\\
		128 & 1.34e-02 & 1.04  & 9.23e-03 & 1.09 & 2.24e-03 & 0.87  & 8.41e-03 & 1.04\\
		256 & 6.53e-03 & 1.04  & 4.45e-03 & 1.06 & 1.27e-03 & 0.82  & 4.20e-03 & 1.01\\
		\hline
	\end{tabular}
	\caption{Compressible Taylor-Couette flow -- Convergence history of the conservative variables (momentum and energy) on the regular meshes using the FCFV method and the second-order CCFV scheme by Ansys Fluent.} 
	\label{tb:convergenceConservativeStructured}
\end{table}
For the FCFV method, conservative variables are the primal variables of computation and first-order accuracy is achieved for both momentum and energy.  The corresponding approximations computed using Ansys Fluent do not inherit the second-order convergence property of velocity and temperature previously observed and only first-order accuracy is achieved.  Moreover, similarly to the previous result,  the FCFV solution slightly outperforms the accuracy of the energy approximation computed by means of Ansys Fluent.

Finally, the accuracy of the approximation of the stress tensor and the heat flux is presented in table~\ref{tb:convergenceDerivativesStructured}. In this case, first-order accuracy is obtained by the FCFV approximation for both quantities. Regarding the CCFV solution by Ansys Fluent, the heat flux converges with a rate above one, providing an approximation more accurate than the FCFV method. Nonetheless,  the convergence rate of the stress tensor rapidly deteriorates and only a suboptimal convergence rate of 0.5 is achieved.
\begin{table}[!ht]
		\centering
	\begin{tabular}{|c||cc|cc||cc|cc|}
		\hline
		\multicolumn{1}{|c||}{} & \multicolumn{4}{c||}{$\norm{E_{\bsigma}}_{\eltwo}$} & \multicolumn{4}{c|}{$\norm{E_\bq}_{\eltwo}$}\\
		\hline
		\multicolumn{1}{|c||}{} & \multicolumn{2}{c|}{FCFV} & \multicolumn{2}{c||}{Fluent} & \multicolumn{2}{c|}{FCFV} & \multicolumn{2}{c|}{Fluent} \\
		\textbf{$\sqrt{\numel}$} & Error & Rate & Error & Rate & Error & Rate & Error & Rate  \\
		\hline
		16 & 4.66e-01 & --  & 1.03e-01 & -- & 2.29e-01 & --  & 6.51e-02 & --\\
		32 & 2.20e-01 & 1.14  & 3.70e-02 & 1.57 & 1.02e-01 & 1.23  & 3.33e-02 & 1.02\\
		64 & 1.12e-01 & 1.01  & 2.12e-02 & 0.82 & 4.73e-02 & 1.15  & 1.48e-02 & 1.20\\
		128 & 5.94e-02 & 0.93  & 1.46e-02 & 0.55 & 2.22e-02 & 1.10  & 6.24e-03 & 1.27\\
		256 & 3.22e-02 & 0.89  & 1.04e-02 & 0.49 & 1.04e-02 & 1.11  & 2.45e-03 & 1.36\\
		\hline
	\end{tabular}
	\caption{Compressible Taylor-Couette flow -- Convergence history of the stress tensor and the heat flux vector on the regular meshes using the FCFV method and the second-order CCFV scheme by Ansys Fluent.} 
	\label{tb:convergenceDerivativesStructured}
\end{table}

%------------------------------------------------------------------------------------------------------------
\subsubsection{Convergence study on distorted meshes}\label{sssc:Couette_convergenceDistorted}

\begin{figure}[!ht]
	\subfloat[FCFV, regular]{\includegraphics[width=0.24\textwidth]{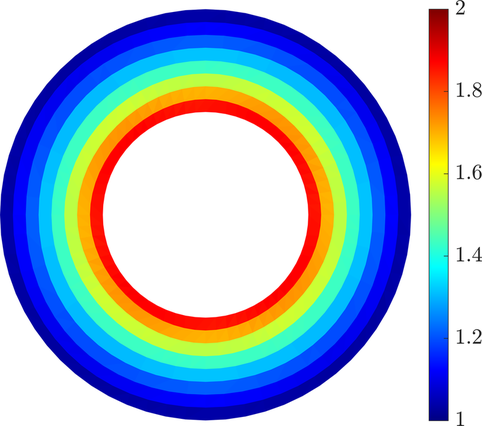}} \hfill
	\subfloat[Fluent, regular]{\includegraphics[width=0.24\textwidth]{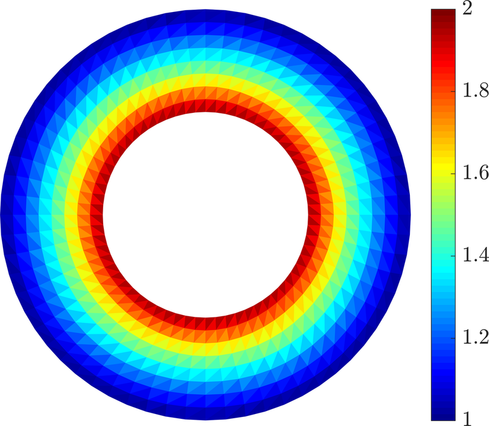}} \hfill
	\subfloat[FCFV, distorted]{\includegraphics[width=0.24\textwidth]{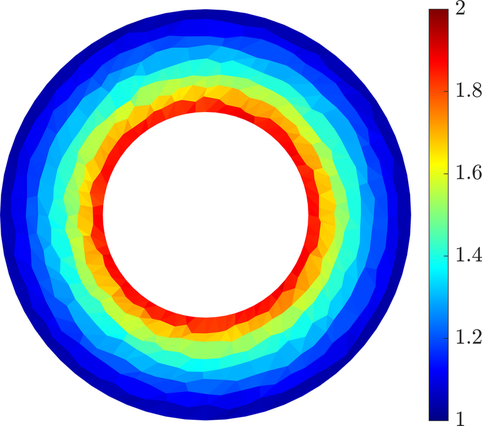}} \hfill
	\subfloat[Fluent, distorted]{\includegraphics[width=0.24\textwidth]{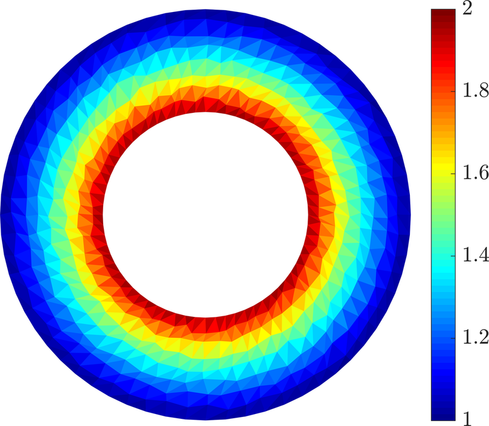}}
	\caption{Compressible Taylor-Couette flow -- Temperature field obtained with the FCFV method and the second-order CCFV scheme by Ansys Fluent on the second grid refinement of the regular and distorted meshes.} 
	\label{fig:Couette_T_all}
\end{figure}

The convergence study presented in the previous section is now repeated using the set of perturbed meshes.  Figure~\ref{fig:Couette_T_all} displays the temperature field obtained on the second mesh refinement of the regular and the distorted meshes,  using both the FCFV method and Ansys Fluent second-order CCFV scheme.  In all cases, the results are comparable, with a slight tendency of Ansys Fluent to present abrupt variations of the solution cell-by-cell.

The convergence of the $\eltwo(\Omega)$ error for the primitive variables (density, velocity, temperature and pressure),  the conservative variables (momentum and energy) and the stress tensor and the heat flux are reported in tables~\ref{tb:convergenceDistorted}, \ref{tb:convergenceConservativeDistorted} and \ref{tb:convergenceDerivativesDistorted}, respectively.
\begin{table}[!ht]
	\centering
	\begin{tabular}{|c||cc|cc|cc|cc|}
		\hline
		\multicolumn{9}{|c|}{FCFV} \\
		\hline
		\multicolumn{1}{|c||}{} & \multicolumn{2}{c|}{$\norm{E_\rho}_{\eltwo}$} & \multicolumn{2}{c|}{$\norm{E_\bv}_{\eltwo}$} & \multicolumn{2}{c|}{$\norm{E_{T}}_{\eltwo}$} & \multicolumn{2}{c|}{$\norm{E_p}_{\eltwo}$}\\
		\textbf{$\sqrt{\numel}$} & Error & Rate & Error & Rate & Error & Rate & Error & Rate \\
		\hline
		16 & 6.13e-02 & -- & 1.68e-01 & -- & 6.35e-02 & -- & 1.34e-02 & --\\
		32 & 3.43e-02 & 0.98 & 7.65e-02 & 1.32 & 3.24e-02 & 1.13 & 7.53e-03 & 0.97\\
		64 & 1.88e-02 & 0.86 & 3.84e-02 & 0.99 & 1.71e-02 & 0.92 & 3.94e-03 & 0.93\\
		128 & 9.41e-03 & 1.10 & 1.92e-02 & 1.10 & 8.35e-03 & 1.14 & 2.23e-03 & 0.91\\
		256 & 4.93e-03 & 0.94 & 9.54e-03 & 1.01 & 4.16e-03 & 1.01 & 1.32e-03 & 0.76\\
		\hline
		\multicolumn{9}{|c|}{Fluent} \\
		\hline
		\multicolumn{1}{|c||}{} & \multicolumn{2}{c|}{$\norm{E_\rho}_{\eltwo}$} & \multicolumn{2}{c|}{$\norm{E_\bv}_{\eltwo}$} & \multicolumn{2}{c|}{$\norm{E_{T}}_{\eltwo}$} & \multicolumn{2}{c|}{$\norm{E_p}_{\eltwo}$}\\
		\textbf{$\sqrt{\numel}$} & Error & Rate & Error & Rate & Error & Rate & Error & Rate \\
		\hline
		16 & 7.72e-02 & -- & 3.35e-02 & -- & 9.21e-03 & -- & 7.36e-02 & --\\
		32 & 3.24e-02 & 1.46 & 1.14e-02 & 1.81 & 5.55e-03 & 0.85 & 2.77e-02 & 1.64\\
		64 & 1.47e-02 & 1.13 & 4.83e-03 & 1.23 & 2.61e-03 & 1.08 & 1.25e-02 & 1.15\\
		128 & 6.93e-03 & 1.20 & 2.35e-03 & 1.14 & 1.11e-03 & 1.35 & 6.15e-03 & 1.12\\
		256 & 3.25e-03 & 1.10 & 1.12e-03 & 1.07 & 5.02e-04 & 1.16 & 2.96e-03 & 1.06\\
		\hline
	\end{tabular}
	\caption{Compressible Taylor-Couette flow -- Convergence history of the primitive variables (density, velocity, temperature and pressure) on the distorted meshes using the FCFV method and the second-order CCFV scheme by Ansys Fluent.} 
	\label{tb:convergenceDistorted}
\end{table}
\begin{table}[!ht]
	\centering
	\begin{tabular}{|c||cc|cc||cc|cc|}
		\hline
		\multicolumn{1}{|c||}{} & \multicolumn{4}{c||}{$\norm{E_{\rho \bv}}_{\eltwo}$} & \multicolumn{4}{c|}{$\norm{E_{\rho E}}_{\eltwo}$}\\
		\hline
		\multicolumn{1}{|c||}{} & \multicolumn{2}{c|}{FCFV} & \multicolumn{2}{c||}{Fluent} & \multicolumn{2}{c|}{FCFV} & \multicolumn{2}{c|}{Fluent} \\
		\textbf{$\sqrt{\numel}$} & Error & Rate & Error & Rate & Error & Rate & Error & Rate  \\
		\hline
		16 & 1.68e-01 & --  & 6.77e-02 & -- & 1.50e-02 & --  & 7.34e-02 & --\\
		32 & 7.45e-02 & 1.36  & 3.24e-02 & 1.23 & 8.31e-03 & 0.99  & 2.78e-02 & 1.63\\
		64 & 3.87e-02 & 0.94  & 1.56e-02 & 1.06 & 4.55e-03 & 0.87  & 1.25e-02 & 1.15\\
		128 & 1.90e-02 & 1.13  & 7.23e-03 & 1.22 & 2.52e-03 & 0.94  & 6.17e-03 & 1.13\\
		256 & 9.68e-03 & 0.98  & 3.42e-03 & 1.08 & 1.48e-03 & 0.77  & 2.96e-03 & 1.06\\
		\hline
	\end{tabular}
	\caption{Compressible Taylor-Couette flow -- Convergence history of the conservative variables (momentum and energy) on the distorted meshes using the FCFV method and the second-order CCFV scheme by Ansys Fluent.} 
	\label{tb:convergenceConservativeDistorted}
\end{table}
\begin{table}[!ht]
		\centering
	\begin{tabular}{|c||cc|cc||cc|cc|}
		\hline
		\multicolumn{1}{|c||}{} & \multicolumn{4}{c||}{$\norm{E_{\bsigma}}_{\eltwo}$} & \multicolumn{4}{c|}{$\norm{E_\bq}_{\eltwo}$}\\
		\hline
		\multicolumn{1}{|c||}{} & \multicolumn{2}{c|}{FCFV} & \multicolumn{2}{c||}{Fluent} & \multicolumn{2}{c|}{FCFV} & \multicolumn{2}{c|}{Fluent} \\
		\textbf{$\sqrt{\numel}$} & Error & Rate & Error & Rate & Error & Rate & Error & Rate  \\
		\hline
		16 & 7.56e-01 & --  & 1.12e-01 & -- & 3.67e-01 & --  & 6.47e-02 & --\\
		32 & 4.30e-01 & 0.95  & 4.52e-02 & 1.52 & 1.86e-01 & 1.14  & 3.65e-02 & 0.96\\
		64 & 2.52e-01 & 0.77  & 2.61e-02 & 0.79 & 9.86e-02 & 0.92  & 1.74e-02 & 1.07\\
		128 & 1.49e-01 & 0.83  & 1.84e-02 & 0.55 & 5.10e-02 & 1.05  & 9.42e-03 & 0.97\\
		256 & 8.69e-02 & 0.78  & 1.42e-02 & 0.38 & 2.63e-02 & 0.96  & 6.99e-03 & 0.43\\
		\hline
	\end{tabular}
	\caption{Compressible Taylor-Couette flow -- Convergence history of the stress tensor and the heat flux vector on the distorted meshes using the FCFV method and the second-order CCFV scheme by Ansys Fluent.} 
	\label{tb:convergenceDerivativesDistorted}
\end{table}

First of all, the FCFV scheme maintains the first-order accuracy in the approximation of both the primitive variables (i.e., density, velocity, temperature and pressure) in table~\ref{tb:convergenceDistorted} and the conservative variables (i.e., momentum and energy) in table~\ref{tb:convergenceConservativeDistorted},  even in the presence of cell distortion.  Similarly, the CCFV approximation computed using Ansys Fluent maintains the first-order accuracy for density and pressure (cf. table~\ref{tb:convergenceDistorted}), as well as for momentum and energy (cf. table~\ref{tb:convergenceConservativeDistorted}). Nonetheless, the second-order convergence of velocity and temperature observed on regular grids (cf. table~\ref{tb:convergenceStructured}) is lost in this case and only first-order accuracy is achieved for these variables, confirming the sensitivity of the reconstruction strategy of CCFV methods to the quality of the employed meshes~\citep{Diskin2010,Diskin2011}. 

Finally, the convergence history in table~\ref{tb:convergenceDerivativesDistorted} confirms the robustness of the FCFV method to mesh distortion, with almost first-order accuracy achieved by both the stress tensor and the heat flux. On the contrary, Ansys Fluent CCFV scheme shows a suboptimal behaviour in the approximation of these quantities, with a convergence rate stagnating around 0.4 for both variables.

The results of the Taylor-Couette flow demonstrate the accuracy properties of the FCFV method and its robustness to different meshes types. Indeed, the method provides first-order accuracy for all the flow variables (primitive and conservative),  as well as for the viscous stress tensor and the heat flux,  using both regular and distorted meshes. This is in contrast with the results provided by Ansys Fluent. Although the CCFV method outperforms the FCFV scheme on regular grids achieving second-order accuracy for velocity and temperature,  the remaining quantities (density, pressure, momentum, energy and heat flux) only converge with order one and the viscous stress tensor experiences a suboptimal behaviour.  In addition the CCFV scheme by Ansys Fluent displays a strong sensitivity to mesh distortion,  with a deterioration of the convergence order of velocity and temperature to order one and of the viscous stress tensor and the heat flux to order $0.4$ in the presence of perturbed meshes.

%------------------------------------------------------------------------------------------------------------
\subsection{Viscous laminar flow cases over a NACA 0012 aerofoil}

A set of viscous laminar flows over a NACA 0012 aerofoil is studied next.  Three different cases corresponding to subsonic~\citep{Swanson2016,Bassi-BR:1997b,Mavriplis1990}, transonic~\citep{Cambier1987} and supersonic viscous laminar flows at zero angle of attack are analysed, imposing adiabatic wall conditions on the aerofoil surface. The flow conditions are described in table~\ref{tb:NACAviscous_FlowConditions}. The purpose of these tests is to examine the capability of the FCFV method to provide accurate results of aerodynamic quantities of interest in various flow conditions, comparing them with reference numerical solutions available in the literature and with the outcome of first and second-order CCFV simulations using Ansys Fluent.
\begin{table}[!ht]
	\centering
	\begin{tabular}{|c|c|c|}
		\hline
		Subsonic case & Transonic case & Supersonic case\\
		\hline
		$\Minf = 0.5$ & $\Minf = 0.85$ & $\Minf = 2$ \\
		$\Rey = 5,000$ & $\Rey = 500$ & $\Rey = 10,000$ \\
		\hline
	\end{tabular}
	\caption{Viscous laminar flow cases over a NACA 0012 aerofoil -- Flow conditions.} 
	\label{tb:NACAviscous_FlowConditions}
\end{table}

Given the viscous nature of the flows under analysis, non-uniform meshes with highly stretched cells in the boundary layer region are employed.
\hl{In particular,} a C-type computational domain is considered, with the far-field boundary located at least at 15 chord units from the aerofoil. A set of four structured meshes consisting of $128\times128$, $256\times 256$, $512\times512$ and $1,024\times1,024$ quadrilateral cells is employed.  \hl{The specifics of the meshes are summarised in table~\ref{tb:NACAviscous_MeshDetails}, with details on the height $h_{\min}$ of the first cell, the geometric growth rate in the direction normal to the wall and the number $\numbl$ of mesh layers within the boundary layer, defined as the region up to a distance of 0.05 chord units from the aerofoil surface.
}
%\JV{The grids feature, respectively, a geometric growth rate of a $5 \%$, $2 \%$, $0.75 \%$ and $0.25 \%$ on the normal direction to the wall. As a result, the boundary layer, considered within 0.05 chord units from the aerofoil surface, contains 60, 120, 220 and 372 elements, respectively, being the corresponding height of their first element $1.4\cdot10^{-4}$, $1.1\cdot10^{-4}$, $9.1\cdot10^{-5}$ and $8.2\cdot10^{-5}$ chord units.}
%
\begin{table}[!ht]
	\centering
	\begin{tabular}{|c||c|c|c|c|}
		\hline
		Mesh & $\sqrt{\numel}$ & $h_{\min}$ & Growth rate & $\numbl$ \\
		\hline
		1 & 128 & 1.4e-04 & 0.50e-01 & 60  \\
		2 & 256 & 1.1e-04 & 0.20e-01 & 120  \\
		3 & 512 & 9.1e-05 & 0.75e-02 & 220  \\
		4 & 1,024 & 8.2e-05 & 0.25e-02 & 372  \\				
		\hline
	\end{tabular}
	\caption{\hl{Viscous laminar flow cases over a NACA 0012 aerofoil  -- Specifics of the meshes, including number of cells $\numel$,  height $h_{\min}$ of the first cell,  geometric growth rate in the direction normal to the wall and number $\numbl$ of cell layers in the boundary layer.}}
	\label{tb:NACAviscous_MeshDetails}
\end{table}

\begin{figure}[!ht]
	\subfloat[Mesh 1: 128$\times$128]{\includegraphics[width=0.24\textwidth]{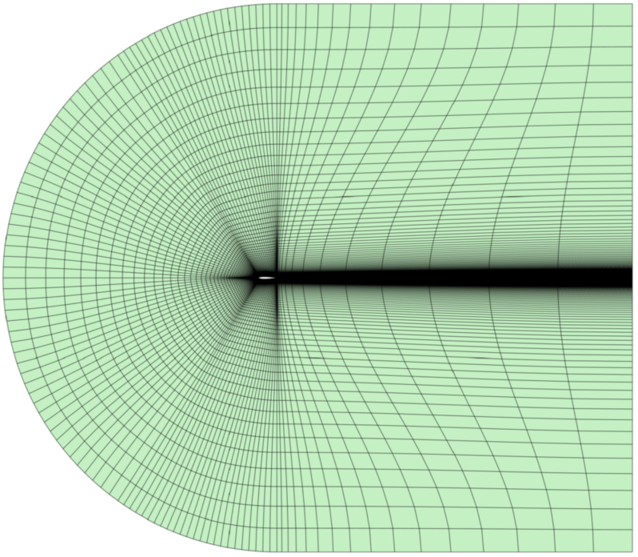}} \hfill
	\subfloat[Mesh 2: 256$\times$256]{\includegraphics[width=0.24\textwidth]{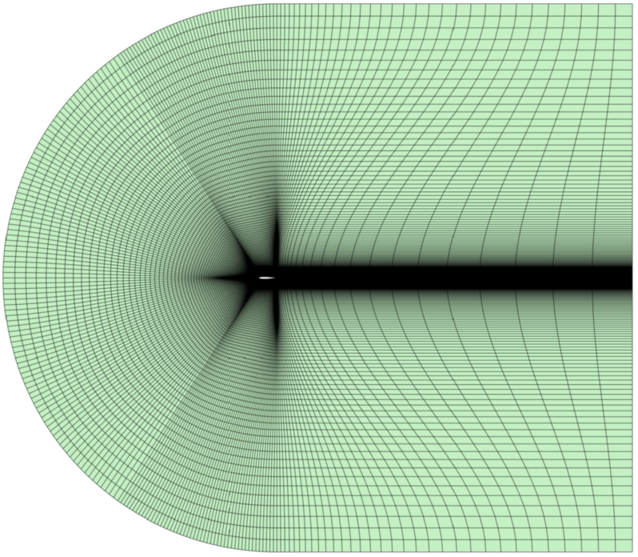}} \hfill
	\subfloat[Mesh 3: 512$\times$512]{\includegraphics[width=0.24\textwidth]{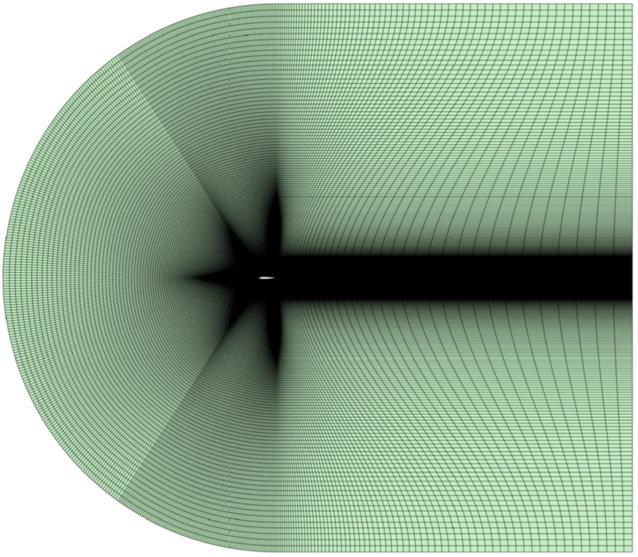}} \hfill
	\subfloat[Mesh 4: 1,024$\times$1,024]{\includegraphics[width=0.24\textwidth]{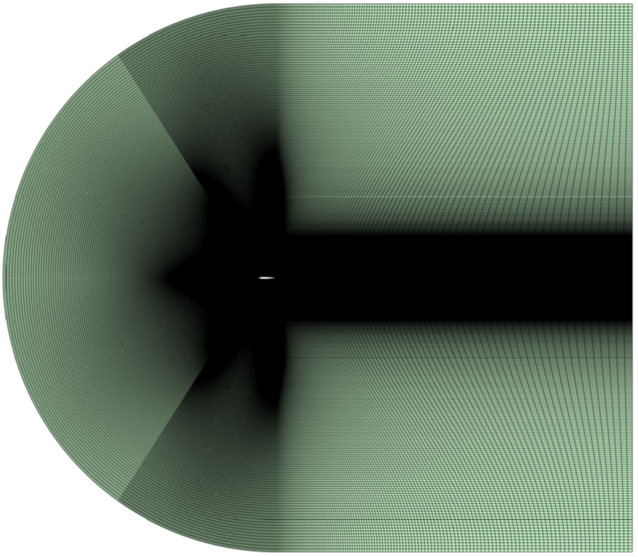}} \\
	\subfloat[Mesh 1, detail]{\includegraphics[width=0.24\textwidth]{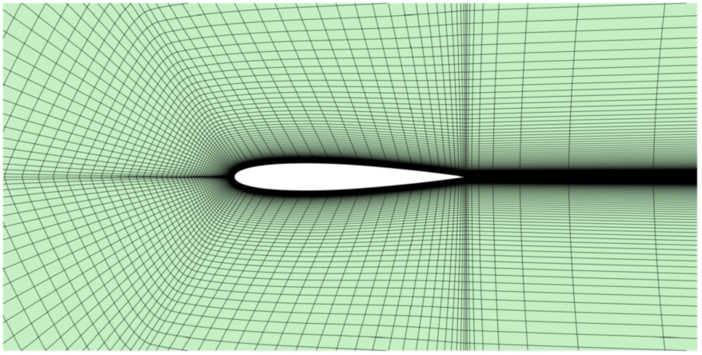}} \hfill
	\subfloat[Mesh 2, detail]{\includegraphics[width=0.24\textwidth]{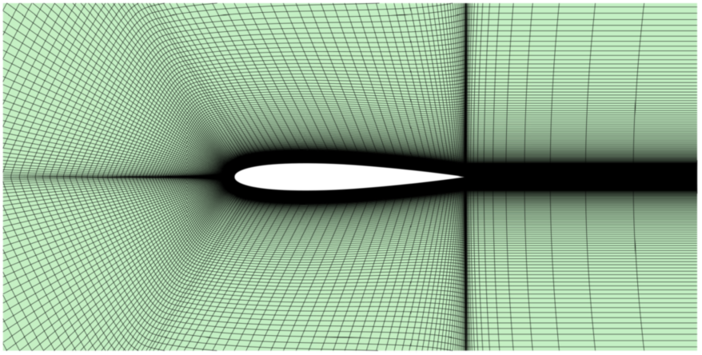}} \hfill
	\subfloat[Mesh 3, detail]{\includegraphics[width=0.24\textwidth]{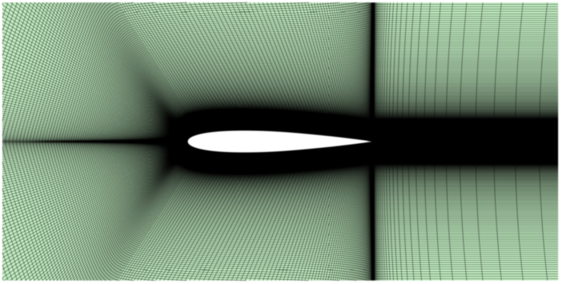}} \hfill
	\subfloat[Mesh 4, detail]{\includegraphics[width=0.24\textwidth]{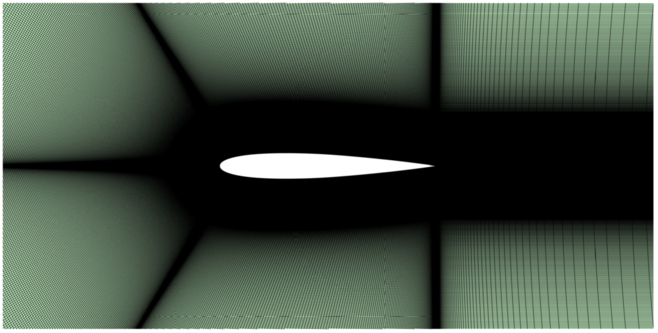}} \\
	\caption{Viscous laminar flow cases over a NACA 0012 aerofoil -- Set of meshes.} 
	\label{fig:NACAviscous_Meshes}
\end{figure}

\begin{figure}[!ht]
\captionsetup{justification=centering}
	\subfloat[Subsonic\\ $\Ma_{\! \infty}=0.5$, $\Rey=5,000$]{\includegraphics[width=0.32\textwidth]{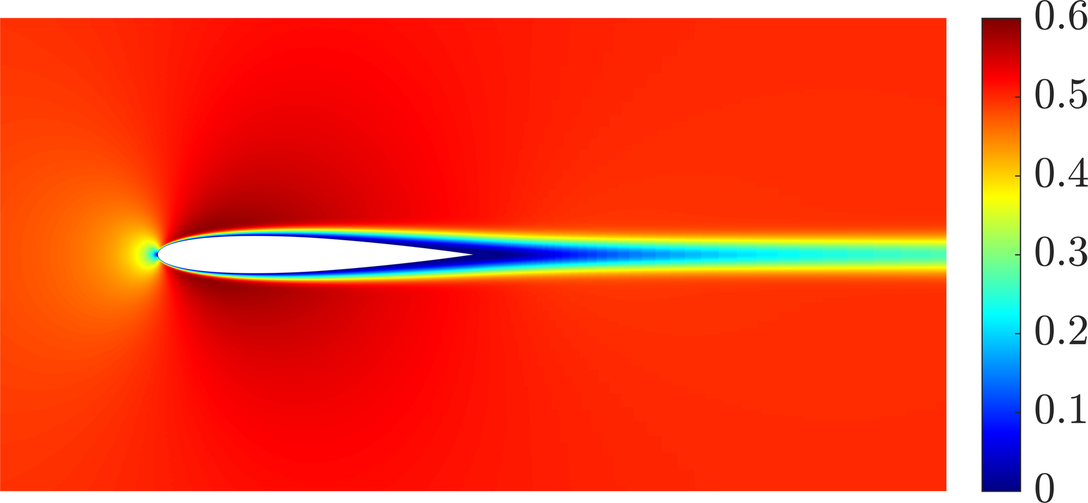}} \hfill
	\subfloat[Transonic\\ $\Ma_{\! \infty}=0.85$, $\Rey=500$]{\includegraphics[width=0.32\textwidth]{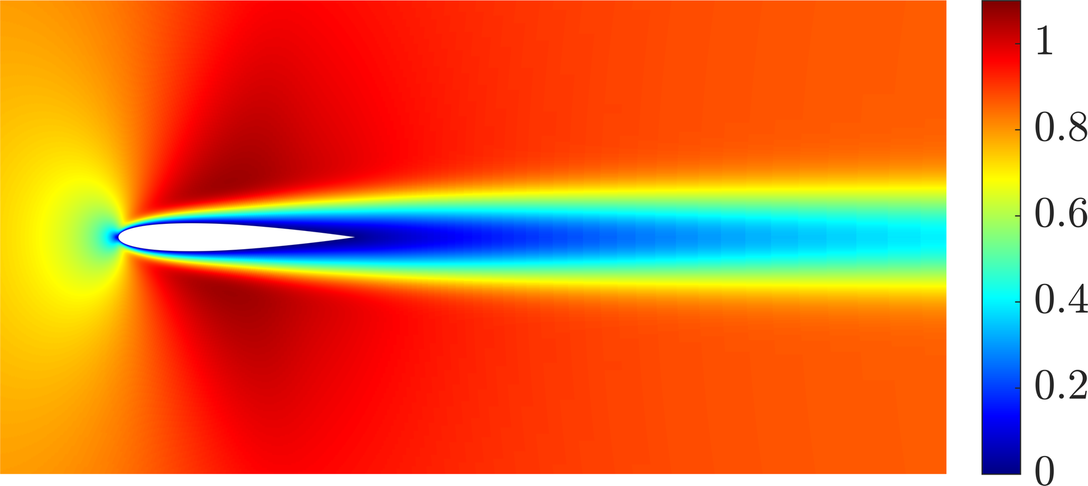}} \hfill
	\subfloat[Supersonic\\ $\Ma_{\! \infty}=2$, $\Rey=10,000$]{\includegraphics[width=0.32\textwidth]{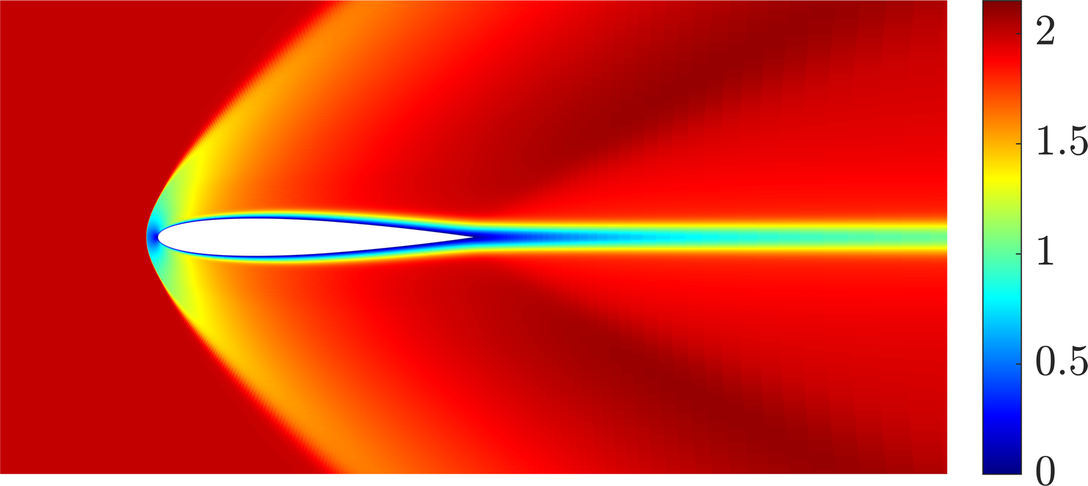}}
\captionsetup{justification=justified}
	\caption{Viscous laminar flow cases over a NACA 0012 aerofoil -- Mach number distribution obtained with the FCFV method in the finest mesh for the three cases of the study.} 
	\label{fig:NACAviscous_Solutions}
\end{figure}

Figure~\ref{fig:NACAviscous_Meshes} displays the four meshes with a close-up view near the aerofoil, highlighting the refinement in the boundary layer region,  with an aspect ratio between $10^3$ and $10^4$. Moreover, additional refinement is introduced near the leading and trailing edges in order to capture the high gradients of the flow and to avoid numerical issues due to the geometric singularity. The FCFV approximation of the Mach number distribution around the aerofoil is depicted in figure~\ref{fig:NACAviscous_Solutions}.

\begin{figure}[!ht]
	\subfloat[Pressure coeff., subsonic]{\includegraphics[width=0.32\textwidth]{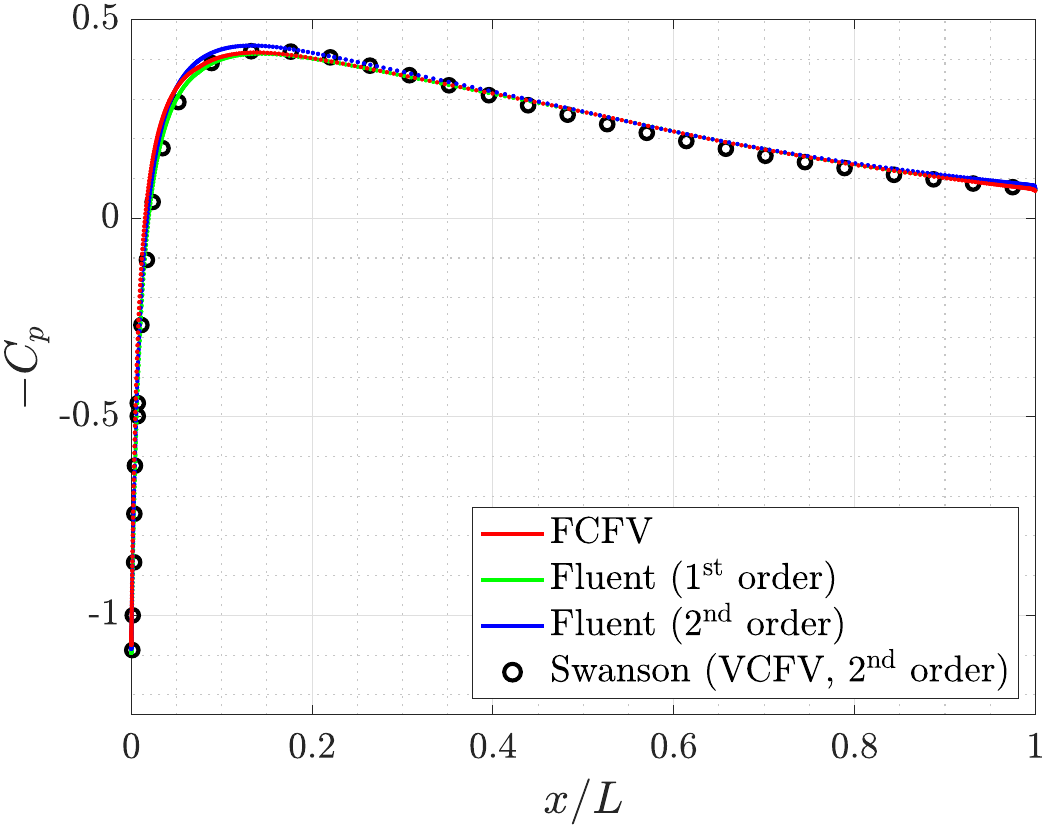}} \hfill
	\subfloat[Pressure coeff., transonic]{\includegraphics[width=0.32\textwidth]{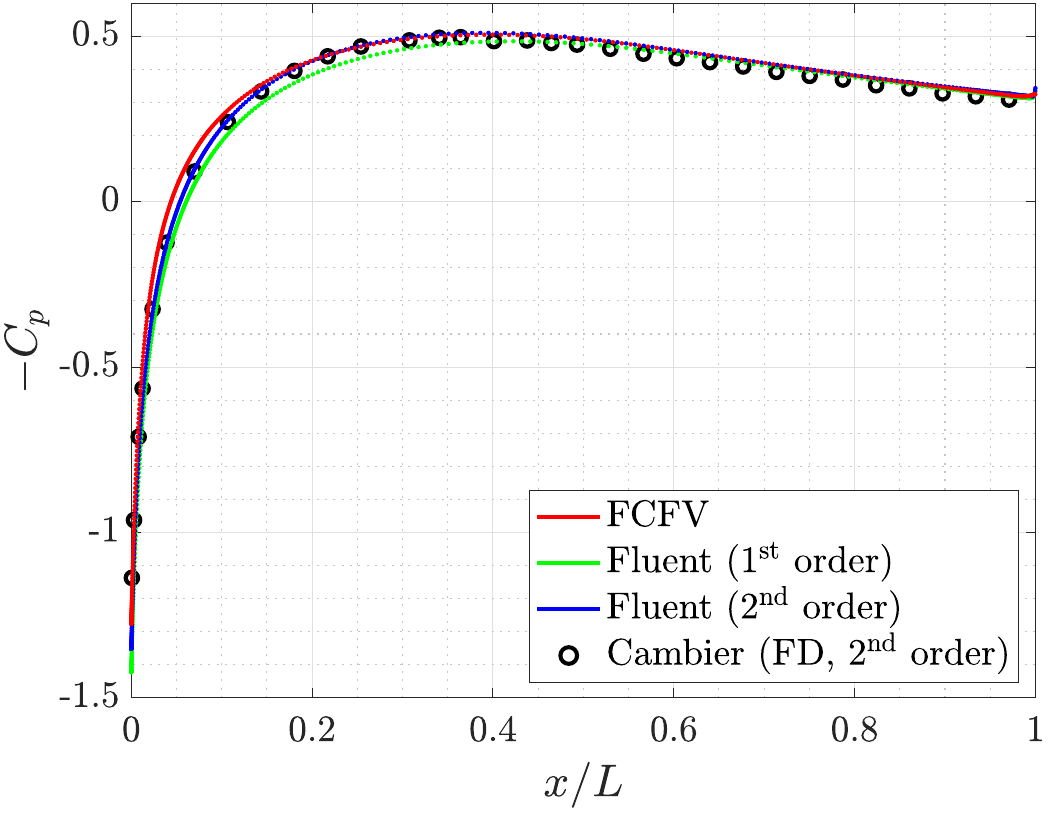}} \hfill
	\subfloat[Pressure coeff.,  supersonic]{\includegraphics[width=0.32\textwidth]{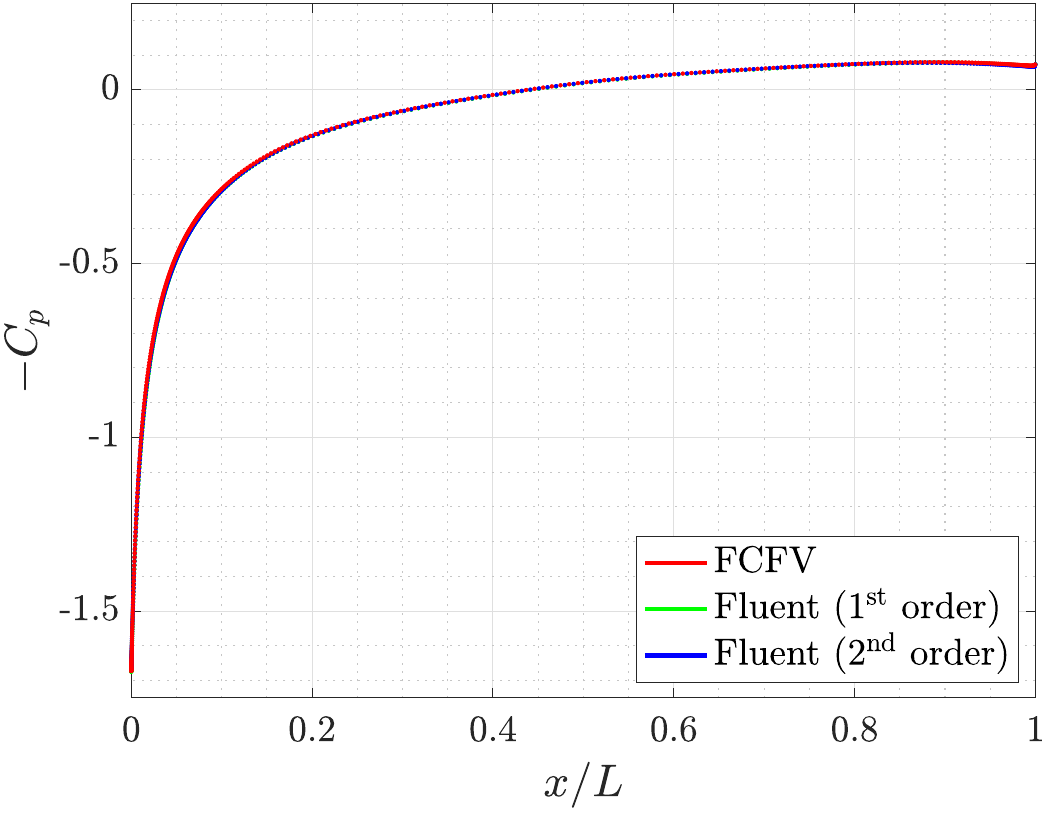}} \\
	\subfloat[Skin friction, subsonic]{\includegraphics[width=0.32\textwidth]{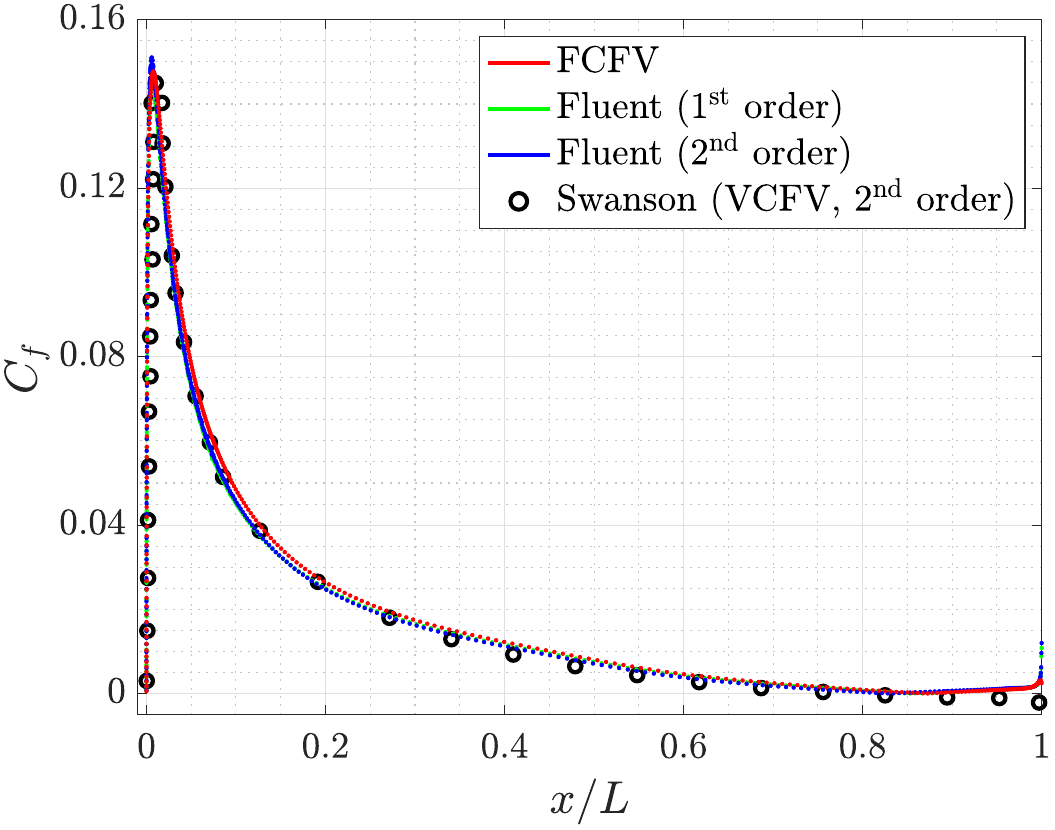}} \hfill
	\subfloat[Skin friction,  transonic]{\includegraphics[width=0.32\textwidth]{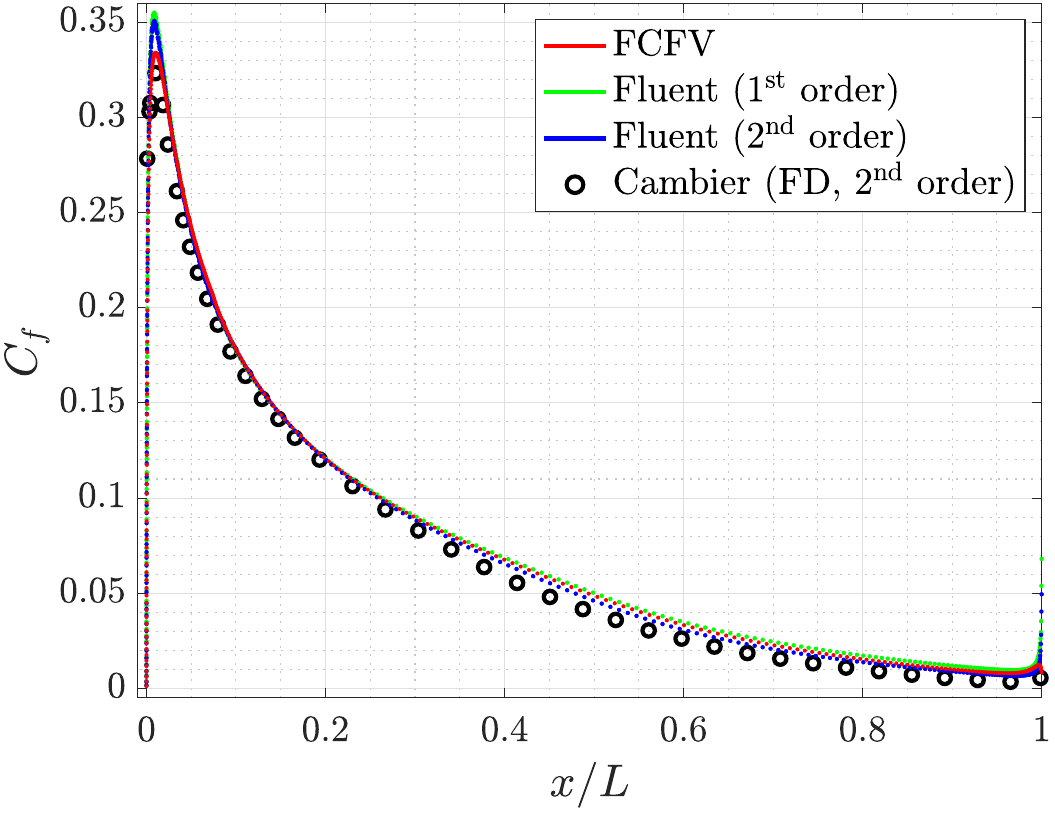}} \hfill
	\subfloat[Skin friction,  supersonic]{\includegraphics[width=0.32\textwidth]{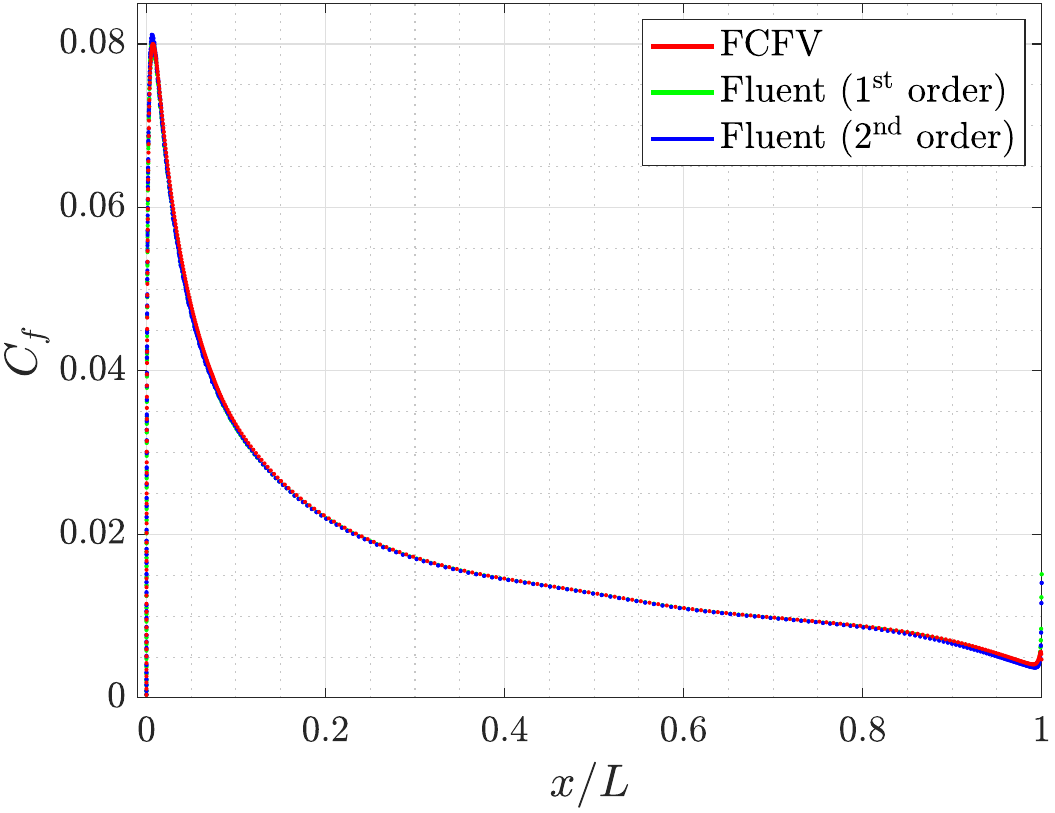}}
	\caption{Viscous laminar flow cases over a NACA 0012 aerofoil -- Pressure and skin friction coefficients on the aerofoil surface for the subsonic, transonic and supersonic flows, computed on the finest mesh refinement using the FCFV and the CCFV methods by Ansys Fluent.} 
	\label{fig:NACAviscous_Coefficients}
\end{figure}

In order to evaluate the accuracy of the FCFV method in predicting the pressure and the viscous contributions of the aerodynamic forces, the pressure and skin friction coefficients obtained with the finest mesh are presented in figure~\ref{fig:NACAviscous_Coefficients} for the three different cases. Available reference solutions of these coefficients for the subsonic~\citep{Swanson2016} and transonic~\citep{Cambier1987} examples are included for comparison, as well as the results provided by Ansys Fluent CCFV solvers. In the three cases,  only minimal differences are observed among the FCFV and the CCFV results and all numerical solutions are in excellent agreement with the references available in the literature. These results certify the robustness of the FCFV scheme, even in the presence of highly stretched meshes. Moreover,  they showcase the capability of the method to accurately predict relevant metrics in viscous laminar flows,  highlighting its competitiveness with respect to existing commercial CCFV solvers.
%
%\begin{figure}[!ht]
%	\subfloat[Subsonic]{\includegraphics[width=0.45\textwidth]{NACAsubsonic_Cp}} \hfill
%	\subfloat[Subsonic]{\includegraphics[width=0.45\textwidth]{NACAsubsonic_Cf}}\\
%	\subfloat[Transonic]{\includegraphics[width=0.45\textwidth]{NACAtransonic_Cp}} \hfill
%	\subfloat[Transonic]{\includegraphics[width=0.45\textwidth]{NACAtransonic_Cf}} \\
%	\subfloat[Supersonic]{\includegraphics[width=0.45\textwidth]{NACAsupersonic_Cp}} \hfill
%	\subfloat[Supersonic]{\includegraphics[width=0.45\textwidth]{NACAsupersonic_Cf}}
%	\caption{Viscous laminar flow cases over a NACA 0012 aerofoil -- Pressure (left) and skin friction (right) coefficients along the aerofoil surface for the subsonic (top), transonic (middle) and supersonic (bottom) examples, obtained for the FCFV method and the CCFV approaches by Fluent on the finest mesh refinement.} 
%	\label{fig:NACAviscous_Coefficients}
%\end{figure}
%

%To quantitatively compare the accuracy of the numerical schemes under analysis, the mesh convergence of the pressure and the viscous contributions of the drag coefficient computed using the FCFV and Ansys Fluent first and second-order CCFV methods is presented in table~\ref{tb:NACAviscous_Drag}. The results display only minor differences among the methods, all showing good agreement with the reference values available in the literature. In particular, Ansys Fluent second-order CCFV scheme achieves convergence even on coarse meshes and it provides accurate approximations of both the individual contributions --when available-- and the total drag force.  The predictions computed using the FCFV and the first-order CCFV method also match the reference value of the total drag force. Nonetheless,  comparing the estimated pressure drags in the subsonic case with the value determined by the second-order scheme,  FCFV underpredicts this quantity, whereas the first-order CCFV tends to overestimate the result. It is worth noticing that the error introduced by the two methods is comparable, with a discrepancy of approximately 30 drag counts in both cases. Finally,  the results for the viscous drag in the subsonic case display a higher accuracy of the first-order CCFV scheme by Ansys Fluent which shows a closer behaviour to the second-order scheme than the FCFV method.

\hl{
To quantitatively compare the accuracy of the numerical schemes under analysis,  table~\ref{tb:NACAviscous_Drag} reports the pressure and the viscous contributions of the drag coefficient, computed on different meshes, using the FCFV and Ansys Fluent first and second-order CCFV methods. The results display only minor differences among the methods, all showing good agreement with the reference values available in the literature. In particular, Ansys Fluent second-order CCFV scheme achieves convergence even on coarse meshes and it provides accurate approximations of both the individual contributions --when available-- and the total drag force.  The predictions computed using the FCFV and the first-order CCFV method also match the reference value of the total drag force. Nonetheless,  some differences are identified when the pressure and viscous contributions are studied independently.  On the one hand,  considering the pressure drag estimated by the second-order scheme as a reference value, FCFV underpredicts this quantity, whereas the first-order CCFV tends to overestimate the result. It is worth noticing that the error introduced by the two methods is however comparable, with a discrepancy of approximately 30 drag counts with respect to the reference value in both cases.  On the other hand,  in the prediction of the viscous drag,  the first-order CCFV scheme by Ansys Fluent outperforms the FCFV method, providing more accurate results,  closer to the reference values of the second-order approximation.
}
\begin{table}[!ht]
	\centering
	\hspace*{-25pt}\begin{tabular}{|c|ccc|ccc|ccc|}
		\hline
		\multicolumn{10}{|c|}{\textbf{Subsonic}} \\
		\multicolumn{10}{|c|}{\citep{Swanson2016,Bassi-BR:1997b,Mavriplis1990}} \\
		\hline
		\multirow{2}{*}{Mesh} & \multicolumn{3}{c|}{$C_{d_{p}}$} & \multicolumn{3}{c|}{$C_{d_{v}}$} & \multicolumn{3}{c|}{$C_d$} \\
		& FCFV & Fluent-1 & Fluent-2 & FCFV & Fluent-1 & Fluent-2 & FCFV & Fluent-1 & Fluent-2 \\
		\hline
		1 & 0.0392 & 0.0450 & 0.0231 & 0.0470 & 0.0390 & 0.0332 & 0.0862 & 0.0840 & 0.0563 \\
		2 & 0.0291 & 0.0361 & 0.0226 & 0.0431 & 0.0360 & 0.0330 & 0.0722 & 0.0721 & 0.0556 \\
		3 & 0.0234 & 0.0305 & 0.0225 & 0.0416 & 0.0350 & 0.0329 & 0.0650 & 0.0655 & 0.0554 \\
		4 & 0.0197 & 0.0256 & 0.0226 & 0.0391 & 0.0335 & 0.0333 & 0.0588 & 0.0591 & 0.0559 \\
		\hline
		Refs.  & \multicolumn{3}{c|}{$[0.0196,0.0288]$} & \multicolumn{3}{c|}{$[0.0305,0.0344]$} & \multicolumn{3}{c|}{$[0.0501,0.0632]$} \\
		\hline		\hline
		\multicolumn{10}{|c|}{\textbf{Transonic}} \\
		\multicolumn{10}{|c|}{\citep{Cambier1987}} \\		
		\hline
		\multirow{2}{*}{Mesh} & \multicolumn{3}{c|}{$C_{d_{p}}$} & \multicolumn{3}{c|}{$C_{d_{v}}$}  & \multicolumn{3}{c|}{$C_d$} \\
		& FCFV & Fluent-1 & Fluent-2 & FCFV & Fluent-1 & Fluent-2 & FCFV & Fluent-1 & Fluent-2 \\
		\hline
		1 & 0.0921 & 0.1055 & 0.0866 & 0.1733 & 0.1613 & 0.1455 & 0.2654 & 0.2668 & 0.2321 \\
		2 & 0.0840 & 0.0964 & 0.0863 & 0.1654 & 0.1535 & 0.1453 & 0.2494 & 0.2499 & 0.2316 \\
		3 & 0.0800 & 0.0923 & 0.0862 & 0.1619 & 0.1501 & 0.1452 & 0.2419 & 0.2424 & 0.2314 \\
		4 & 0.0780 & 0.0917 & 0.0865 & 0.1576 & 0.1512 & 0.1458 & 0.2356 & 0.2429 & 0.2323 \\
		\hline
		Ref.  & \multicolumn{3}{c|}{} & \multicolumn{3}{c|}{} & \multicolumn{3}{c|}{$[0.2176,0.2420]$} \\
		\hline		\hline
		\multicolumn{10}{|c|}{\textbf{Supersonic}} \\
		\hline
		\multirow{2}{*}{Mesh} & \multicolumn{3}{c|}{$C_{d_{p}}$} & \multicolumn{3}{c|}{$C_{d_{v}}$}  & \multicolumn{3}{c|}{$C_d$} \\
		& FCFV & Fluent-1 & Fluent-2 & FCFV & Fluent-1 & Fluent-2 & FCFV & Fluent-1 & Fluent-2 \\
		\hline
		1 & 0.0988 & 0.1006 & 0.0962 & 0.0369 & 0.0354 & 0.0342 & 0.1357 & 0.1360 & 0.1304 \\
		2 & 0.0980 & 0.0986 & 0.0958 & 0.0356 & 0.0348 & 0.0341 & 0.1336 & 0.1334 & 0.1299 \\
		3 & 0.0959 & 0.0974 & 0.0959 & 0.0353 & 0.0345 & 0.0341 & 0.1312 & 0.1319 & 0.1300 \\
		4 & 0.0951 & 0.0962 & 0.0957 & 0.0349 & 0.0342 & 0.0341 & 0.1300 & 0.1304 & 0.1298 \\
		\hline
	\end{tabular}
	\caption{Viscous laminar flow cases over a NACA 0012 aerofoil -- Mesh convergence of the pressure ($C_{d_{p}}$) and viscous ($C_{d_{v}}$) contributions of the drag coefficient for the subsonic, transonic and supersonic examples, computed using the FCFV and the first-order (Fluent-1) and second-order (Fluent-2) CCFV solvers by Ansys Fluent.} 
	\label{tb:NACAviscous_Drag}
\end{table}

\hl{
Finally, the convergence of the error of the estimated drag and lift coefficients upon mesh refinement is presented for the three cases under analysis (Fig.~\ref{fig:NACAviscous_DragLift}).  
In the top row, the percent error in the drag coefficient computed by the different numerical methods is reported,  considering the reference value provided by the converged second-order CCFV solution by Ansys Fluent on the finest mesh. The figures clearly display that the second-order CCFV scheme converges to the reference solution even on coarse meshes.  The FCFV and the first-order CCFV approximations show comparable performance, achieving errors of $5\%$ or less in all flow regimes. It is worth mentioning that higher errors are displayed in the subsonic case, whereas solutions of the supersonic problem feature the lowest levels of error. In the transonic case, the FCFV method outperforms the first-order CCFV scheme, showing better convergence towards the reference solution.
The bottom row of figure~\ref{fig:NACAviscous_DragLift} presents the percent error in the computed lift coefficient for the three cases. It is worth recalling that, given the symmetry of the flows under analysis,  a zero lift reference value is considered.  Hence, the estimated lift coefficient provides a measure of the discretisation error introduced by the different numerical schemes.  In the performed studies, errors in the lift below $0.1\%$ are achieved by all methods.  In particular, the FCFV scheme produces the most accurate approximations of the lift coefficient in the three regimes, with errors between $10^{-3}\%$ and $10^{-6}\%$.
\begin{figure}[!ht]
	\subfloat[Drag coefficient, subsonic]{\includegraphics[width=0.32\textwidth]{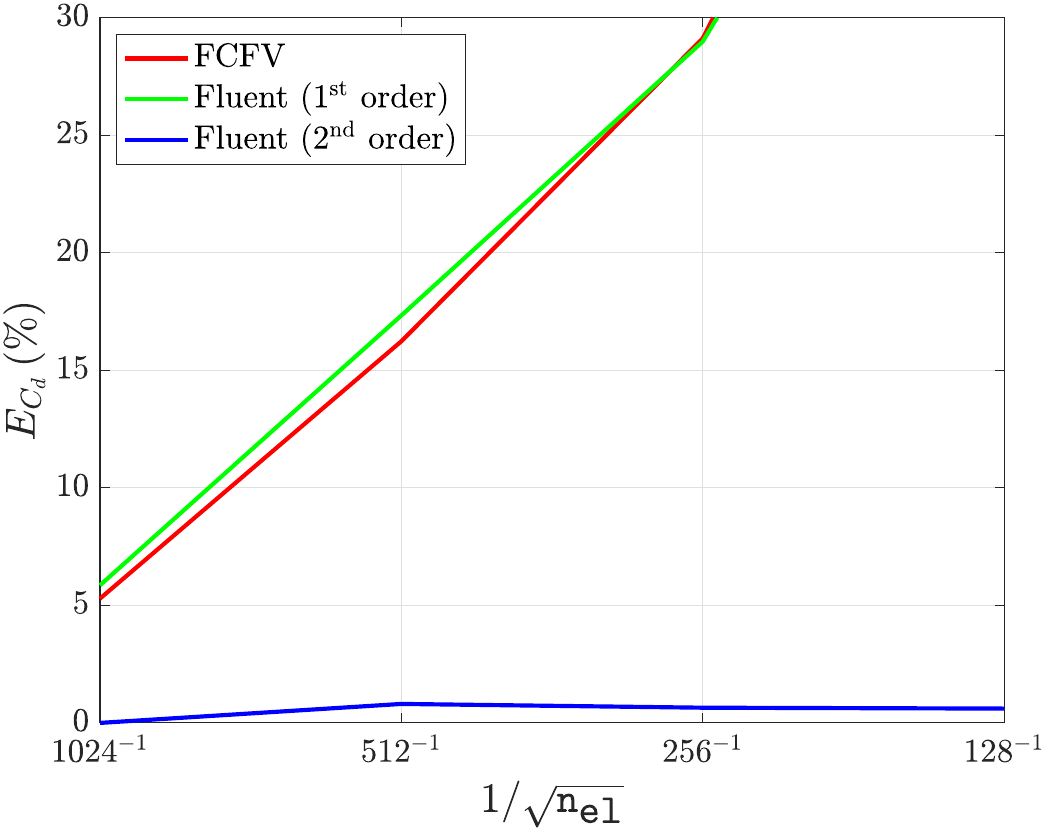}} \hfill
	\subfloat[Drag coefficient, transonic]{\includegraphics[width=0.32\textwidth]{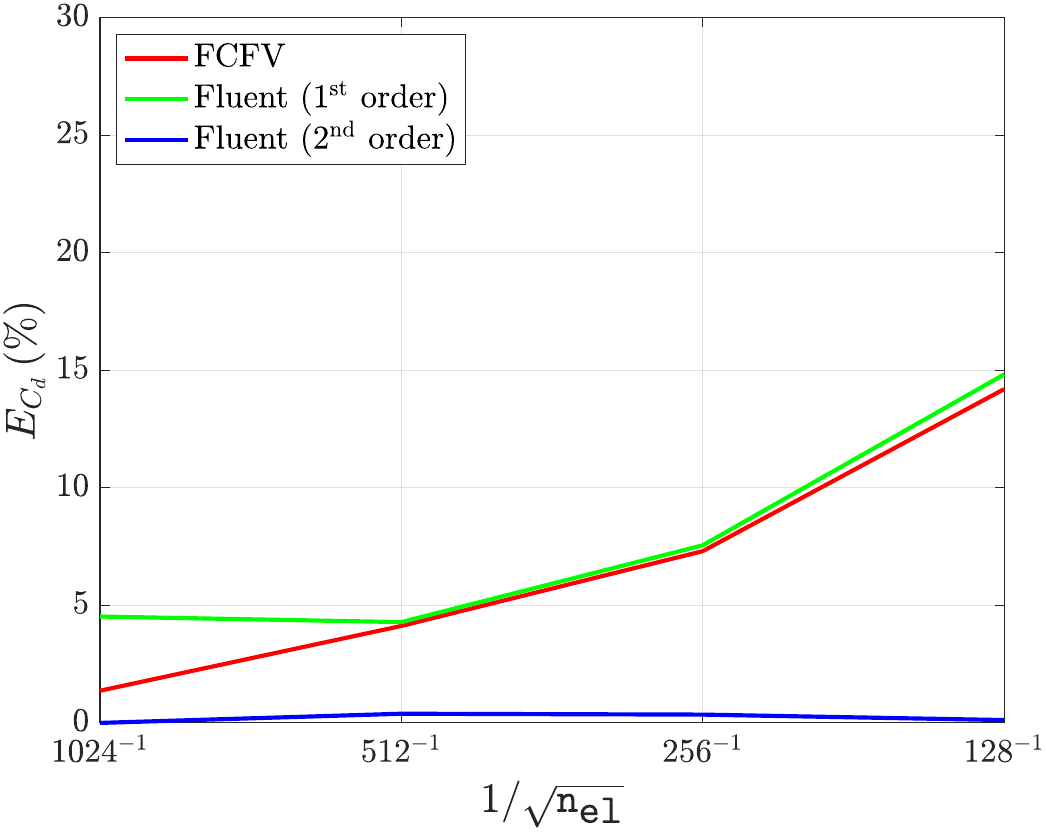}} \hfill
	\subfloat[Drag coefficient,  supersonic]{\includegraphics[width=0.32\textwidth]{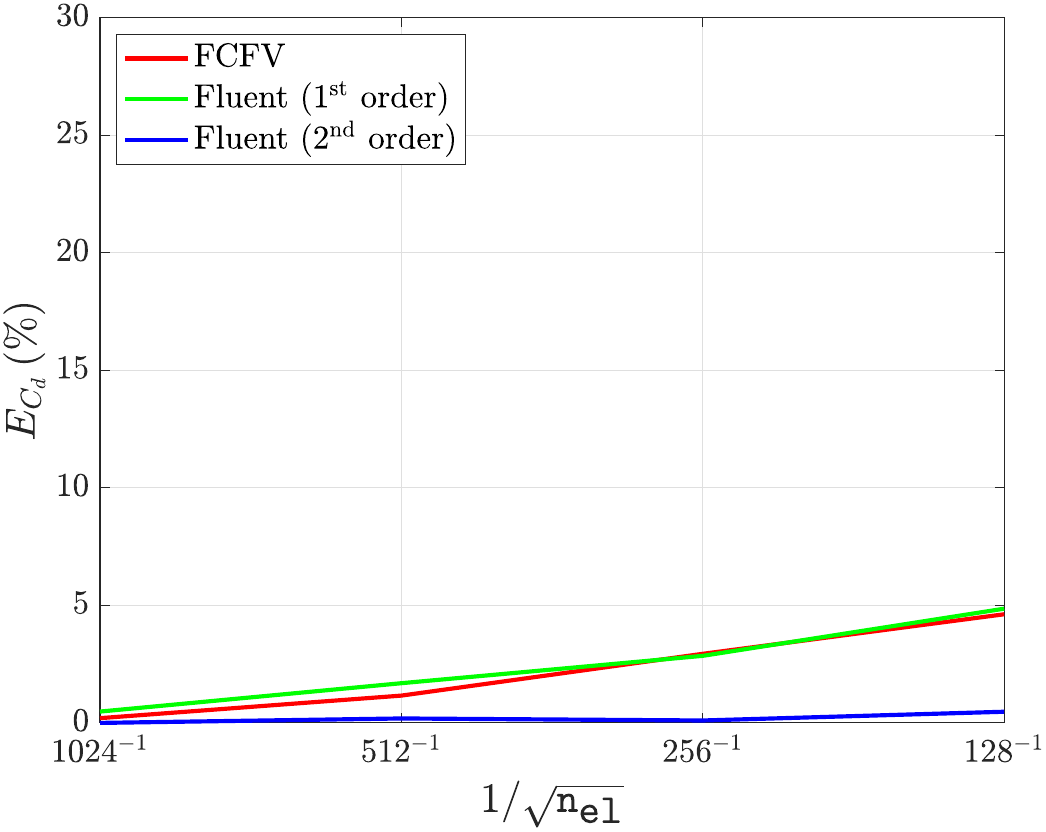}} \\
	\subfloat[Lift coefficient, subsonic]{\includegraphics[width=0.32\textwidth]{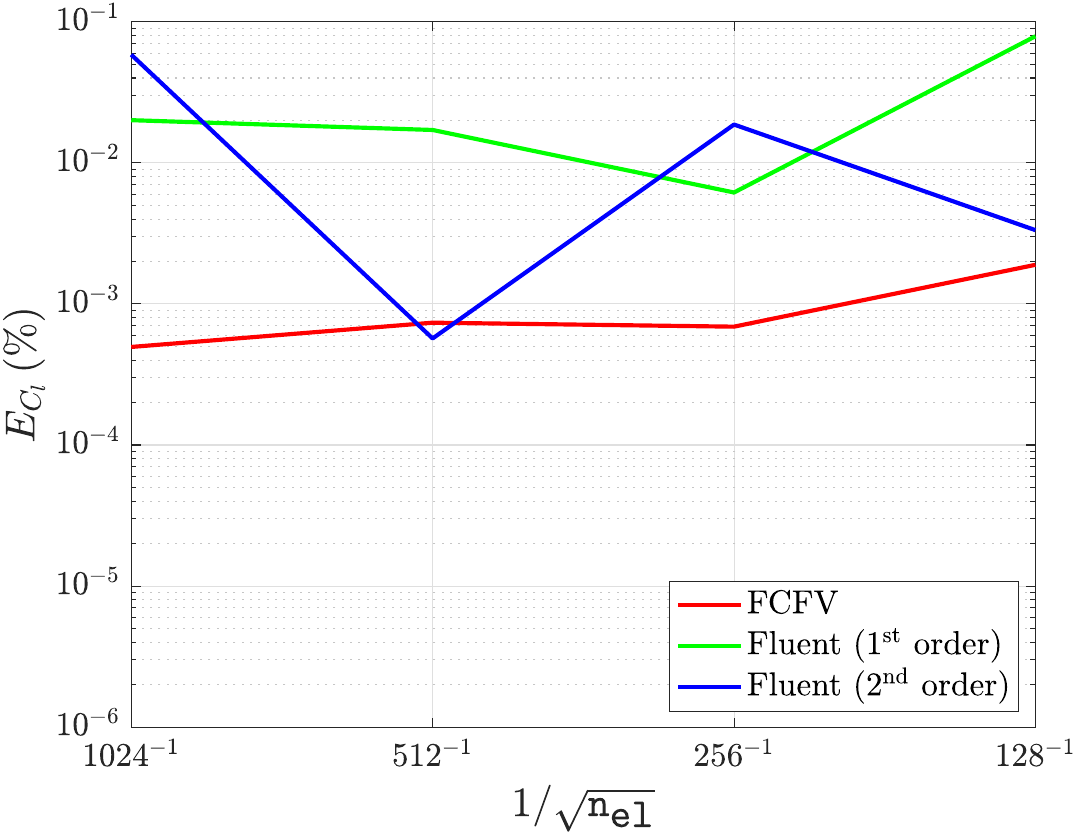}} \hfill
	\subfloat[Lift coefficient,  transonic]{\includegraphics[width=0.32\textwidth]{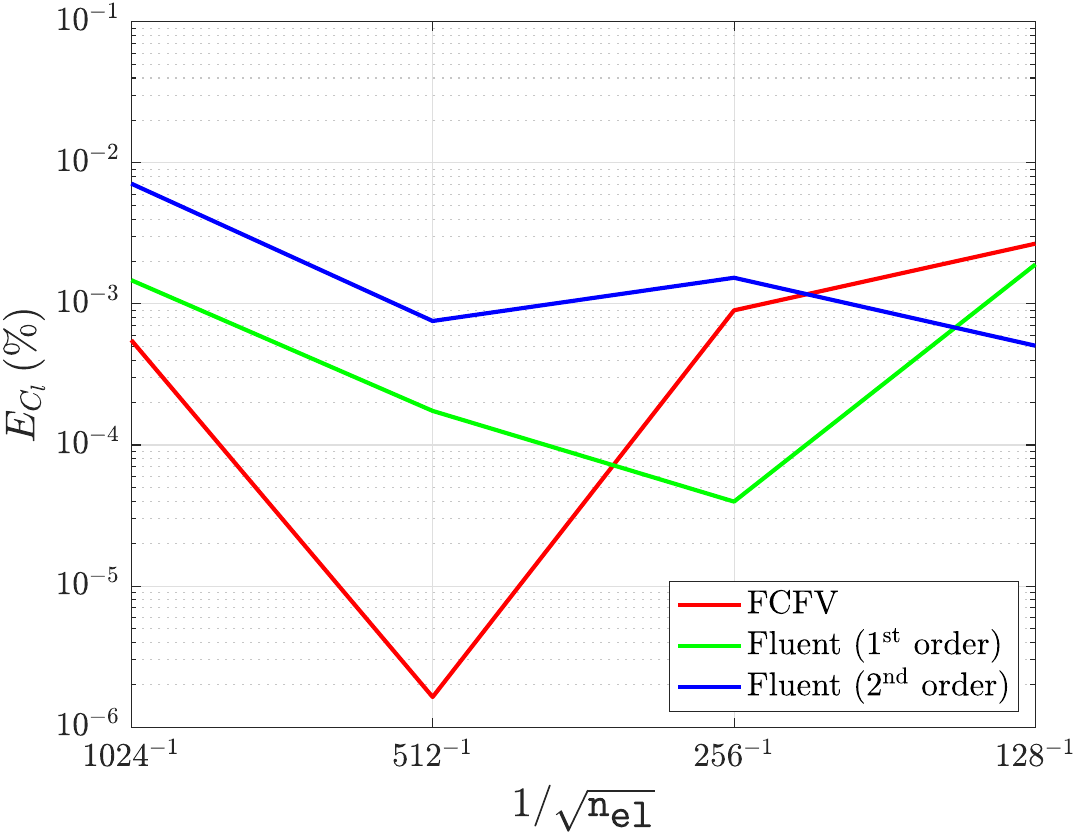}} \hfill
	\subfloat[Lift coefficient,  supersonic]{\includegraphics[width=0.32\textwidth]{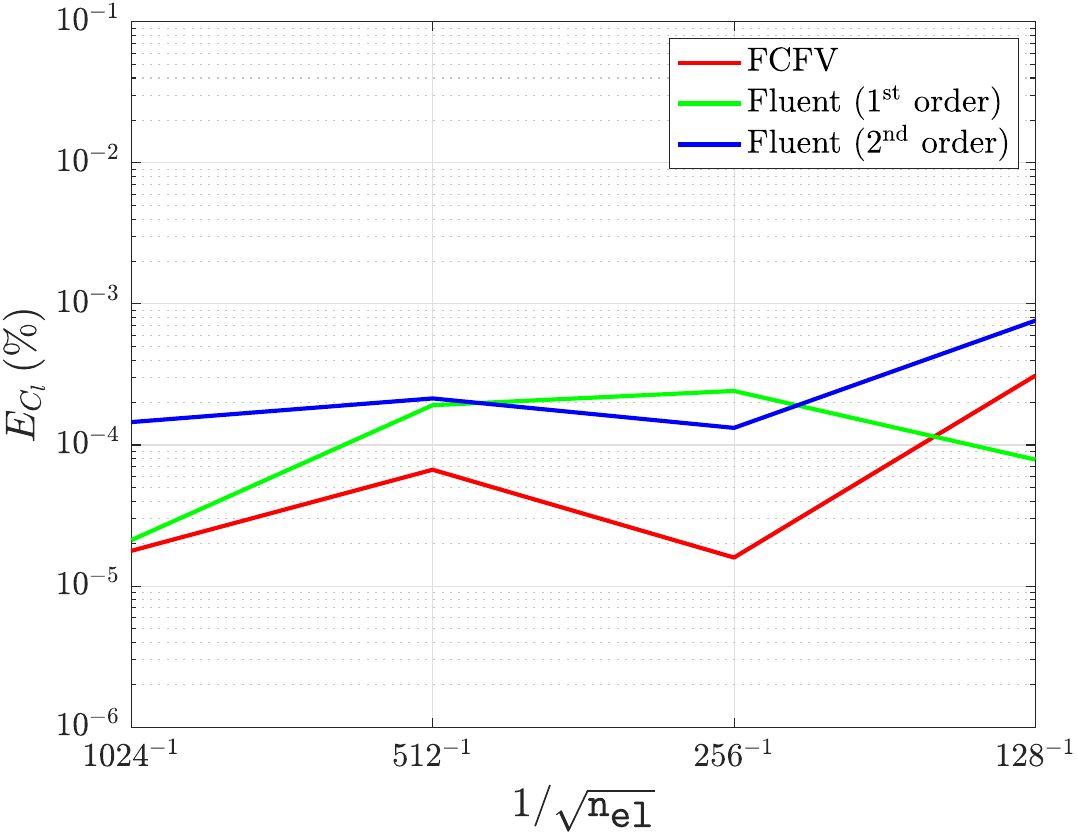}}
	\caption{\hl{Viscous laminar flow cases over a NACA 0012 aerofoil -- Percent error of the drag and lift coefficients for the subsonic, transonic and supersonic flows, computed using the FCFV and the CCFV methods by Ansys Fluent on different mesh refinements. %Reference value for the drag coefficient: solution provided by the second-order CCFV method on the finest mesh. Reference value for the lift coefficient: zero.
	}} 
	\label{fig:NACAviscous_DragLift}
\end{figure}
}

The presented viscous laminar flow cases show the robustness of the FCFV scheme in a variety of flow conditions, from subsonic to supersonic regime, even in the presence of meshes with high aspect ratio. Moreover, the method displays good performance in predicting relevant aerodynamic quantities matching the accuracy of Ansys Fluent first-order CCFV scheme. \hl{However, the second-order CCFV method maintains its superiority in the approximation of viscous flows, providing better estimates of aerodynamic quantities and converging to reference solutions,  even when coarse meshes are employed.}

%------------------------------------------------------------------------------------------------------------
\subsection{Inviscid flow cases over a NACA 0012}

In this section,  three cases of inviscid flow over a  NACA 0012 aerofoil at different angles of attack are presented, in subsonic~\citep{Sevilla-SHM:2013,Nogueira2009}, transonic~\citep{Sevilla-SHM:2013,Thibert-TGO:1979,Balan2015,Balan2012,Yano2012} and supersonic~\citep{Balan2015,Persson-PP:2006} regimes.  The details of the flow conditions are presented in table~\ref{tb:NACAinviscid_FlowConditions}. The objective of these tests is to assess the robustness of the FCFV method in purely inviscid flows, ranging from smooth to discontinuous solutions with shocks, and to compare the results with high-order reference solutions and CCFV results computed using Ansys Fluent.
\begin{table}[!ht]
	\centering
	\begin{tabular}{|c|c|c|}
		\hline
		Subsonic case & Transonic case & Supersonic case\\
		\hline
		$\Minf = 0.63$ & $\Minf = 0.8$ & $\Minf = 1.5$ \\
		$\alpha = 2^\circ$ & $\alpha = 1.25^\circ$ & $\alpha = 0^\circ$ \\
		\hline
	\end{tabular}
	\caption{Inviscid flow cases over a NACA 0012 aerofoil -- Flow conditions.} 
	\label{tb:NACAinviscid_FlowConditions}
\end{table}

\begin{figure}[!ht]
	\centering
	\subfloat[Mesh]{\includegraphics[width=0.33\textwidth]{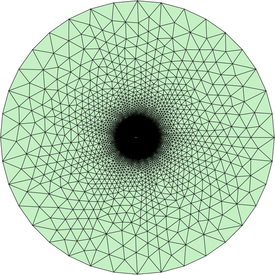}} \qquad
	\subfloat[Close-up view around the aerofoil]{\includegraphics[width=0.35\textwidth]{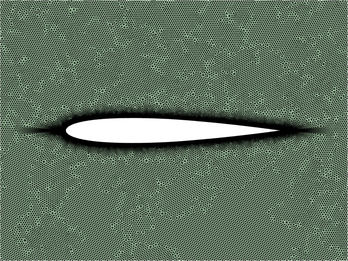}}
	\caption{\hl{Inviscid flow cases over a NACA 0012 aerofoil -- Mesh.}} 
	\label{fig:NACAinviscid_Meshes}
\end{figure}

The aerofoil is embedded in a computational domain which extends up to 15 chord lengths from the surface. Inviscid wall conditions are imposed on the aerofoil surface, whereas far-field conditions are enforced on the outer boundary by means of the Riemann solver. The unstructured mesh designed for inviscid flow simulations is depicted in figure~\ref{fig:NACAinviscid_Meshes} and it consists of \hl{236,178} triangular cells, with non-uniform refinement on the surface of the aerofoil and at the leading and trailing edges. \hl{The resulting mesh, designed for different flow regimes without any case-dependent grid adaptation,  is capable of capturing strong and weak shocks near the aerofoil as well as detached bow shocks,  making it suitable for subsonic, transonic and supersonic simulations.}

\begin{figure}[!ht]
	\centering
	\subfloat[Subsonic]{\includegraphics[width=0.32\textwidth]{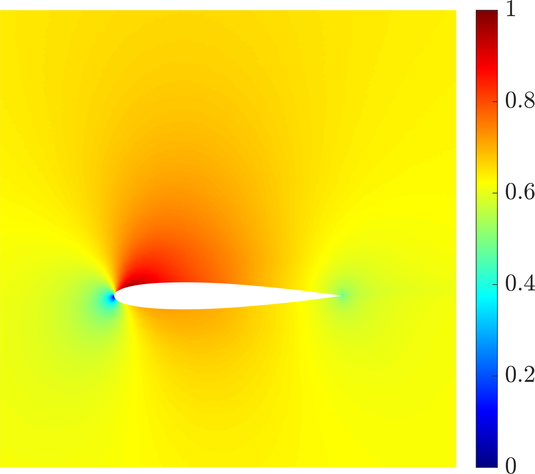}} \hfill
	\subfloat[Transonic]{\includegraphics[width=0.32\textwidth]{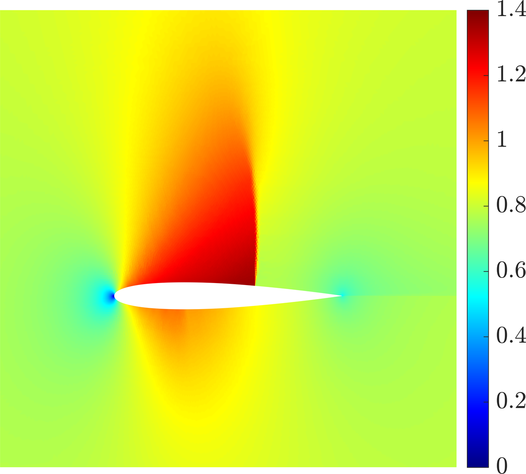}} \hfill	
	\subfloat[Supersonic]{\includegraphics[width=0.32\textwidth]{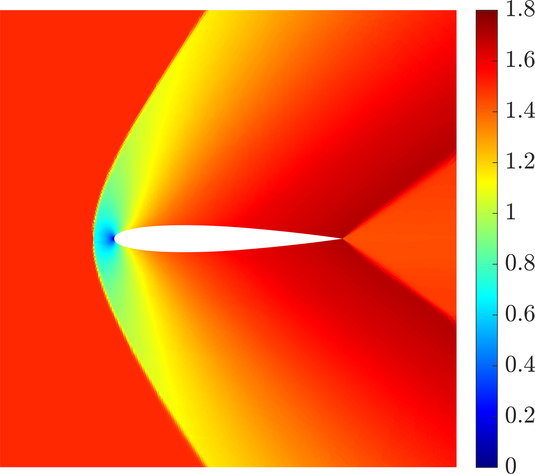}}
	\caption{\hl{Inviscid flow cases over a NACA 0012 aerofoil -- Mach number distribution obtained with the FCFV method for the three cases of the study.}}
	\label{fig:NACAinviscid_FCFV}
\end{figure}

The FCFV solver shows good performance in the solution of all three problems and a detail of the computed Mach number distributions is displayed in figure~\ref{fig:NACAinviscid_FCFV}.

\begin{figure}[!ht]
	\subfloat[FCFV, transonic]{\includegraphics[width=0.32\textwidth]{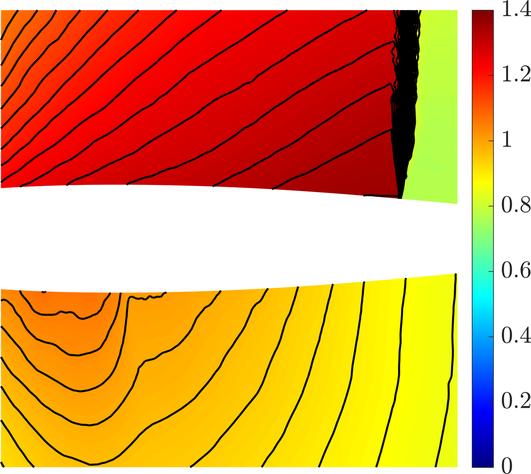}\label{fig:NACAinviscid_TransonicFCFV}} \hfill
	\subfloat[Fluent-1, transonic]{\includegraphics[width=0.32\textwidth]{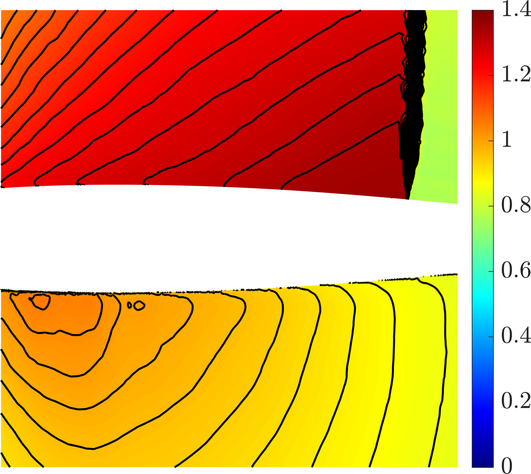}} \hfill
	\subfloat[Fluent-2, transonic]{\includegraphics[width=0.32\textwidth]{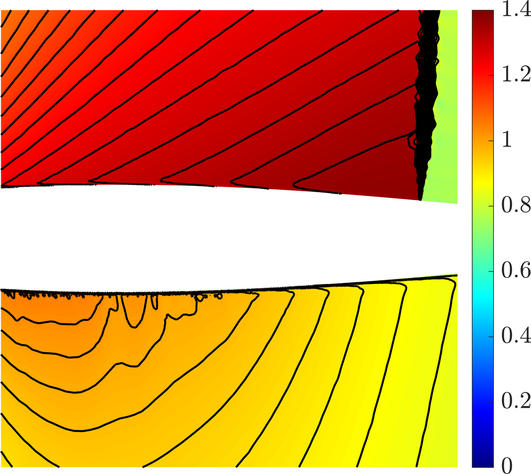} \label{fig:NACAinviscid_TransonicFluent2}}

	\subfloat[FCFV, supersonic]{\includegraphics[width=0.32\textwidth]{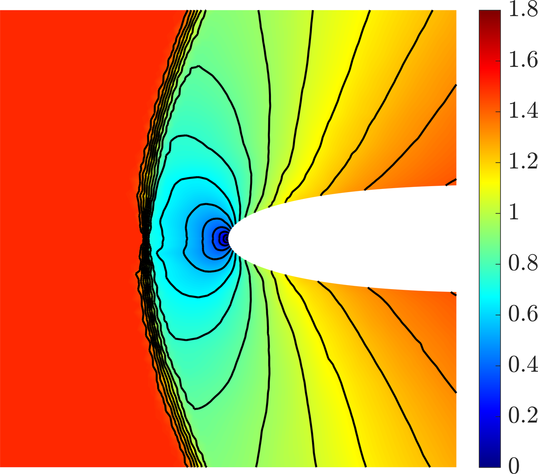}
\label{fig:NACAinviscid_SupersonicFCFV}} \hfill
	\subfloat[Fluent-1, supersonic]{\includegraphics[width=0.32\textwidth]{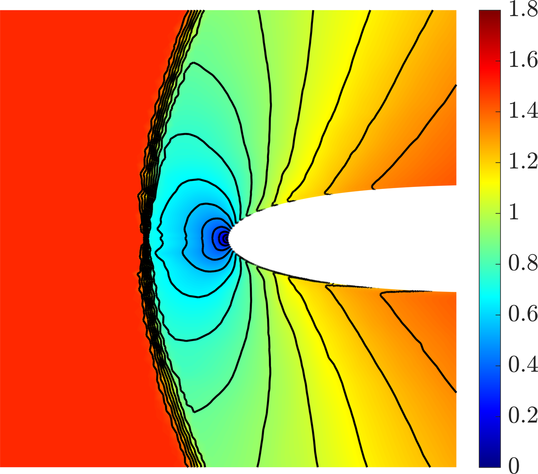}\label{fig:NACAinviscid_SupersonicFluent1}} \hfill
	\subfloat[Fluent-2, supersonic]{\includegraphics[width=0.32\textwidth]{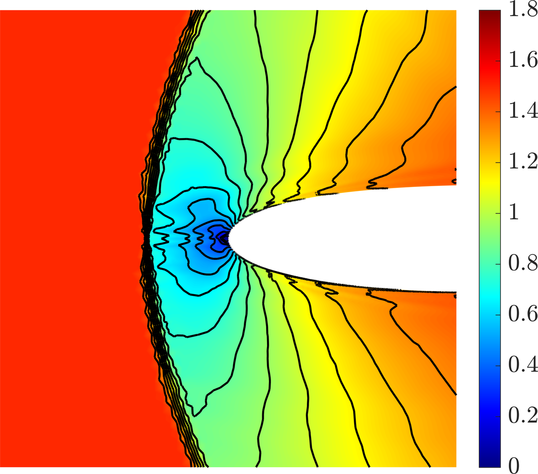} \label{fig:NACAinviscid_SupersonicFluent2}}
	\caption{\hl{Inviscid flow cases over a NACA 0012 aerofoil -- Close-up view of the contour plots of the Mach number distribution around the shock for the transonic and supersonic cases computed using the FCFV, the first-order (Fluent-1) and second-order (Fluent-2) CCFV solvers by Ansys Fluent.}}
	\label{fig:NACAinviscid_All}
\end{figure}

On the contrary,  the CCFV method by Ansys Fluent shows an excellent behaviour in simulating the subsonic flow but it suffers in the transonic and supersonic cases, see figure~\ref{fig:NACAinviscid_All}. More precisely, both approaches struggle to achieve a steady-state solution and localised oscillations appear, especially in the vicinity of strong shock waves and when the second-order scheme is employed, as visible in  figure~\ref{fig:NACAinviscid_SupersonicFluent2}. Such a difficulty of the second-order CCFV method to compute a smooth approximation is responsible for both the deterioration of the convergence of the solver and the overall loss of quality of the flowfield (cf.  figure~\ref{fig:NACAinviscid_TransonicFluent2}).  The first-order CCFV approach remedies some of the above issues and it displays reasonable performance,  with smoother approximation of the contour lines of the Mach number distribution in both the transonic and the supersonic case. Nonetheless,  figure~\ref{fig:NACAinviscid_SupersonicFluent1} shows that slight perturbations and instabilities are still present along the stagnation line of the supersonic flow, in the region between the bow shock and the leading edge. \hl{In this region, the FCFV method achieves an accuracy similar to the first-order CCFV approximation, whereas overall smoother representations of the flowfield are obtained in the remainder of the domain for all regimes. Indeed,  it is worth noticing that the contour lines of the Mach number distribution computed using Ansys Fluent solvers tend to experience a very localised variation near the wall, as opposed to the clear intersection with the wall surface shown by the FCFV solution in figure~\ref{fig:NACAinviscid_TransonicFCFV}.  In this figure, it can also be appreciated the ability of the FCFV method to capture the weak shock on the lower side of the aerofoil, which is not detected by the CCFV solvers.}

\begin{figure}[!ht]
	\centering
	\subfloat[Subsonic]{\includegraphics[width=0.32\textwidth]{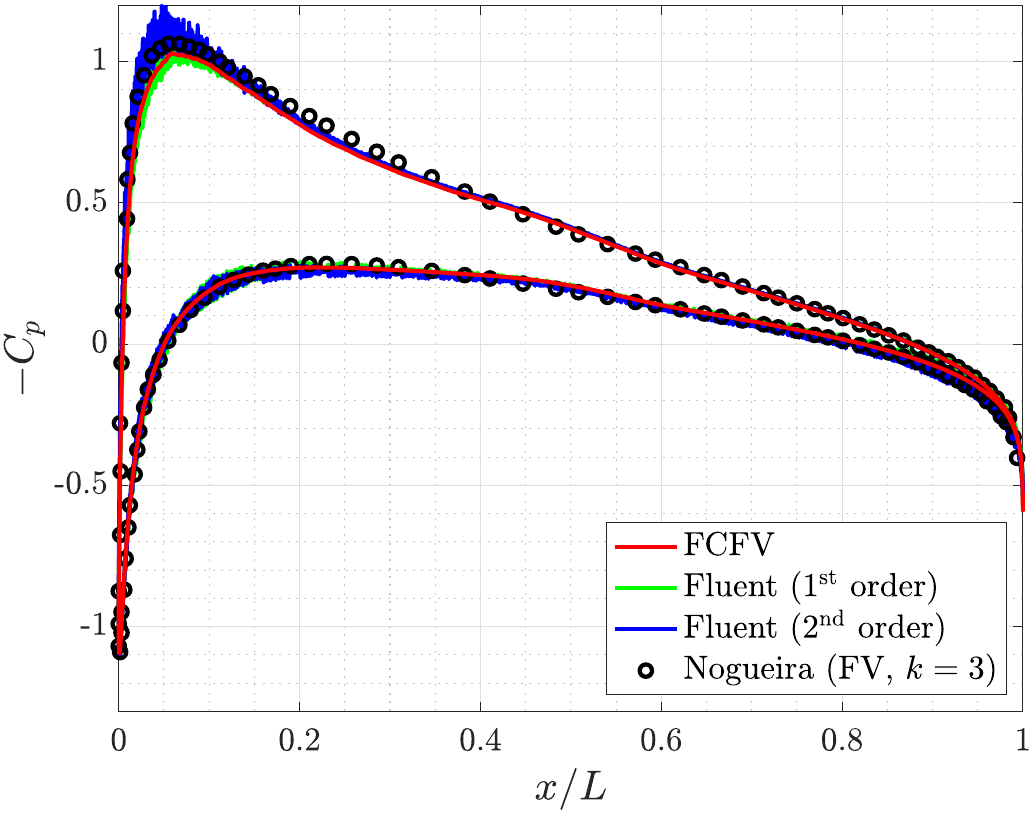}
\label{fig:NACAinviscid_Cp_sub}} \hfill
	\subfloat[Transonic]{\includegraphics[width=0.32\textwidth]{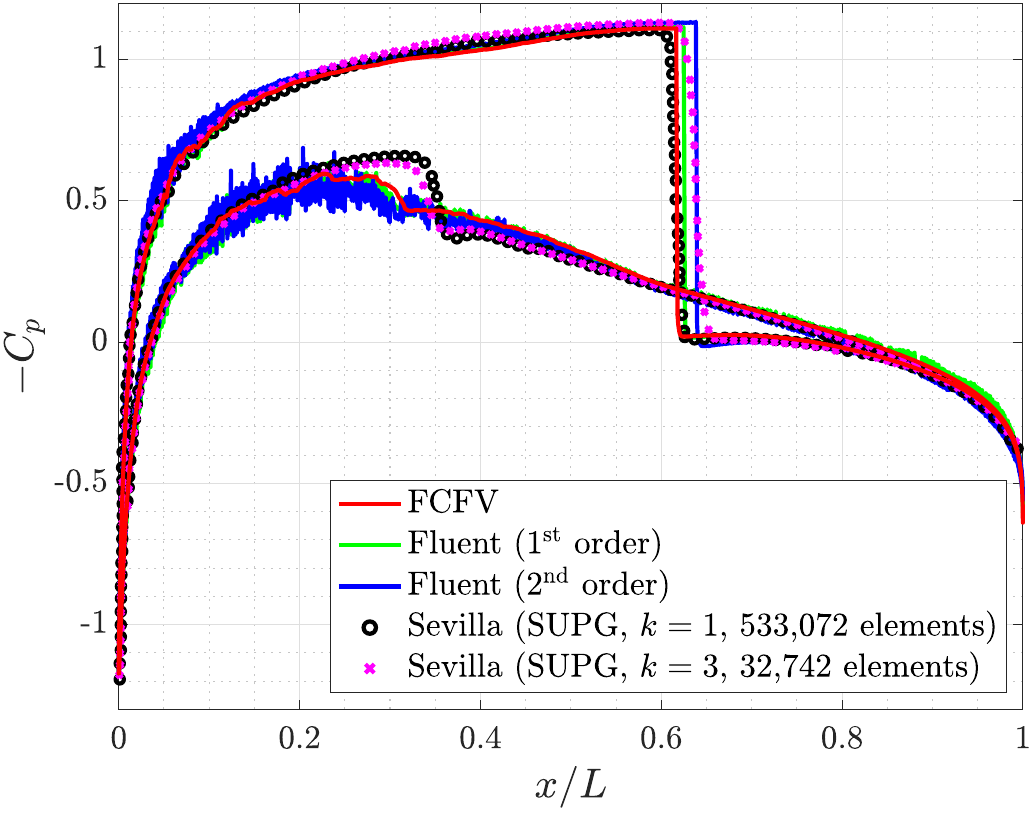}
\label{fig:NACAinviscid_Cp_trans}} \hfill
	\subfloat[Supersonic]{\includegraphics[width=0.32\textwidth]{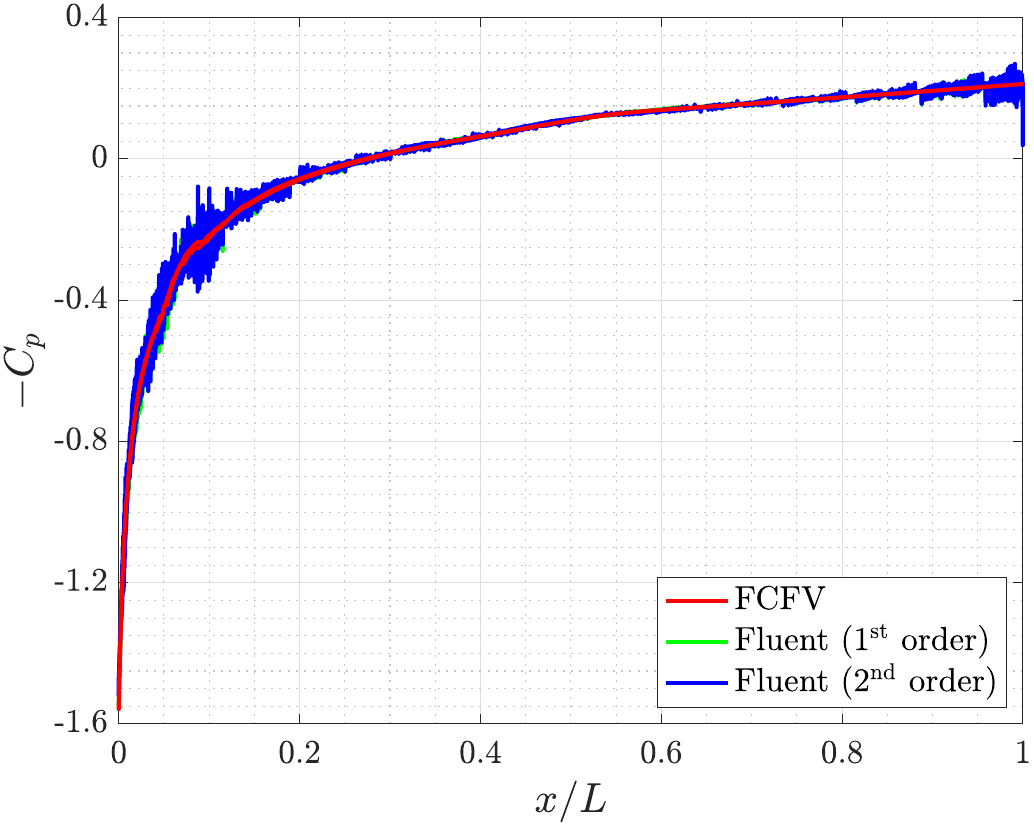}
\label{fig:NACAinviscid_Cp_super}}
	\caption{\hl{Inviscid flow cases over a NACA 0012 aerofoil -- Pressure coefficient along the aerofoil surface for the subsonic, transonic and supersonic examples, obtained using the FCFV and the CCFV methods by Ansys Fluent.}} 
	\label{fig:NACAinviscid_Cp}
\end{figure}

In order to assess the accuracy of the FCFV method in evaluating aerodynamic quantities of interest,  figure~\ref{fig:NACAinviscid_Cp} depicts the pressure coefficient obtained for the three inviscid flows under analysis, comparing it with available reference solutions and Ansys Fluent CCFV results.  In all cases, the FCFV solution provides a smooth approximation of the pressure coefficient,  whereas the first and second-order CCFV approaches present an oscillatory description of \hl{$C_p$}. Moreover, in both the subsonic and the supersonic case, the FCFV curve lies on top of the first and second-order solutions provided by Ansys Fluent, which show only minor differences from one another. Concerning the subsonic case in figure~\ref{fig:NACAinviscid_Cp_sub}, the FCFV approximation also displays excellent agreement with the reference solution computed using a high-order FV scheme~\citep{Nogueira2009}.  Figure~\ref{fig:NACAinviscid_Cp_trans} reports the more complex transonic regime. In this case, all tested FV methods are unable to accurately capture the position of the weak shock, as testified by the comparison with the high-order reference solution in~\citep{Sevilla-SHM:2013}. Concerning the approximation of the strong shock on the upper surface of the aerofoil, the FCFV method shows good agreement with a reference solution computed with polynomial approximation of degree 1 on a mesh of $533,072$ elements~\citep{Sevilla-SHM:2013}. Nonetheless, a discrepancy is observed when the pressure coefficient is compared to a high-order reference distribution obtained using polynomial approximation of degree 3 on a mesh of 32,742 elements. This is a well-known issue due to the role of geometric error in the production of nonphysical entropy and the need for accurate geometry approximation~\citep{Bassi-BR:97,Sevilla-SFH-11}. It is worth noticing that in the transonic case, the first and second-order CCFV solvers only present minor differences from one another and they both provide a more accurate prediction of the position of the strong shock wave than the FCFV scheme, although they tend to estimate a more vertical shock line than the high-order reference.

\hl{
A more quantitative assessment is reported in table~\ref{tb:NACAinviscid_Forces}, with a comparison of the computed drag and lift coefficients with high-order reference solutions. The results display that the FCFV method is capable of providing accurate predictions of the drag coefficient in all cases, whereas it yields larger errors in the approximation of the lift coefficient. More precisely,  the $C_d$ prediction for the transonic case lies 7 drag counts above the range of reference values, see~\citep{Sevilla-SHM:2013,Thibert-TGO:1979,Balan2015,Balan2012,Yano2012}, whereas in the supersonic case an error of only 4 drag counts is achieved with respect to the unique reference~\citep{Balan2015}.  Concerning the lift coefficient, the FCFV scheme introduces an error of 23-26 and 35-43 lift counts in the subsonic and transonic case, respectively. On the contrary, the first-order CCFV solution tends to experience larger errors in the approximation of the drag coefficient and increased accuracy in the $C_l$ predictions: an error in $C_d$ of 31-45 drag counts is achieved in the transonic case and it lowers to 14 drag counts in the supersonic case; for the lift coefficient,  the first-order CCFV scheme underpredicts the value by 14-17 and 14-22 lift counts in the subsonic and transonic case, respectively. Despite the large oscillations observed in figures~\ref{fig:NACAinviscid_SupersonicFluent2} and~\ref{fig:NACAinviscid_Cp},  the quantitative results in the table show that the second-order method by Ansys Fluent outperforms the accuracy of the first-order one, providing an estimate of $C_d$ within the range of published values in the literature for the transonic case and achieving an error of 7 drag counts in the supersonic case. The corresponding $C_l$ predictions overestimate the reference results by 3-6 and 12-20 lift counts for the subsonic and transonic case, respectively. Finally, the drag coefficient in the subsonic case and the lift coefficient in the supersonic case have zero reference value and can thus be interpreted as measures of the approximation error.  In this context, although a precision of $10^{-3}$ in the subsonic case and $10^{-4}$ in the supersonic case is achieved by all methods, the second-order CCFV scheme provides more accurate results than the FCFV solution in the subsonic problem, whereas the FCFV method outperforms Ansys Fluent solvers in the supersonic case.}
%
%\begin{table}[!ht]
%	\centering
%	\begin{tabular}{|l|ccc|ccc|}
%		\hline
%		\multicolumn{1}{|c|}{} & \multicolumn{3}{c|}{$C_{d}$} & \multicolumn{3}{c|}{$C_{l}$} \\
%		%		\hline
%		%		\multicolumn{1}{|c|}{Reference} & \multicolumn{3}{c|}{0.0228} & \multicolumn{3}{c|}{0.0328} \\
%		\hline 
%		Case & FCFV & Fluent-1 & Fluent-2 & FCFV & Fluent-1 & Fluent-2  \\
%		\hline
%		Subsonic & 0.0066 & 0.0057 & 0.0059 & 0.302 & 0.329 & 0.330\\
%		Transonic & 0.0223 & 0.0265 & 0.0265 & 0.313 & 0.366 & 0.366\\
%		Supersonic & 0.0969 & 0.0982 & 0.0982 & $-2\cdot 10^{-5}$ & 0.001 & 0.001\\ %Cl_FCFV = -2.09e-5, but usually only 3 lift counts reported
%		\hline
%	\end{tabular}
%	\caption{Inviscid flow cases over a NACA 0012 aerofoil -- Drag and lift coefficients computed for the subsonic, transonic and supersonic cases, using the FCFV and the first-order (Fluent-1) and second-order (Fluent-2) CCFV solvers by Ansys Fluent.} 
%	\label{tb:NACAinviscid_Forces}
%\end{table}
%
\begin{table}[!ht]
	\centering
		\hspace*{-12pt}\begin{tabular}{|l|c|lll|c|}
		\hline 
		Case & QoI & FCFV & Fluent-1 & Fluent-2 & Reference value \\
		\hline
		\multirow{2}{*}{Subsonic} & $C_{d}$ & 0.0059 & 0.0070 & 0.0010 & 0 \\
		& $C_{l}$ & 0.304 & 0.313 & 0.333 & $[0.327,0.330]$ \\
		\hline 
		Refs. & \multicolumn{5}{c|}{\citep{Sevilla-SHM:2013,Nogueira2009}} \\
		\hline \hline
		\multirow{2}{*}{Transonic} & $C_{d}$ & 0.0236 & 0.0260 & 0.0224 & $[0.0215,0.0229]$  \\
		& $C_{l}$  & 0.310 & 0.331 & 0.365 & $[0.345,0.353]$ \\
		\hline 
		\multirow{2}{*}{Refs. } & \multicolumn{5}{c|}{\citep{Sevilla-SHM:2013,Thibert-TGO:1979}} \\
		& \multicolumn{5}{c|}{\citep{Balan2015,Balan2012,Yano2012}} \\
		\hline \hline
		\multirow{2}{*}{Supersonic} & $C_{d}$ & 0.0967 & 0.0977 & 0.0956 & 0.0963 \\
%		& $C_{l}$ & $1.45\cdot 10^{-4}$ & $8.49\cdot 10^{-4}$ & $2.68\cdot 10^{-4}$ & 0 \\ 
		& $C_{l}$ & 0.145e-03 & 0.849e-03 & 0.268e-03 & 0 \\ 
%Cl_FCFV = -2.09e-5, but usually only 3 lift counts reported
		\hline
		Ref. & \multicolumn{5}{c|}{\citep{Balan2015}} \\
		\hline
	\end{tabular}
	\caption{\hl{Inviscid flow cases over a NACA 0012 aerofoil -- Drag ($C_d$) and lift ($C_l$) coefficients computed for the subsonic, transonic and supersonic cases, using the FCFV, the first-order (Fluent-1) and second-order (Fluent-2) CCFV solvers by Ansys Fluent.}}
	\label{tb:NACAinviscid_Forces}
\end{table}

\hl{
The test cases discussed above confirm the robustness of the FCFV scheme across a wide range of regimes of inviscid flows.  In addition, the method showcases a particular superiority with respect to Ansys Fluent CCFV solvers in supersonic flows with strong shocks.  For subsonic and transonic cases with smooth flowfields or less abrupt variations,  the best accuracy is provided by CCFV schemes, especially by the second-order solver.
}

%------------------------------------------------------------------------------------------------------------
\subsection{Nearly incompressible viscous laminar flow over a flat plate}

In this section,  the viscous laminar flow over a flat plate at zero angle of attack with $\Ma_{\! \infty} = 0.1$ and $\Rey=10^5$ is analysed to assess the robustness and the accuracy of the FCFV method in the nearly incompressible limit, benchmarking the results using Blasius' law~\citep{Blasius1908} and comparing the performance with Ansys Fluent pressure-based CCFV solver.

\begin{figure}[!ht]
	\centering
	\includegraphics[width=0.75\textwidth]{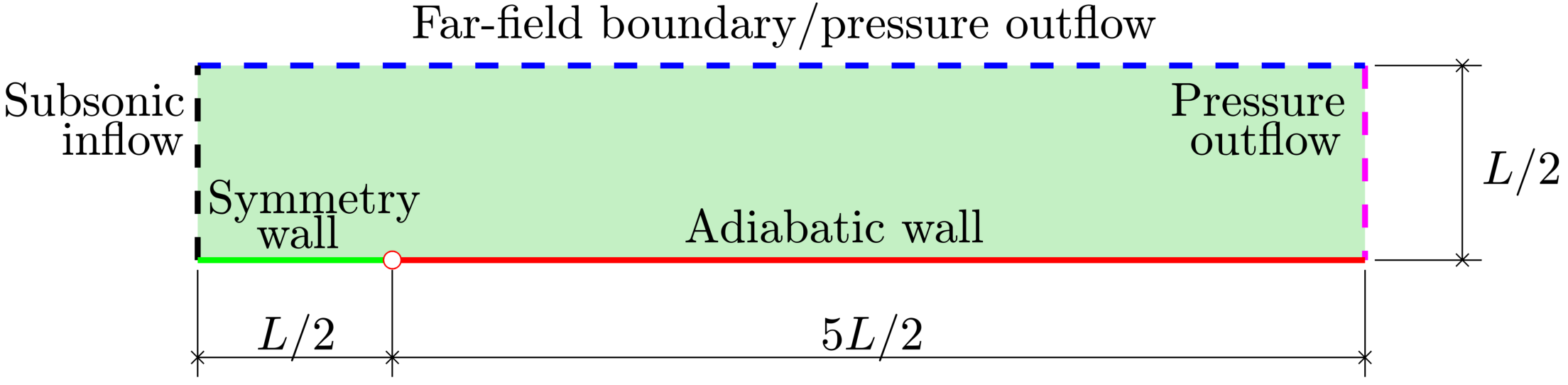}
	\caption{Nearly incompressible viscous laminar flow over a flat plate -- Domain and boundary conditions.} 
	\label{fig:Flatplate_domain}
\end{figure}

\hl{An adiabatic flat plate of length $2.5L$ is embedded in the rectangular domain in figure~\ref{fig:Flatplate_domain}. Here, $L$ denotes the characteristic length used for the nondimensionalisation of the problem, leading to the aforementioned value of the Reynolds number. It is worth recalling that, for a smooth flat plate at zero angle of attack,  turbulent effects start to appear at a critical Reynolds number $\Rey_{cr} = 5 \cdot 10^5$~\citep{Schlichting-BL:2016},  that is, at a distance of $5L$ from the leading edge. In order to guarantee that the flow remains laminar along the entire plate,  a total length of $2.5L$ is considered.}

A symmetry condition is imposed upstream of the leading edge,  whereas the top boundary is modelled by means of a far-field condition in Ansys Fluent, imposing the free-stream Mach number $\Ma_{\! \infty}$,  and a pressure outflow in the FCFV method, setting the corresponding free-stream pressure $p_\infty$. Finally, subsonic inflow and pressure outflow conditions are prescribed on the left and right boundaries, respectively. In particular, the condition at the inlet is enforced via the Riemann solver and the pressure $p_\infty$ guarantees a zero pressure drop at the outlet.

\begin{figure}[!ht]
	\centering
	\subfloat[Structured mesh \label{fig:Flatplate_MeshStruct}]{\includegraphics[width=0.5\textwidth]{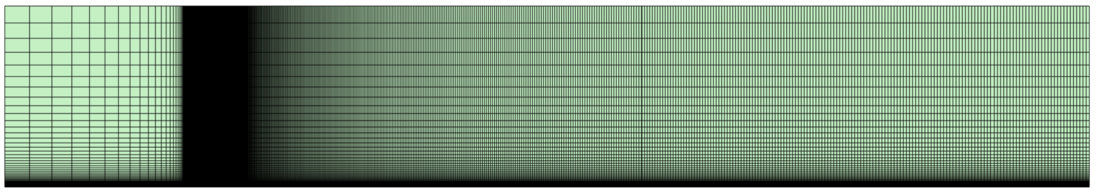}} 
	\subfloat[Unstructured mesh \label{fig:Flatplate_MeshUnstruct}]{\includegraphics[width=0.5\textwidth]{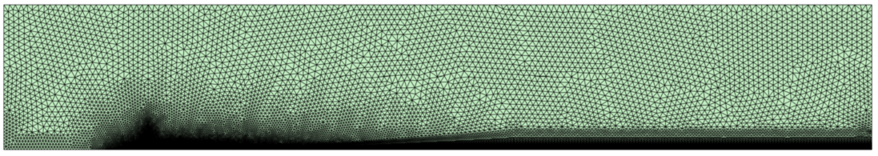}}

	\subfloat[Boundary layer,  \\ structured mesh \label{fig:Flatplate_MeshStructBL}]{\includegraphics[width=0.25\textwidth]{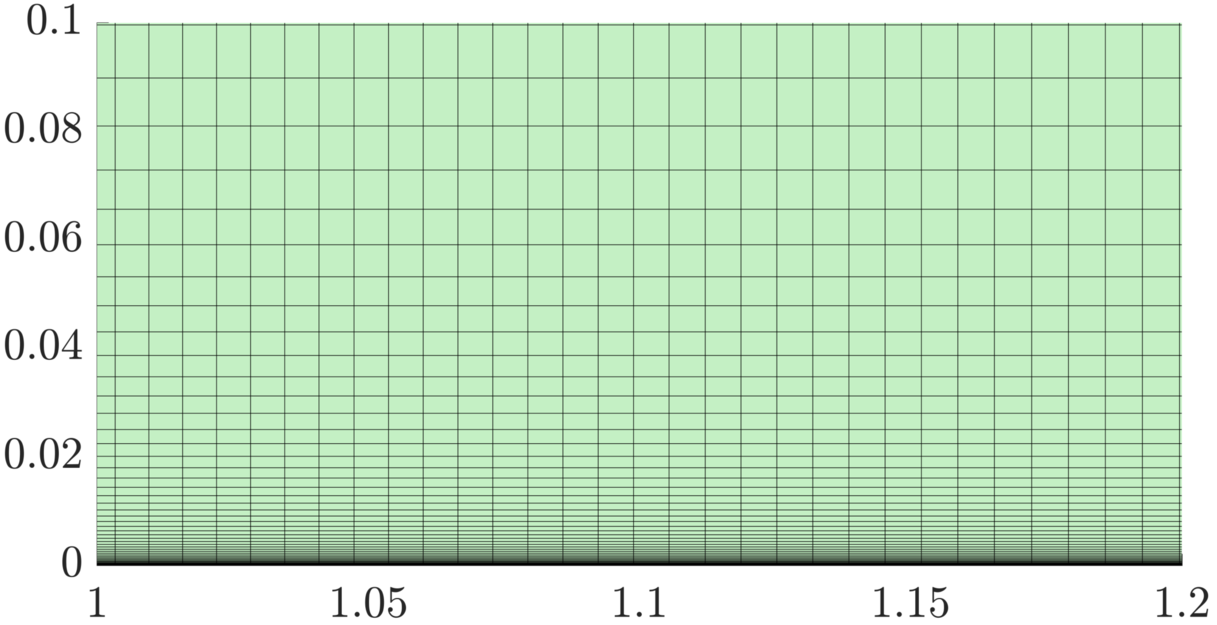}}	
	\subfloat[Leading edge,  \\ structured mesh \label{fig:Flatplate_MeshStructLE}]{\includegraphics[width=0.25\textwidth]{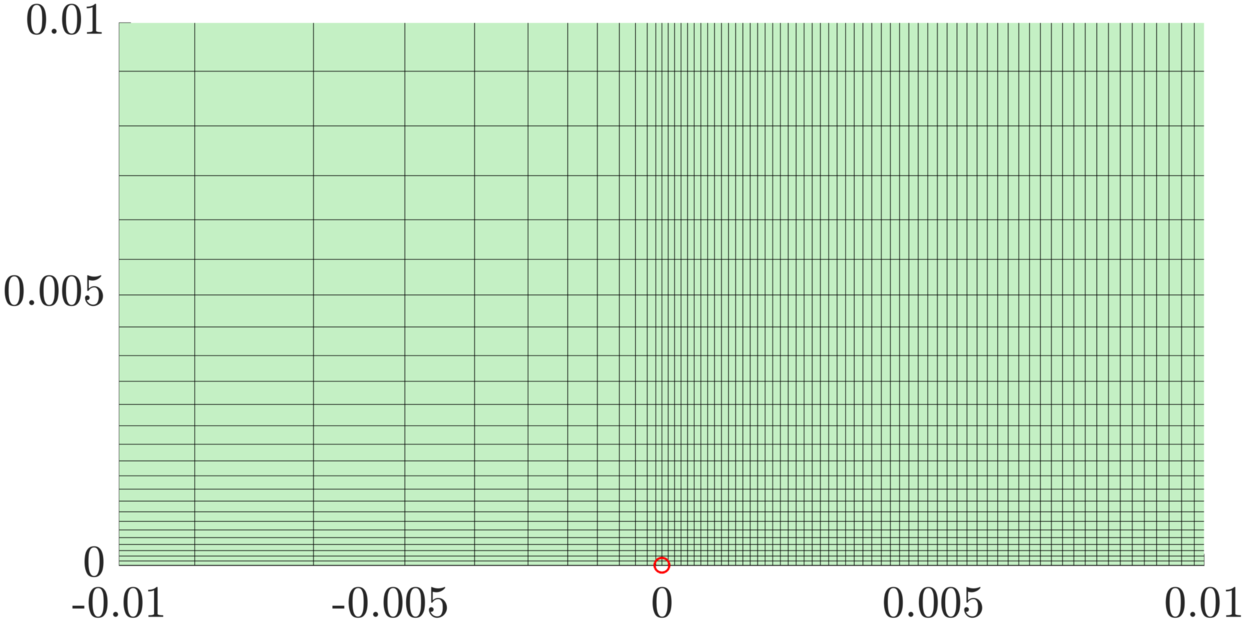}} 
	\subfloat[Boundary layer,  \\ unstructured mesh \label{fig:Flatplate_MeshUnstructBL}]{\includegraphics[width=0.25\textwidth]{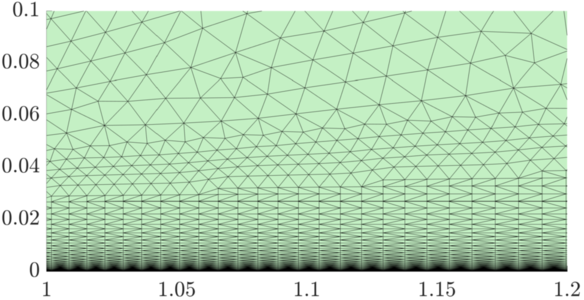}}
	\subfloat[Leading edge,  \\ unstructured mesh \label{fig:Flatplate_MeshUnstructLE}]{\includegraphics[width=0.25\textwidth]{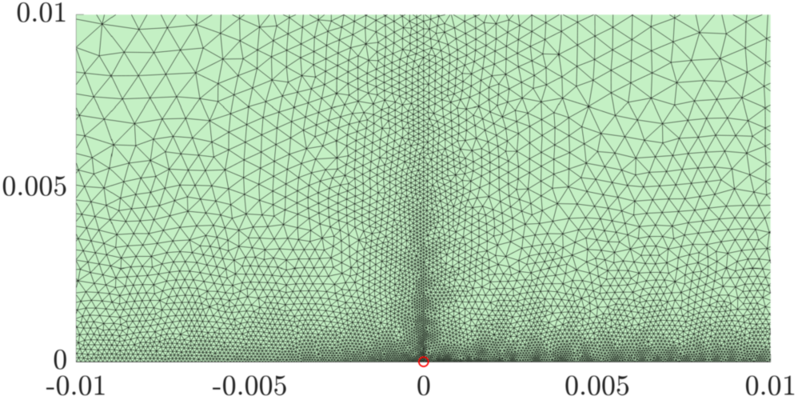}} 
	\caption{Nearly incompressible viscous laminar flow over a flat plate -- (a) Structured mesh of quadrilaterals, (b) unstructured mesh of triangles and (c-f) details of the boundary layer and leading edge regions.} 
	\label{fig:Flatplate_Mesh}
\end{figure}

The structured mesh of quadrilateral cells and the unstructured mesh of triangles displayed in figure~\ref{fig:Flatplate_Mesh} are employed for this study. In both cases, the size of the first layer of cells is of order $10^{-4} < \Rey^{-3/4}$, with an aspect ratio of approximately 100. The resulting structured grid in figure~\ref{fig:Flatplate_MeshStruct} consists of 80,500 cells, with a uniform stretching in the normal direction to the plate and towards the leading edge, as shown in figures~\ref{fig:Flatplate_MeshStructBL} and~\ref{fig:Flatplate_MeshStructLE}.  Figure~\ref{fig:Flatplate_MeshUnstruct} depicts the unstructured mesh composed by 101,751 triangular cells: it features a structured region in the boundary layer (Fig.~\ref{fig:Flatplate_MeshUnstructBL}), whereas a specific refinement is performed in the vicinity of the leading edge to capture the geometric singularity, as visible in figure~\ref{fig:Flatplate_MeshUnstructLE}.
%
%\begin{figure}[!ht]
%	\centering
%	\subfloat[Mesh, structured \label{fig:Flatplate_MeshStruct}]{\includegraphics[width=0.7\textwidth]{FlatPlateStructuredMesh}} \\
%	\subfloat[Leading edge, structured \label{fig:Flatplate_MeshStructLE}]{\includegraphics[width=0.325\textwidth]{FlatPlateStructuredMeshLE}} \qquad
%	\subfloat[Boundary layer, structured \label{fig:Flatplate_MeshStructBL}]{\includegraphics[width=0.325\textwidth]{FlatPlateStructuredMeshBL}}
%	\caption{Laminar flow over a flat plate -- Structured mesh of quadrilaterals considered for the study.} 
%	\label{fig:Flatplate_MeshStructured}
%\end{figure}
%%
%\begin{figure}[!ht]
%	\centering
%	\subfloat[Mesh, unstructured \label{fig:Flatplate_MeshUnstruct}]{\includegraphics[width=0.7\textwidth]{FlatPlateUnstructuredMesh}}\\
%	\subfloat[Leading edge, unstructured \label{fig:Flatplate_MeshUnstructLE}]{\includegraphics[width=0.325\textwidth]{FlatPlateUnstructuredMeshLE}} \qquad
%	\subfloat[Boundary layer, unstructured \label{fig:Flatplate_MeshUnstructBL}]{\includegraphics[width=0.325\textwidth]{FlatPlateUnstructuredMeshBL}}
%	\caption{Laminar flow over a flat plate -- Unstructured mesh of triangles considered for the study.} 
%	\label{fig:Flatplate_MeshUnstructured}
%\end{figure}
%
%

\begin{figure}[!ht]
	\centering
	\includegraphics[width=0.675\textwidth]{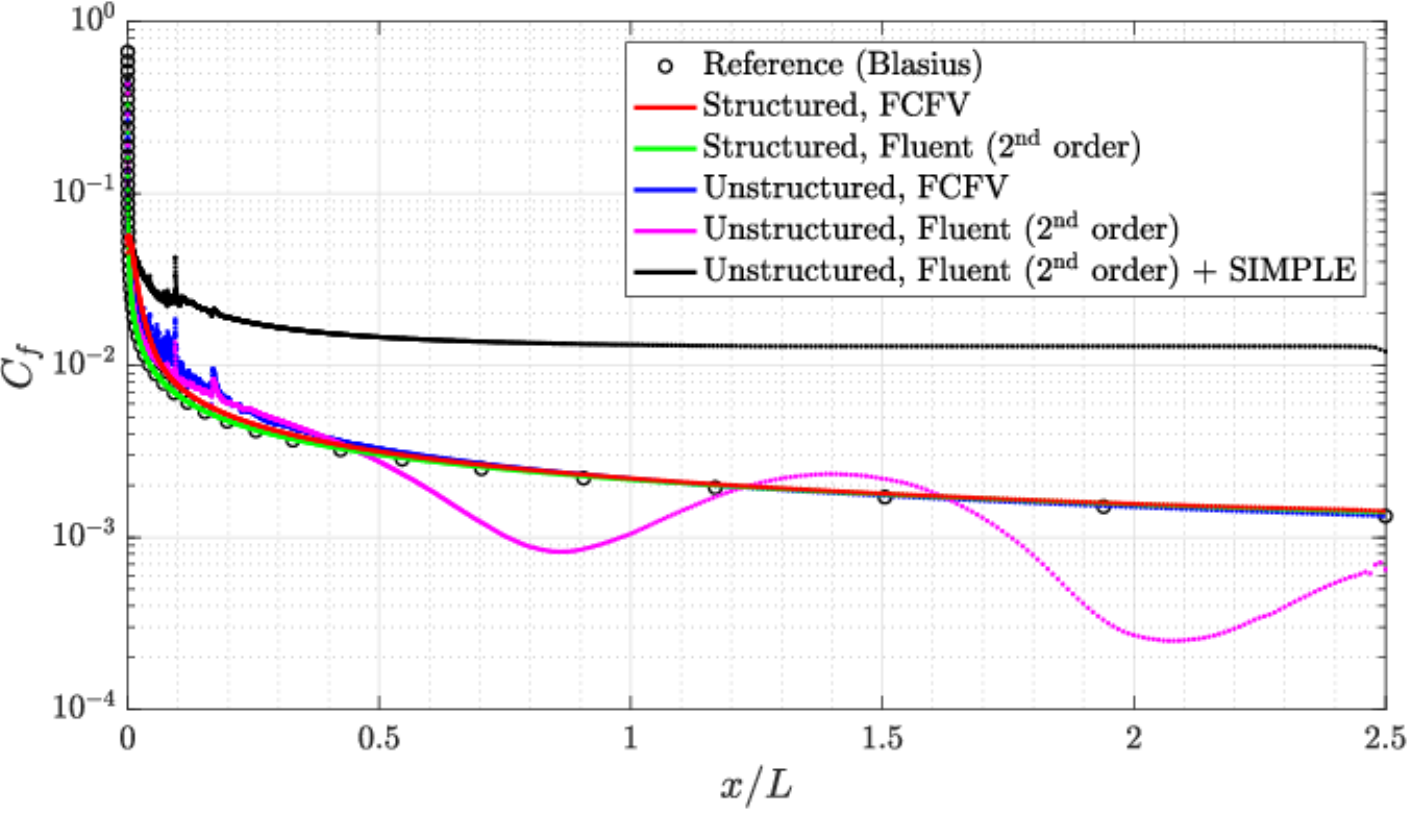}
	\caption{Nearly incompressible viscous laminar flow over a flat plate -- Skin friction coefficient along the flat plate, obtained for the FCFV and the CCFV methods by Ansys Fluent.} 
	\label{fig:Flatplate_Cf}
\end{figure}

The skin friction coefficient along the flat plate is compared to the reference value given by the analytical boundary layer solution by Blasius for incompressible flows~\citep{Blasius1908}. The results in figure~\ref{fig:Flatplate_Cf} show that both the FCFV method and the second-order CCFV scheme by Ansys Fluent provide an accurate description of the viscous boundary layer using the structured grid. More precisely,  almost comparable accuracy is achieved by the two approaches, with the CCFV method slightly outperforming the FCFV scheme in the neighbourhood of the leading edge. 
Using the unstructured mesh, the FCFV method is still able to compute a reasonable approximation of the skin friction coefficient, although some oscillations appear in the region following the leading edge.  Indeed, the unstructured nature of the mesh in this region is responsible for the distance between the centroid of the first layer of cells and the plate to be non-uniform. Since the skin friction coefficient is computed using the mixed variable, defined at the centroid of the cells,  and this information is not extrapolated to the boundary,  slight oscillations appear in this area. On the contrary, Ansys Fluent second-order CCFV scheme is unable to converge to a stable steady-state solution on the unstructured mesh, showing the diverging oscillatory behaviour reported in figure~\ref{fig:Flatplate_Cf} for a fixed iteration. In order to remedy this issue, the Ansys Fluent solver resorts to the SIMPLE algorithm, introducing a velocity-pressure splitting to achieve convergence. Nonetheless, the solution obtained in this case overestimates the skin friction coefficient, approximately by one order of magnitude.

The results of the flow over a flat plate show the robustness of the FCFV solver in the incompressible limit, independently of the cell type and the quality of the employed mesh. The FCFV scheme is able to provide an accurate approximation of aerodynamic quantity of interest in all tested cases, whereas the CCFV method by Ansys Fluent is restricted to structured meshes of quadrilaterals. More precisely, the CCFV solver is unable to converge to a steady-state solution when the unstructured mesh is employed, requiring a pressure correction based on the SIMPLE algorithm. Nonetheless, the results provided by Ansys Fluent in this case display an error of one order of magnitude with respect to Blasius' reference solution.

%------------------------------------------------------------------------------------------------------------
\subsection{Supersonic viscous laminar flow over a cylinder}

The final test case considers the supersonic viscous laminar flow over a cylinder at $\Ma_{\! \infty}=4$ and $\Rey=10^4$~\citep{Barter2007}. This example represents a particularly challenging benchmark for viscous compressible flow solvers due to the presence of a strong bow shock, together with a thin boundary layer producing important heat transfer effects near the cylinder wall.

\begin{figure}[!ht]
	\centering
	\subfloat[Domain and boundary conditions]{\includegraphics[width=0.6\textwidth]{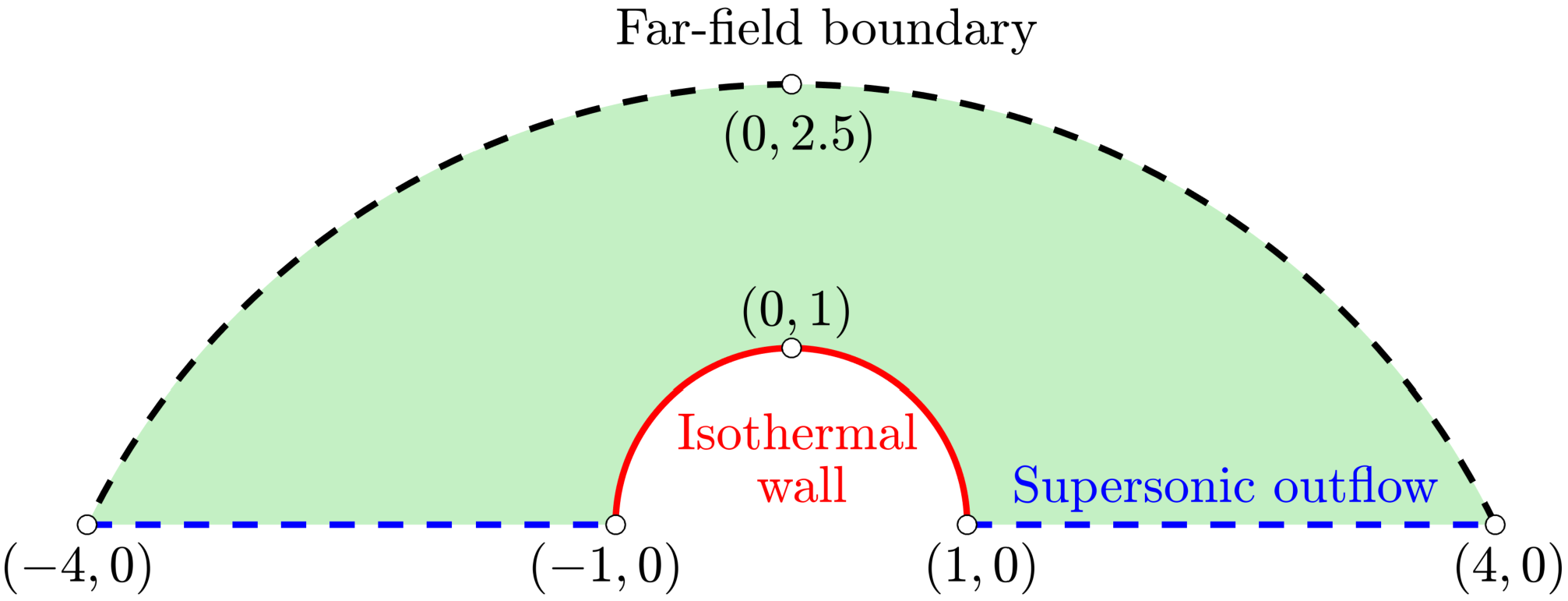} \label{fig:Cylinder_Domain}}
	
	\subfloat[Mesh]{\includegraphics[width=0.6\textwidth]{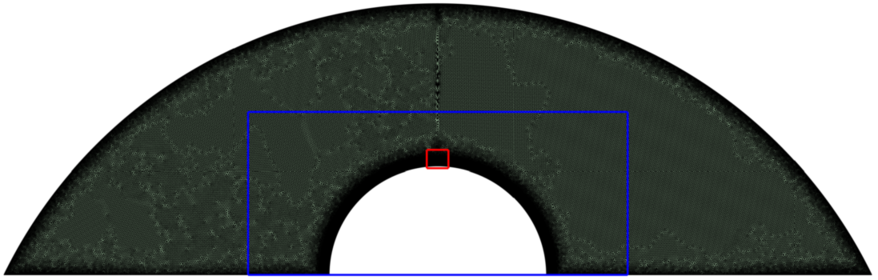} \label{fig:Cylinder_Meshes}}
	
	\subfloat[Close-up view around the cylinder]{\includegraphics[width=0.415\textwidth]{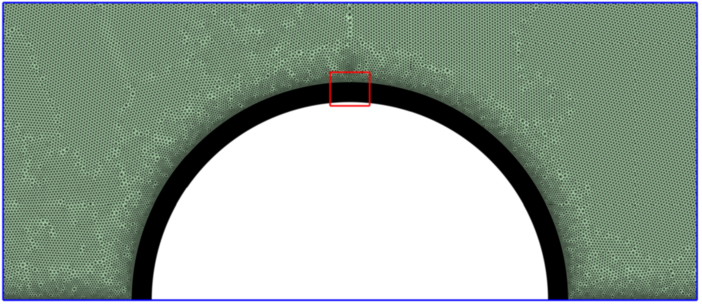}} \qquad
	\subfloat[Boundary layer]{\includegraphics[width=0.215\textwidth]{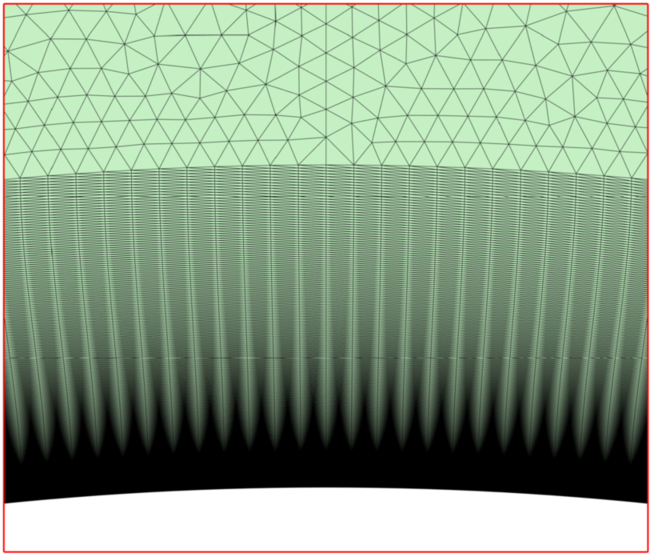}}
	\caption{Supersonic viscous laminar flow over a cylinder -- (a) Domain and boundary conditions,  (b) mesh and (c-d) details of the unstructured and structured regions.} 
	\label{fig:Cylinder_Supersonic}
\end{figure}

The domain features a half-cylinder of radius $L$ (the characteristic length of the problem), with isothermal boundary conditions imposing a wall temperature of 2.5 times the free-stream temperature.  Far-field conditions are enforced at the inflow, whereas a supersonic flow is assumed at the outflow, as indicated in figure~\ref{fig:Cylinder_Domain}.
An unstructured mesh of 468,854 triangular cells with a structured boundary layer region is constructed (Fig.~\ref{fig:Cylinder_Meshes}).
%
%\begin{figure}[!ht]
%	\centering
%	\includegraphics[width=0.65\textwidth]{HCsupersonic_Domain}
%	\caption{Supersonic viscous laminar flow over a cylinder -- Domain and boundary conditions.} 
%	\label{fig:Cylinder_Domain}
%\end{figure}
%%
%\begin{figure}[!ht]
%	\centering
%	\subfloat[Computational mesh.]{\includegraphics[width=0.65\textwidth]{HCsupersonic_mesh}} \\
%	\subfloat[Close-up view around the cylinder.]{\includegraphics[width=0.425\textwidth]{HCsupersonic_meshZoom1}} \qquad
%	\subfloat[Boundary layer region.]{\includegraphics[width=0.215\textwidth]{HCsupersonic_meshZoom2}}\\
%	\caption{Supersonic viscous laminar flow over a cylinder -- Computational mesh and details of the unstructured and structured regions.} 
%	\label{fig:Cylinder_Meshes}
%\end{figure}

\begin{figure}[!ht]
	\subfloat[FCFV, Mach]{\includegraphics[width=0.48\textwidth]{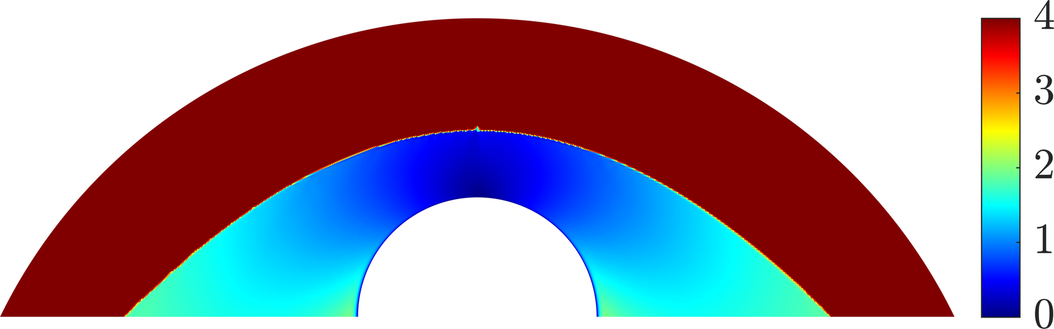}} \hfill
	\subfloat[Fluent-1, Mach]{\includegraphics[width=0.48\textwidth]{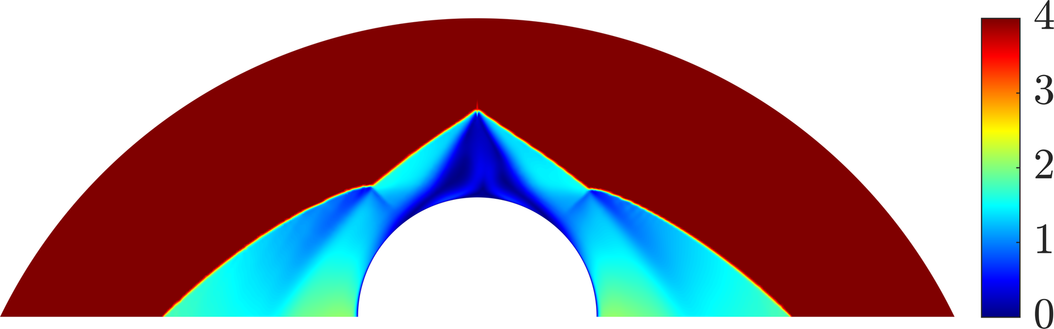}} \\
	\subfloat[FCFV, pressure]{\includegraphics[width=0.48\textwidth]{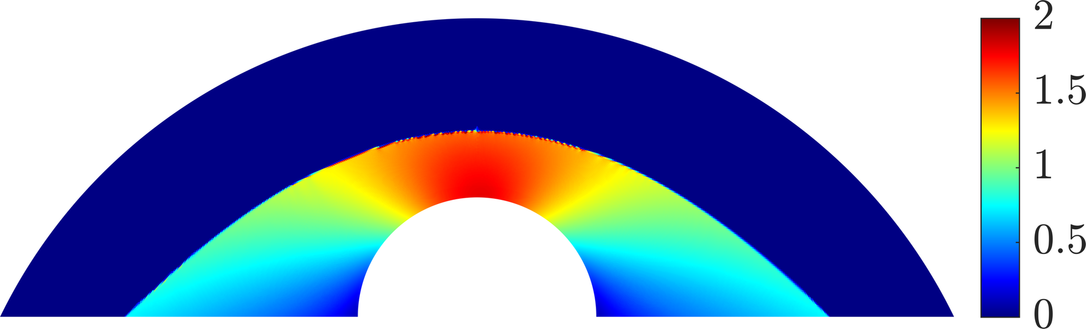}} \hfill
	\subfloat[Fluent-1, pressure]{\includegraphics[width=0.48\textwidth]{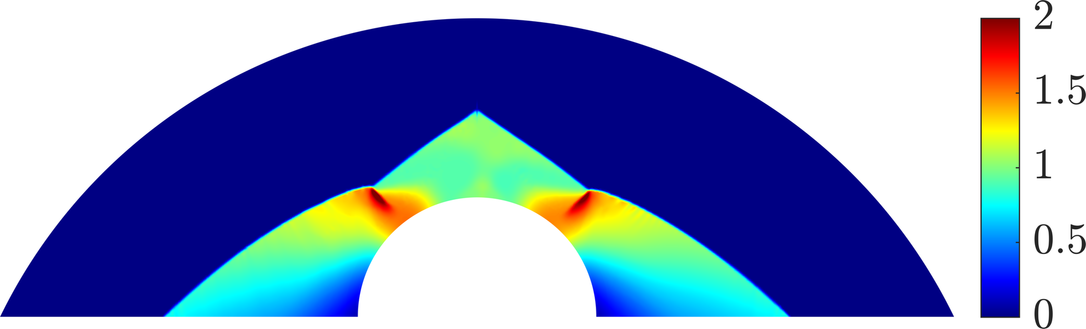}} \\
	\subfloat[FCFV, temperature]{\includegraphics[width=0.48\textwidth]{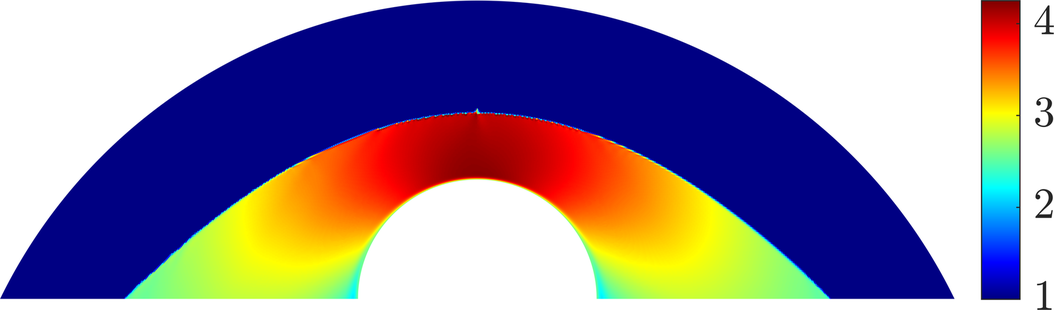}} \hfill
	\subfloat[Fluent-1, temperature]{\includegraphics[width=0.48\textwidth]{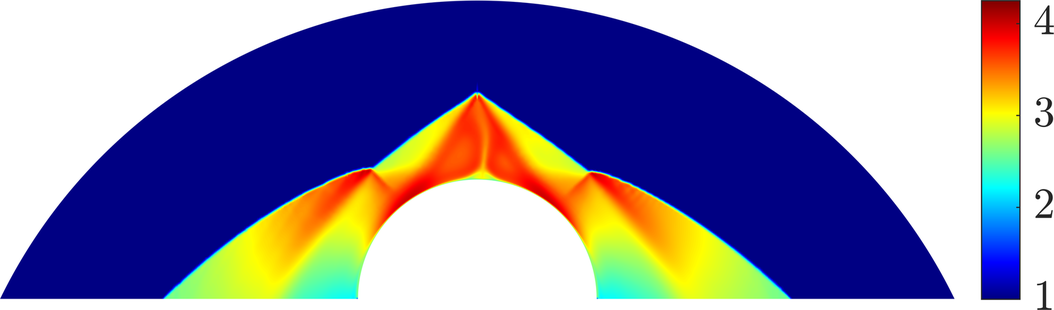}}
	\caption{Supersonic viscous laminar flow over a cylinder -- Mach number, pressure and temperature distributions computed using the FCFV and the first-order (Fluent-1) CCFV solver by Ansys Fluent.} 
	\label{fig:Cylinder_Fields}
\end{figure}

The problem is solved using the FCFV method and the first-order density-based CCFV solver by Ansys Fluent and the results are compared to a reference solution consisting of a high-order discontinuous Galerkin (DG) approximation of polynomial degree 3, computed on a structured mesh of 16,000 triangular elements~\citep{Barter2007}. Both FV approaches are equipped with positivity-preserving Riemann solvers, namely an HLL flux for FCFV and an advection upstream splitting method (AUSM) flux for CCFV.  The FCFV results (Fig.~\ref{fig:Cylinder_Fields}, left) show a good qualitative description of the flowfield for all physical quantities, capturing both the strong bow shock and the steep temperature gradient near the wall. It is worth noticing that the use of unstructured meshes such as the one in figure~\ref{fig:Cylinder_Meshes} is particularly challenging for high Mach number flows because numerical artifacts leading to asymmetries of the solution tend to appear, due to the curvature of the streamlines along the stagnation line~\citep{Nompelis2004,Gnoffo2004,Ching2019}. The FCFV method is able to preserve the symmetry of the solution, confirming its accuracy and robustness, even for $\Ma_{\! \infty}=4$. 
For Ansys Fluent CCFV solver,  a relaxation approach based on an artificial time step,  with CFL number between 0.1 and 0.5,  is employed to achieve a steady-state solution.  The first-order solver with the relaxation approach, the positivity-preserving AUSM scheme and a flux limiter provides the steady-state solution on the right of figure~\ref{fig:Cylinder_Fields},  exhibiting a loss of accuracy due to the carbuncle phenomenon. This is caused by a lack of dissipation of the numerical discretisation, leading to numerical instabilities in the shock, with a nonphysical peak in the normal region to the bow shock~\citep{Elling2009,Kitamura2012,Pandolfi2001,Chauvat2005,Kitamura2013}. A common approach to remedy this issue is to employ different numerical fluxes in the simulation. Nonetheless, the alternative option provided by Ansys Fluent consists of the Roe flux, which is prone to develop the carbuncle phenomenon in the presence of strong bow shocks.  Indeed, different combinations of CCFV solvers with the Roe flux and various time-stepping strategies, not reported here for brevity, were tested, all leading to unstable results. Moreover, it is worth noticing that the second-order CCFV scheme was unable to converge in this problem.

\begin{figure}[!ht]
	\subfloat[Pressure coefficient]{\includegraphics[width=0.32\textwidth]{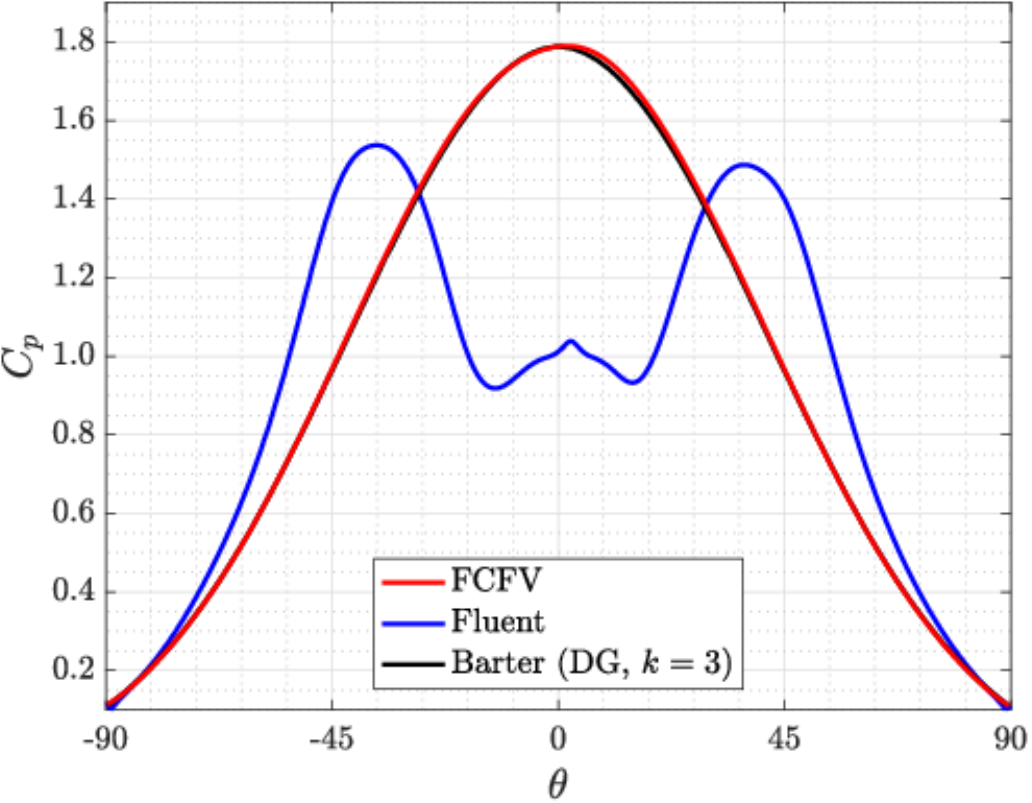} \label{fig:Cylinder_CoefficientPressure}} \hfill
	\subfloat[Skin friction coefficient]{\includegraphics[width=0.32\textwidth]{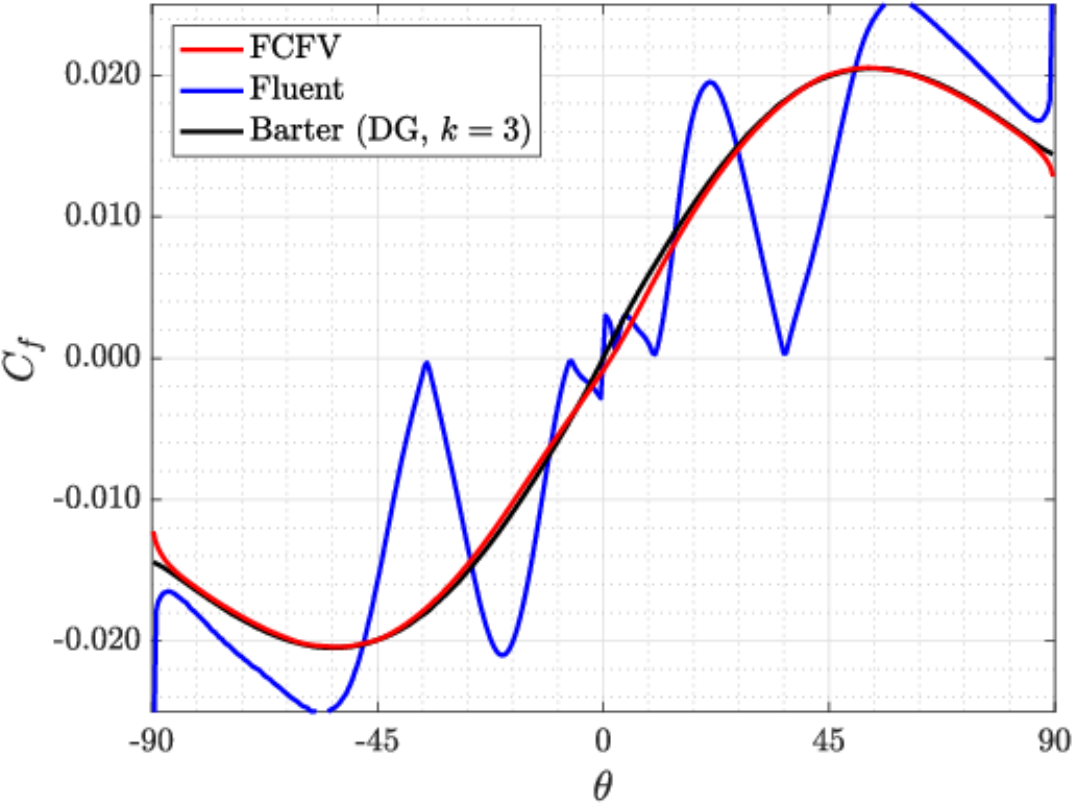} \label{fig:Cylinder_CoefficientSkinFrict}} \hfill
	\subfloat[Stanton number]{\includegraphics[width=0.32\textwidth]{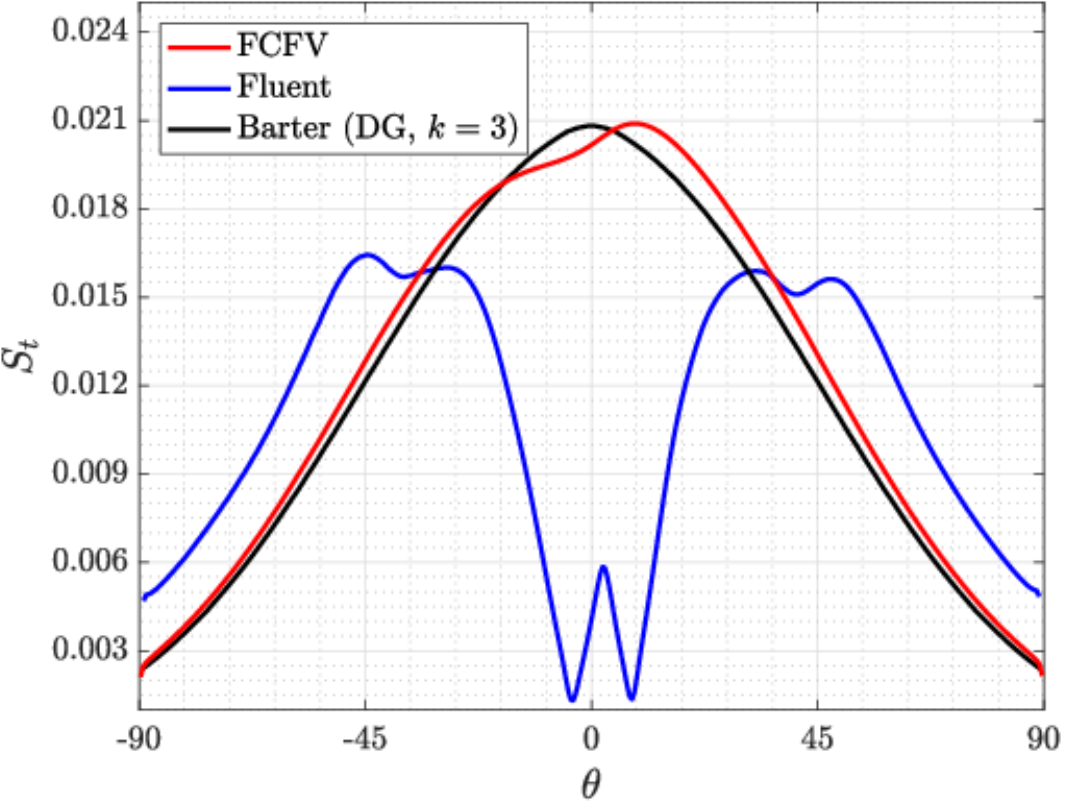} \label{fig:Cylinder_CoefficientStanton}}\\
	\caption{Supersonic viscous laminar flow over a cylinder -- Pressure, skin friction and Stanton/heat transfer coefficients along the cylinder surface obtained for the FCFV and the first-order CCFV approach by Ansys Fluent.} 
	\label{fig:Cylinder_Coefficients}
\end{figure}

Finally, a quantitative assessment of the accuracy of the FCFV solution is performed comparing the wall quantities with the reference results in~\citep{Barter2007}. Figures~\ref{fig:Cylinder_CoefficientPressure} and~\ref{fig:Cylinder_CoefficientSkinFrict} showcase excellent agreement of the FCFV approximation of the pressure and the skin friction coefficient with the reference solution, with $\eltwo$ errors of $0.7 \%$ and $2.9 \%$, respectively. Moreover, the Stanton number (or heat transfer coefficient) is displayed in figure~\ref{fig:Cylinder_CoefficientStanton}: for this quantity, the FCFV method provides an overall similarity of the profile but a certain discrepancy is observed in the neighbourhood of the leading edge. In particular, the asymmetry of the FCFV solution is probably influenced by the unstructured nature of the mesh in the region outside the boundary layer, see figure~\ref{fig:Cylinder_Meshes}, whereas the reference solution is computed using a symmetric grid. Nonetheless, the relative $\eltwo$ error of the overall approximation is $5.0 \%$. On the contrary, the carbuncle phenomenon in the results yielded by Ansys Fluent leads to completely erroneous predictions of all the quantities of interest under analysis.

%------------------------------------------------------------------------------------------------------------
%------------------------------------------------------------------------------------------------------------
\section{Concluding remarks}
\label{sc:Conclusions}

An extensive benchmark study of the FCFV method was presented for compressible laminar flows across a variety of regimes, from inviscid to viscous laminar flows, from subsonic to transonic and supersonic flows. The assessment of the numerical properties of the method was performed via a qualitative analysis of the distribution of the physical variables and a quantitative analysis of aerodynamic quantities of interest, including drag, lift, pressure, skin friction and heat transfer coefficients. The FCFV results were compared to reference values published in the literature, as well as to the solution provided by the CCFV solvers available in the commercial CFD software Ansys Fluent.

More precisely, the compressible Taylor-Couette flow was employed to assess the convergence properties of the FCFV method using both regular and distorted meshes. The results display that the method provides first-order convergence of the stress tensor, the heat flux, the primitive variables (density, velocity, temperature and pressure) and the conservative variables (momentum and energy), showing the insensitiveness of this approach to cell distortion. On the contrary, the CCFV solver by Ansys Fluent outperforms the FCFV method in the approximation of the velocity and the temperature using structured regular meshes thanks to the reconstruction of the gradient. Nonetheless, such an advantage is lost when distorted grids are employed and the accuracy of the CCFV scheme deteriorates, especially in the approximation of the stress tensor and the heat flux. This is particularly critical in the evaluation of quantities of engineering interest, such as drag, lift and heat transfer coefficients, which involve the gradient of the flow variables.

To further study the robustness and accuracy properties of the FCFV method, a set of viscous and inviscid flows over a NACA 0012 aerofoil was presented. On the one hand, the viscous cases confirmed the capability of the FCFV scheme to accurately predict quantities of engineering interest,  even in the presence of highly stretched meshes in the boundary layer region. Indeed, the method provided results comparable to Ansys Fluent first-order CCFV solver, whereas the second-order one outperformed the remaining two approaches on coarse meshes.  On the other hand, the results of the inviscid tests highlighted a stronger sensitivity of the FV methods under analysis to the value of the Mach number. \hl{More precisely, although for subsonic and transonic flows Ansys Fluent CCFV solvers outperformed the FCFV method, the latter provided more accurate and robust results in the presence of strong bow shocks appearing in supersonic flows.}

The superiority of the FCFV method for high Mach number simulations was also observed in the viscous case, with the laminar flow over a cylinder at Mach 4. In this context, the FCFV method showed its superior performance in terms of accuracy and robustness by providing an accurate description of the flowfield and a prediction of the engineering quantities of interest with errors below $5 \%$, even when unstructured meshes were employed outside the boundary layer. This problem is especially challenging because the combination of unstructured meshes and high Mach number is known to yield the appearance of numerical artifacts due to the carbuncle phenomenon suffered by many discretisation methods, including Ansys Fluent CCFV solvers.

Finally,  the FCFV method was shown to be robust also in the incompressible limit, independently of the type of computational mesh. 
Although classical CCFV schemes provide excellent results using structured grids, the quality of the approximation greatly deteriorates when unstructured meshes are employed, leading to oscillatory solutions unable to converge to a steady-state result. To remedy this issue the solver provided by Ansys Fluent relies on pressure correction techniques to retrieve stable solutions. Nonetheless, the resulting approximation displays an error of one order of magnitude with respect to Blasius' solution, whereas the FCFV method provides excellent agreement with the analytical solution.

To summarise, the FCFV method demonstrates a robust performance in a wide variety of flow conditions, providing accurate solutions on general unstructured meshes, insensitively to cell distortion and stretching. The presented results showcase the suitability of the method to treat industrial flow problems with complex geometries,  relaxing the restrictions of mesh quality imposed by existing FV solvers and alleviating the need for time-consuming manual mesh generation procedures performed by specialised technicians.  Future studies will investigate tailored solution strategies for the FCFV simulation of large-scale systems and turbulent phenomena.

%------------------------------------------------------------------------------------------------------------
\section*{Acknowledgements}
%------------------------------------------------------------------------------------------------------------
This work was supported by the Spanish Ministry of Science and Innovation and the Spanish State Research Agency MCIN/AEI/10.13039/501100011033 (PID2020-113463RB-C33 to M.G., PID2020-113463RB-C32 to A.H., CEX2018-000797-S to A.H. and M.G.).  M.G. also acknowledges the support of the Generalitat de Catalunya through the Serra H\'unter Programme.

%------------------------------------------------------------------------------------------------------------
%------------------------------------------------------------------------------------------------------------
%------------------------------------------------------------------------------------------------------------
%\bibliographystyle{abbrv}
%\bibliographystyle{unsrt}
\bibliographystyle{apsr}
\bibliography{FCFV-Fluent}
%%\printbibliography

%------------------------------------------------------------------------------------------------------------
\appendix
%------------------------------------------------------------------------------------------------------------

%------------------------------------------------------------------------------------------------------------
%------------------------------------------------------------------------------------------------------------
%\section{}

\end{document}